\documentclass[]{aa}

\usepackage{graphicx}
\usepackage{txfonts}
\usepackage[svgnames]{xcolor}
\usepackage[colorlinks=true, linkcolor = blue, urlcolor = magenta,hyperfootnotes=false]{hyperref}

\newcommand{\TROY}{$\mathcal{T}\kern-0.12em\mathcal{R}\kern-0.12em\mathcal{O}\kern-0.12em\mathcal{Y}$}

\usepackage{rotating}
\usepackage{aalongtable}
\usepackage{lineoaa}
\usepackage{tablefootnote}
\usepackage{bm}
\usepackage{multirow}
\usepackage{comment}

\usepackage{pdflscape}

\usepackage{changepage}

\hypersetup{colorlinks,citecolor=blue}
\usepackage{natbib}
\bibpunct{(}{)}{;}{a}{}{,} 
\usepackage{subfigure}

\usepackage[cal=zapfc]{mathalfa}

\begin{document}

 \title{The \TROY project\\ III. Exploring co-orbitals around low-mass stars\thanks{Table \ref{tab:rvs} is only available in electronic form at the CDS via anonymous ftp to cdsarc.u-strasbg.fr (130.79.128.5) or via \url{http://cdsarc.u-strasbg.fr/viz-bin/qcat?J/A+A/}.}}

 \author{O.~Balsalobre-Ruza
  \inst{\ref{cab}}
  \and
  J.~Lillo-Box\inst{\ref{cab}}
  \and
  D.~Barrado\inst{\ref{cab}}
  \and
  A.~Correia\inst{\ref{cfisuc}, \ref{imcee}}
  \and
  J.~P.~Faria\inst{\ref{geneve}}
  \and
  P.~Figueira\inst{\ref{geneve}, \ref{porto}}
  \and
  A.~Leleu\inst{\ref{geneve}}
  \and
  P.~Robutel\inst{\ref{imcee}}
  \and
  N.~Santos\inst{\ref{porto}}
  \and
  E.~Herrero-Cisneros\inst{\ref{cab2}}
  }

 \institute{Centro de Astrobiología (CAB), CSIC-INTA, Camino Bajo del Castillo s/n, 28692, Villanueva de la Cañada, Madrid, Spain\\
 e-mail: {\tt obalsalobre@cab.inta-csic.es}\label{cab}
 \and
 CFisUC, Departamento de F\'isica, Universidade de Coimbra, 3004-516 Coimbra, Portugal\label{cfisuc}
 \and
 IMCCE, UMR8028 CNRS, Observatoire de Paris, PSL Universit\'e, 77 Av. Denfert-Rochereau, 75014\label{imcee}
 \and
 Observatoire Astronomique de l’Universit\'{e} de Gen\`{e}ve, Chemin Pegasi 51b, 1290 Versoix, Switzerland\label{geneve}
 \and
 Instituto de Astrof\'{i}sica e Ci\^{e}ncias do Espa\c{c}o, Universidade do Porto, CAUP, Rua das Estrelas, 4150-762 Porto, Portugal\label{porto}
 \and
 Centro de Astrobiología (CAB), CSIC-INTA, Crta. Ajalvir km 4, E-28850 Torrejón de Ardoz, Madrid, Spain\label{cab2}
 }
 \date{Received ; accepted }

  \abstract
 {Co-orbital objects, also known as trojans, are frequently found in simulations of planetary system formation. In these configurations, a planet shares its orbit with other massive bodies. It is still unclear why there have not been any co-orbitals  discovered thus far in exoplanetary systems (exotrojans) or even pairs of planets found in such a 1:1 mean motion resonance. Reconciling observations and theory is an open subject in the field.}
 {The main objective of the \TROY project is to conduct an exhaustive search for exotrojans using diverse observational techniques. In this work, we analyze the radial velocity time series informed by transits, focusing the search around low-mass stars.}
 {We employed the $\alpha$-test method on confirmed planets searching for shifts between spectral and photometric mid-transit times. This technique is sensitive to mass imbalances within the planetary orbit, allowing us to identify non-negligible co-orbital masses.}
 {Among the 95 transiting planets examined, we find one robust exotrojan candidate with a significant 3-$\sigma$ detection. Additionally, 25 exoplanets show compatibility with the presence of exotrojan companions at a 1-$\sigma$ level, requiring further observations to better constrain their presence. For two of those weak candidates, we find dimmings in their light curves within the predicted Lagrangian region. We established upper limits on the co-orbital masses for either the candidates and null detections.} 
 {Our analysis reveals that current high-resolution spectrographs effectively rule out co-orbitals more massive than Saturn around low-mass stars. This work points out to dozens of targets that have the potential to better constraint their exotrojan upper mass limit with dedicated radial velocity observations. We also explored the potential of observing the secondary eclipses of the confirmed exoplanets in our sample to enhance the exotrojan search, ultimately leading to a more accurate estimation of the occurrence rate of exotrojans.}

 \titlerunning{The \TROY project. III.}
 \maketitle
%

\section{Introduction}
\label{sec:intro}

Co-orbital configurations abound in the Solar System, where two massive bodies share their orbital path around the star. However, their detection beyond our system remains elusive. Two different mechanisms have been proposed to explain their formation, which are also expected to apply to exoplanetary systems: 1) in situ from the same material as protoplanets (e.g., \citealt{2007A&A...463..359B}) or 2) via resonant captures at more advanced stages (e.g., \citealt{2018camN}). Recent ALMA observations of three protoplanetary disks may be the first observational hints in favor of their in situ assembly, since they exhibit substructures that could be interpreted as an accumulation of material within the  L$_{\rm 4}$ \& L$_{\rm 5}$\,Lagrangian points of forming planets. In HD~163296 and LkCa~15, it has been found asymmetric emissions with shapes that are reminiscent of the dust traps that appear in the process of trojan formation, as seen in hydrodynamical simulations (\citealt{2018ApJ...869L..49I}; \citealt{2022ApJ...937L...1L}). An equivalent unresolved emission was found also in PDS~70 promisingly matching the L$_{\rm 5}$\,region of the protoplanet PDS~70\,b (\citealt{2023A&A...675A.172B}).

As posited by \citet{2002AJ....124..592L}, two co-orbiting similar-mass planets can be stable in the long term. Such stability remains possible as long as the total mass of the planet and the trojan does not surpass 3.7\% of their host star mass (i.e., $[\rm{m_p}+\rm{m_t}]/M_{\star}<1/27$). This constraint allows for multiple co-orbital configurations to remain longstanding, including pairs of equal-mass planets. Although this exotic configuration is absent in the Solar System, its theoretically hypothesized stability encourages the detectability of exotrojan pairs for the first time. Such co-orbital systems could be accessible using current instrumentation, even from the ground (e.g., \citealt{2006ApJ...652L.137F}; \citealt{2013CeMDA.117...75H}; \citealt{2015ApJ...811....1H}).

The absence of detections in exotrojan studies so far may be attributed to two non-exclusive reasons: i) co-orbital configurations might get disrupted earlier than theoretically predicted and ii) observational biases, as degeneracies with simpler configurations for the observed signals (see e.g., \citealt{2012MNRAS.421..356G}; \citealt{2006ApJ...647..573G}) or the existence of transit timing variations (hereafter TTVs, e.g., \citealt{2011MNRAS.417L..16G}; \citealt{2021A&A...655A..66L}) can prevent their detection. Dedicated studies are therefore needed to ascertain the presence of the 1:1 mean motion resonance (MMR) in planetary system. This is the purpose of the \TROY project\footnote{\url{www.troy-project.com}}, which this work builds on, following \citet{2018A&A...609A..96L, 2018A&A...618A..42L}.
Here, we extend the search for trojans accompanying confirmed exoplanets through a combination of radial velocities (RVs) and transit information. Former studies of these series were aimed at initiating  searches in systems where the expected signal of a co-orbital pair is maximized and therefore easily distinguishable from a single-planet configuration. That preliminary search allowed us to  estimate the upper limit of the occurrence of trojans for short-period (P$~<~$5\,d) giant planets. Also, it yielded nine candidates, although none proved significant. Here, we did not restrict the search to hot Jupiters, but we included any type of exoplanet orbiting low-mass stars (in the M and late-K spectral type regime). Despite hot Jupiters being easier targets from a detectability perspective, they are not necessarily the most suitable systems to host trojans due to the tidal forces of the star disrupting these configurations (\citealt{2021CeMDA.133...37C}; \citealt{2022Icar..38515087D}).

In Sect.~\ref{sec:sample} we explain the selection process for the new sample comprising 84 low-mass stars. Section~\ref{sec:meth} outlines the method used to test the co-orbital scenario, namely, the so-called $\alpha$-test. Section~\ref{sec:results} presents the results, followed by a discussion in Sect.~\ref{sec:dis}. The conclusions are provided in Sect.~\ref{sec:concl}.

\section{Target selection and data retrieval}
\label{sec:sample}

\begin{figure*}[h!]
  \begin{center}
 \subfigure{\includegraphics[width=86mm]{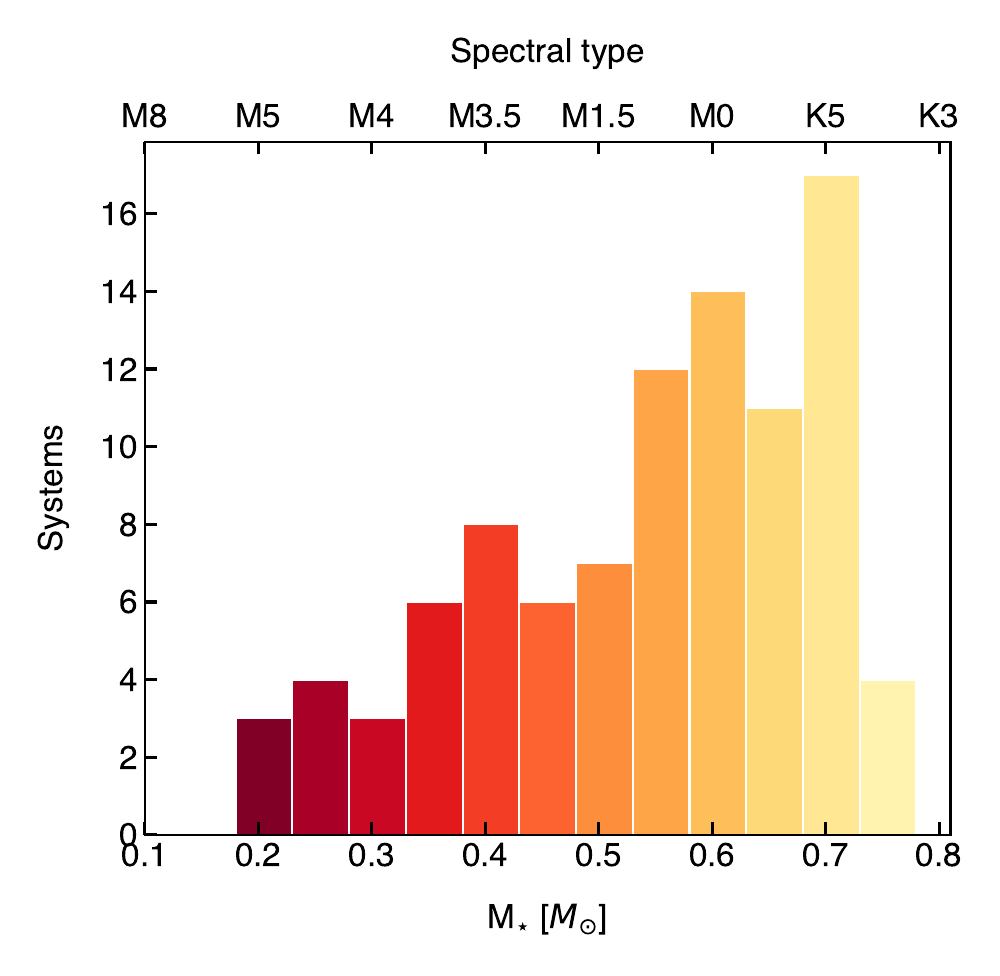}}
 \subfigure{\includegraphics[width=91mm]{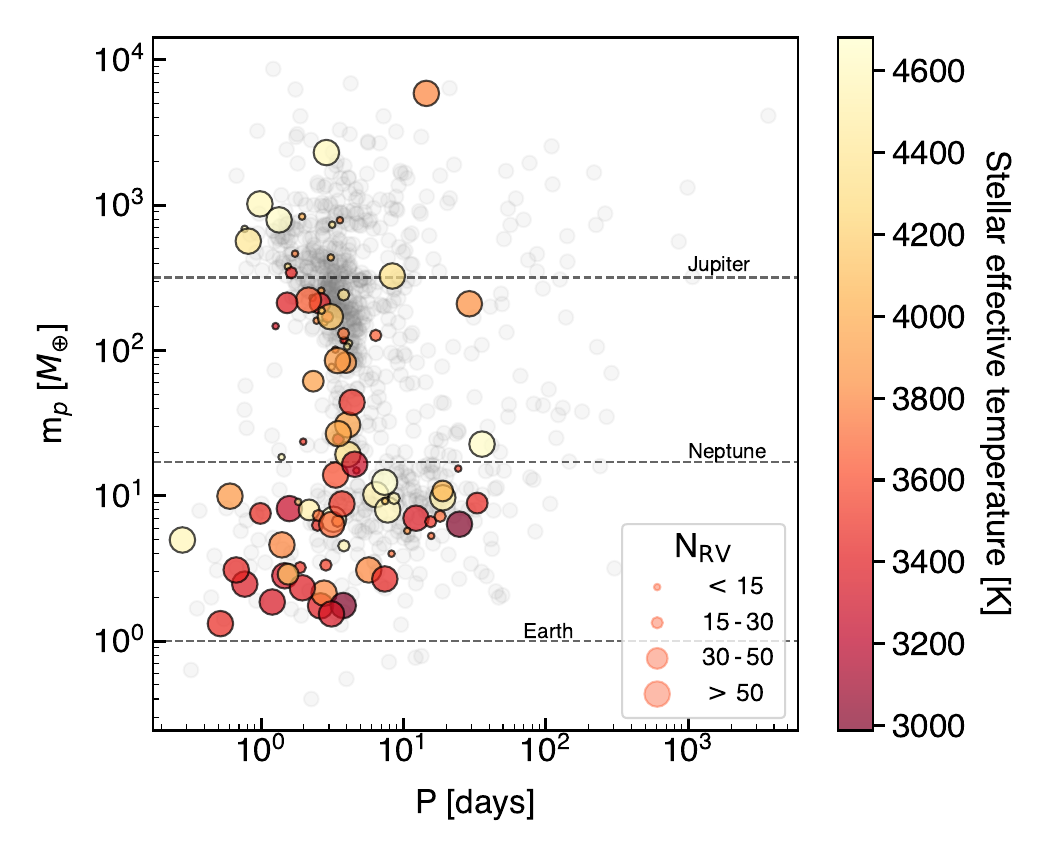}}
  \end{center}
 \caption{Sample of the study. \textit{Left:} Distribution of the stellar spectral type (color coded by {the effective temperature}). \textit{Right:} Period-mass diagram for all confirmed planets excluding those discovered by the imaging technique. Colored symbols correspond to the targets from our sample selected as explained in Sect.~\ref{sec:sample}, {where the color code indicates the effective temperature of the host star. The size of the colored dots informs on the number of RV measurements available for this work as indicated in the legend.} The three horizontal lines show the masses of Jupiter, Neptune, and Earth for reference.}
 \label{fig:sample}
\end{figure*}

Our sample selection is based on the NASA Exoplanet Archive (NEA) planetary systems table\footnote{\url{https://exoplanetarchive.ipac.caltech.edu/cgi-bin/TblView/nph-tblView?app=ExoTbls&config=PS}} (\citealt{2013PASP..125..989A}) in November 2023, which encompasses all the "confirmed" planets to date.
We applied the following criteria to select the sample.

First, the stellar effective temperature was set to be below 4650~K to focus the survey on low-mass stars. This corresponds with spectral types later than K4\,V. Then, the system hosting up to two planets to avoid highly complex RV signals. This would reduce the number of free parameters and, consequently, the degree of degeneracy. Next, at least one of the planets must transit the star. This requirement relies on the methodology, as it exclusively applies to transiting planets. Finally, the transiting planets also have to be  detected by radial velocities, since our method is based on the analysis of publicly available RVs.

A total of 88 systems fulfill these criteria. Seven of them were excluded for various reasons. According to the discovery papers, \object{GJ~1132} and \object{TOI~1266} require three Keplerians to reproduce the RV signal (\citealt{2018A&A...618A.142B}; \citealt{2024MNRAS.527.5464C}), indicating an additional planet candidate not listed in the NEA table. \object{Kepler-16} is a binary star hosting a circumbinary planet that transits both stars (\citealt{2022MNRAS.511.3561T}); hence, the detection method we used cannot be applied to this system (see Sect.\,\ref{sec:meth}). \object{Kepler-91} is ascending the red giant branch and therefore is out of our regime of study (but see \citealt{2014A&A...562A.109L}, where the presence of a trojan for this system is studied). Also noteworthy are the cases of \object{LHS~1815}, \object{TOI-2136}, and \object{Wendelstein-2}, which have insufficient RV observations; therefore, the planet was undetectable and the candidate was only validated (\citealt{2020AJ....159..160G}; \citealt{2022AJ....163..286B}; \citealt{2020A&A...639A.130O}). These cases highlight the need for establishing a protocol that standardizes the requirements for considering a planet as confirmed (Lillo-Box et al., in prep.), which would highly facilitate archival works. 

Finally, we added \object{K2-199}, \object{TOI-532}, and \object{TOI-1801} as these targets meet our constrains but at the time of writing are not yet included as detected by RV variations in the NEA. In summary, our final sample is composed of 84 systems hosting a total of 95 transiting planets. We note that our sample contains some targets previously analyzed in \citealt{2018A&A...609A..96L, 2018A&A...618A..42L}. However, we decided to preserve them for homogeneity. In some cases, new RV measurements are available, and we incorporated some changes in the analysis (e.g., we set the eccentricity to be below 0.1, and we fixed the orbital period and the mid-transit time to the photometric according to the prescription in \citealt{2017A&A...599L...7L}; see Sect.~\ref{sec:meth}).

All the RV datasets used in this work have been obtained from the literature. In Table~\ref{tab:sumrvs_pt0} we summarize them, including the number of measurements (out of transit, see Sect.\,\ref{sec:meth}) and instruments, the time span, and the reference we retrieved the data from. The RV datasets used for all the targets can be found in Table~\ref{tab:rvs}. In Fig.~\ref{fig:sample}, we illustrate the final sample. On the left, the stellar mass distribution showcases the covered spectral domain (spectral types are based on \citealt{2020A&A...642A.115C}, valid for main-sequence low-mass stars). On the right panel, we show the period-mass diagram for all confirmed NEA exoplanets (excluding the ones detected by the imaging technique), {where the color code indicates the stellar effective temperature and the sizes offer information on the number of RV measurements available for our analysis.} 

We conducted an examination of potential selection biases within our sample: (i) only 10\,\% of systems with confirmed planets fall within our target spectral domain (earlier than K4). To our knowledge, no works have shown co-orbital preference for any spectral type over another. However, \citet{2019A&A...631A...6L} and \citet{2019A&A...631A...7C} show that multiple planetary systems in resonant chains tend to retain the co-orbital configurations, two factors that could be more common around low-mass stars systems (e.g., Trappist-1); (ii) three systems (L~98-59, \citealt{2021A&A...653A..41D}; TOI-178, \citealt{2021A&A...649A..26L}; TOI-500, \citealt{2022NatAs...6..736S}) lie outside our sample due to hosting more than two confirmed planets; this means our study comprises 97\,\% of low-mass star planetary systems and, therefore, no bias is anticipated from this constraint. We note that even if these stars are expected to host numerous planets (\citealt{kunimoto20}), they may remain undetectable with current instrumentation; (iii) like any statistical study focusing on transiting planets, our sample is biased towards large-radius planets in edge-on architectures. The former bias is compensated by the fact that low-mass stars host lighter and smaller planets and the signal of such planets in both techniques (RVs and transits) are also enhanced in this spectral domain. Specifically, 51\,\% of our observed planets are below a Neptune mass, 34\,\% fall between Neptune and Jupiter masses, and 15\,\% exceed the Jupiter mass. No effect is therefore expected in our study apart from the difficulty on studying the RV signals of the less massive planets. Regarding the edge-on orientation, as there are not obvious co-orbital transits in their light curves, our sample is biased to non-coplanar co-orbital configurations, or to small radii exotrojans; (iv) the spectral-type domain and the nature of the detection methods also contribute to the fact that the majority of planets in our sample have short periods (85\% of them are below 10\,days). Tidal interactions for those planets could have an impact preventing the long-term co-orbital stability (\citealt{2020A&A...635A..37C}) and this factor is even more critical around low-mass stars (\citealt{2021CeMDA.133...37C}). Therefore, for our sample we would expect to only find exotrojan evidence for the youngest systems or with longer orbital periods.

\section{Methodology}
\label{sec:meth}

A configuration with a single planet and a pair of co-orbitals can be identified using RV time series. Assuming a long time baseline and high cadence for the RV measurements, it is possible to detect the modulation in the main-planet signal as its trojan librates (see e.g., \citealt{2015A&A...581A.128L}). Nonetheless, when the libration of the trojan is negligible as in the case of usual RV monitoring campaigns (spanning few months and with relatively sparse data), the key factor to reveal the presence of a co-orbital configuration is having information about the time of conjunction of the main planet (i.e., $T_{0}$). Even if the RV signal is compatible with a single planet scenario, the inferred transiting time from this technique would correspond with that of the centre of mass of the co-orbitals (\citealt{2006ApJ...652L.137F}). Therefore, if there is an imbalance in the masses of the Lagrangian regions, there would be a shift between the photometric and the RV inferred $T_0$. We note that this is the case as far as an asymmetry exists between the Lagrangian points L$_{\rm 4}$\,and L$_{\rm 5}$: (i) a single trojan leading or trailing the main planet, (ii) two trojans each in its own tadpole (i.e., configuration in which the trojan librates around L$_{\rm 4}$\,or L$_{\rm 5}$) region but unbalanced in mass, or (iii) a trojan in a horseshoe (i.e., the trojan travels from L$_{\rm 4}$\,to L$_{\rm 5}$\,going through L$_{\rm 3}$\,as well) orbit but with a libration period larger than the observation time span. Conversely, this method would be "blind" to equal mass trojans located in L$_{\rm 4}$\,and L$_{\rm 5}$\,regions. We note that all of these scenarios are equivalent when a swarm of asteroids replaces the compact body.

\citet{2017A&A...599L...7L} generalized the RV equation as a function of time ($t$) for this method as follows:
\begin{equation}
    \Delta v(t) = K \left[ \left( \alpha - 2c \right) \cos{nt} - \sin{nt} + c \cos{2nt} + d \sin{2nt}   \right], \label{eq:rv}
\end{equation}
with $K$ as the RV semi-amplitude, and $n$ the mean orbital frequency. Parameters $c$ and $d$ depend on the orbital eccentricity ($e$) and the argument of periastron ($\omega$) as $c = e \cos{\omega}$, and $d = e \sin{\omega}$. The $\alpha$-parameter is the one that holds the trojan information. To the first order in eccentricity, it takes the form:
\begin{equation}
    \alpha\,\simeq\,\rm{m_t}/\rm{m_p}\sin{\zeta}, \label{eq:alp}
\end{equation}
where $\rm{m_t}/\rm{m_p}$ is the trojan-planet mass ratio, and $\zeta$ is the resonant angle that locates the trojan within the orbit (e.g., 60$^{\circ}$/-60$^{\circ}$ for  L$_{\rm 4}$/ L$_{\rm 5}$). Thus, a significantly non-zero value of this parameter would suggest the presence of a massive trojan. As in \citet{2018A&A...609A..96L}, we follow the $\alpha$-test approach to search for trojans accompanying confirmed planets (i.e., transiting planets which mass has been measured through RVs). We note that Eq.~\ref{eq:rv} does not account for the Rossiter-McLaughlin (R-M) effect and for this reason, we removed the RV measurements taken during the transit.

The model and the amount of parameters that need to be explored depend on the number of detected planets ($N_{pla}$), transiting planets ($N_{tra}$), and instruments used to gather the measurements ($N_{ins}$). By construction, our sample is only composed of systems with one or two planets, and at least one of them must transit. The RV dataset of each transiting planet was modeled using Eq.~\ref{eq:rv}, whose free parameters are the RV semi-amplitude ($K$), the orbital architecture ($c$ and $d$), and the $\alpha$ metric. The key to comparing the spectral and photometric mid-transit times relies on fixing the mean motion ($nt = 2\pi\left[t~-~T_0\right]/P$), using the parameters inferred from the transits (collected in Table~\ref{tab:sumrvs_pt0}). We note that for systems composed of two transiting planets, we searched for co-orbitals in both orbits simultaneously, fixing both periods and mid-transit times. Some of the literature values for the $T_0$ parameter were inferred in the corresponding works from a joint analysis considering also the RV data. Nonetheless, there is no conflict with this method since the $T_0$ parameter is mainly derived from the transiting signal as required (the RV dataset is typically not able to constrain it as precisely as the transits, on the order of minutes). Meanwhile, the RV signal of not-transiting planets was modeled with a classic Keplerian\footnote{We use the python module \texttt{RadVel} (\citealt{2018PASP..130d4504F}).}, whose parameters are $K$, $c$, $d$, $T_0$, and the orbital period, $P$. Finally, we added an offset and a jitter per instrument to account for instrumental dependencies and unaccounted systematics, respectively. Thus, we ended up with a total of $5N_{pla}~-~N_{tra}~+~2N_{ins}$ parameters for the modeling.

If the activity of a star is affecting the RV time series, we included a Gaussian process (GP) to account for it. This decision was made when the rotational period ($P_{\rm rot}$) is present in the generalized Lomb-Scargle (GLS) periodogram of at least one spectral activity indicator such as the full width {half} maximum (FWHM) of the cross correlation function (CCF), the differential line width (dLW), or the H$_{\alpha}$ emission line, among others. In some occasions, the RV measurements are also correlated with the activity indicators showing that the same signal is present, and in the most obvious cases, the high RV scatter caused by the activity hinders the detection of the planetary signal. In all the cases, the GP was informed based on an activity proxy, which was chosen for being the indicator that most clearly shows $P_{\rm rot}$ (see Appendix\,\ref{sec:rv_target} for the details on the particular cases). We implemented the GP using the \texttt{george} python package (\citealt{2015ITPAM..38..252A}) opting for a quasi-periodic kernel (QP, e.g., \citealt{2016A&A...588A..31F}) of the form:

\begin{equation}
    \Sigma_{ij} = \eta_1^2\,\rm{exp}\left[ -\frac{(t_i - t_j)^2}{2\eta_2^2} - \frac{2 \rm{sin}^2\left(\frac{\pi (t_i - t_j)}{\eta_3}\right)}{\eta_4^2} \right]. \label{eq:gp}
\end{equation}

Hence, the GP implementation introduces increments into  the number of parameters in $5~+~2N_{ins}$. The proxy and the RV signal need an independent amplitude ($\eta_{1,~\rm{prx}}$ and $\eta_{1,~\rm{RV}}$), there is an aperiodic timescale ($\eta_2$), a correlation period ($\eta_3$), and a periodic scale ($\eta_4$). Additionally, the activity proxy requires a constant ($C$) and a jitter per instrument. {We note that we only consider an amplitude $\eta_{1,~\rm{prx}}$ and $\eta_{1,~\rm{RV}}$ shared for all instruments. Therefore, it represents an average amplitude of the activity signal, which might not be ideal specially when the instruments cover different wavelength ranges}.

To sample the posterior distribution of the model parameters we employ the Markov chain Monte Carlo (MCMC) affine invariant ensemble sampler \texttt{emcee} (\citealt{2013ascl.soft03002F}). We use five (ten when adding a GP) times the number of parameters for the walkers and between $6 \times 10^{4}$ and $2 \times 10^{5}$ steps, following a second run starting the new chains around the maximum a posteriori set of parameters and with half of the steps to speed up the convergence. The number of steps is adapted to ensure that the length  of the chains is at least 50 times the autocorrelation time per parameter. 

In Table~\ref{tab:priors} we show the priors adopted in the analysis. {For most of the priors, we opt to use uniform distributions in order to be uninformative. In the case of the RV semi-amplitude ($K$), we choose the scatter of the RV ($RV_{\rm scatter}\,=\,RV_{\rm max}\,-\,RV_{\rm min}$) to be the upper limit of the uniform distribution. The selection of this prior has proven to be good enough as our posteriors on this parameter converge to values below half of the parameter space (i.e., RV$_{\rm scatter}$/2). We choose equivalent priors for the amplitudes of the GP ($\eta_{\rm 1, prx}$ and $\eta_{\rm 1, RV}$)}. We did constrain $c$ and $d$ with a normal distribution for the planets with detected secondary eclipses by computing their values as Eqs.\,33 and 34 from \citet{2010exop.book...55W} and changing their sign to transform into the planet frame (see Table~\ref{tab:occ}). These are the most reliable candidates since $\alpha$ is degenerated to some extent with the orbital architecture ($e$), being the secondary transit the only means to disentangle them. When no secondary eclipse is available, we need to work under the assumption of low-eccentricity orbits ($e\,<\,0.1$) to be able to apply our methodology since Eq.~\ref{eq:rv} is only valid for eccentricities below this value. Hence, it is important to note that our results are subjected to this scenario. When the orbital period of the planets (either transiting or not) are below 10\,d, we do an additional test setting them to circular orbits ($e\,=\,0$, as based on the tidal circularization criterion, for example, \citealt{2008ApJ...678.1396J}). In the case of systems where one of the planets does not transit, we used normal distributions for the Priors of $P$ and $T_0$ using the data from the bibliography tabulated in Table~\ref{tab:sumrvs_pt0}. The priors for the GP hyperparameters are normal distributions based on literature posteriors if a QP GP has already been implemented for that particular system, or are those shown in Table\,\ref{tab:priors} otherwise.

\setlength{\tabcolsep}{12pt}
\begin{table}[h]
    \centering{
      \caption[]{Prior distributions. The subscripts identify parameters only applicable for transiting ($tra$), and not transiting ($ntr$) planets, and ($occ$) are for the values inferred from occultations.\label{tab:priors}}
    \label{tab:prior}
    \begin{tabular}{@{}lcc@{}}
    \hline \hline
    Parameter & Prior & Units \\ \hline
    $K$ & $\mathcal{U}\left(0,~\rm{RV_{scatter}}\right)$  & m\,s$^{-1}$  \\
    $c$     & $\mathcal{U}\left(-1,~1\right)$ or $\mathcal{G}_t\left(c_{occ},~\Delta c_{occ}\right)$ &    \\
    $d$  & $\mathcal{U}\left(-1,~1\right)$ or $\mathcal{G}_t\left(d_{occ},~\Delta d_{occ}\right)$ &  \\
    $e_{tra}$ & $\mathcal{U}\left(0,~0.1\right)$ & \\
    $P_{ntr}$ & $\mathcal{G}_t\left(P,~\Delta P\right)$ & d\\
    $T_{0,~ntr}$ & $\mathcal{G}_t\left(T_{0},~\Delta T_{0}\right)$ & d \\
    $\alpha$ & $\mathcal{U}\left(-1,~1\right)$  &  \\ 
    jitter & $\mathcal{U}\left(0,~\rm{RV_{scatter}}/4\right)$ & m\,s$^{-1}$ \\ 
    offset & $\mathcal{U}\left(\rm{RV_{\rm{min}}},~\rm{RV_{\rm{max}}}\right)$ & m\,s$^{-1}$ \\ 
    \hline \noalign{\smallskip}
    & GP hyperparameters & \\
    \hline \noalign{\smallskip}
    $C$ & $\mathcal{U}\left(\rm{Proxy_{min},~Proxy_{max}}\right)$ & Proxy units \\
    jitter$_{\rm{prx}}$ & $\mathcal{U}\left(0,\rm{~Proxy_{scatter}/4}\right)$ & Proxy units \\
    $\eta_{1,~\rm{prx}}$ & $\mathcal{U}\left(0,~\rm{Proxy_{scatter}}\right)$ & Proxy units \\
    $\eta_{1,~\rm{RV}}$ & $\mathcal{U}\left(0,~\rm{RV_{scatter}}\right)$ & m\,s$^{-1}$ \\
    $\eta_{2}$ & $\mathcal{U}\left(2P_{rot},~1\,000\right)$ & d \\   
    $\eta_{3}$ & $\mathcal{G}_t\left(P_{rot},~\Delta P_{rot}\right)$ & d \\
    $\eta_{4}$ & $\mathcal{U}\left(0,~5\right)$ &  \\
    \hline
    \end{tabular}}
   \end{table}
   \section{Results}
   \label{sec:results}

   There are 95 transiting planets from the 84 systems for which we can inspect their $\alpha$ searching for co-orbital signs. We test up to four models per system: the slightly eccentric orbit model (where the eccentricity is either below 0.1 or constrained by the secondary eclipse), the circular orbit model (if $P < 10$\,d), and those two models but including a GP if needed ($P_{\rm rot}$ signal present in a spectral indicator). We selected the GP model as the best one (when available) and we decided among the circular and eccentric models by means of the Bayes factor ($\ln \cal{Z}$), computed with the \texttt{bayev} code (\citealt{2016A&A...585A.134D}). {Remarkably, for all of the cases the circular model (simpler) is favored over the eccentric scenario, with $\ln \cal{Z}$ $> 3.5$}.
   
   We classified the sample into four groups based on the inferred $\alpha$. The most relevant are the "strong candidates" (SC) and the "null detections" (ND) groups since they provide conclusive results. The SC are the planets with an $\alpha$ value different from zero within a 99.7\,\% confidence interval (\,$|\alpha|$\,/$\sigma_{\alpha} \geq 3$), therefore being co-orbital detections by the $\alpha$-test. On the other hand, ND are planets with null $\alpha$ values as far as the uncertainty is below 15\,\% ($\sigma_{\alpha} < 0.15$), being observationally very expensive to better constrain the presence of an hypothetical co-orbital and, therefore, we rule them out as candidates. In between, we distinguish the "weak candidates"\ (WC), and the "inconclusive candidates" (INC). For both of them, there is a large uncertainty in terms of their $\alpha$ values that would be required to gather more (or more precise) data until reaching conclusive results. Nonetheless, for the WC, the $\alpha$ differs from zero within 1-$\sigma$ to 3-$\sigma$,  hinting at the presence of trojans as promising targets that should continue to be monitored. In Table\,\ref{tab:groups}, there is a summary for this classification criterion. We note that the sum of the members is 94 since one of the targets (GJ~3090\,b) is rejected for being out of the methodology domain (see Appendix\,\ref{sec:rv_target} for more details).
   
   \setlength{\tabcolsep}{5pt}
   \begin{table}[h]
       \centering{
         \caption[]{Target classification.\label{tab:groups}}
       \begin{tabular}{@{}ccccc@{}}
       $\alpha^a$ & Criterion & Group & Members\\ 
       \hline \hline
       \multirow{-0.3}{*}{$\neq$ 0}  & $|\alpha|/\sigma_\alpha > 3$ & \it{Strong candidate} & 1  \\
       & $1 < |\alpha|/\sigma_\alpha < 3$ & \it{Weak candidate} & 25 \\
       \hline
       \multirow{-0.3}{*}{= 0}  & $\sigma_\alpha > 0.15$ & \it{Inconclusive} & 21 \\
       & $\sigma_\alpha < 0.15$ & \it{Null detection} & 10\\
       \hline
       \hline
        & $\Delta\phi > 0.15$ or  & \multirow{-0.3}{*}{\textit{Sparsely sampled}} & \multirow{-0.3}{*}{37} \\
        & N$_{RVs} < 15$ & & \\
       \hline
       \end{tabular}}
       \tablefoot{\tablefoottext{a}{Within 68.3\,\% of confidence interval (1-$\sigma$).}}
      \end{table}

   A large number of transiting planets in the sample (37, being the 40\,\%) have a poor coverage in the orbital phase space. The criterion we adopted to tag a system as "sparsely sampled" (SS) is that each orbital phase bin of 0.15 contains less than one measurement (i.e., $\Delta\phi > 0.15$, being $\Delta\phi$ the orbital phase difference between consecutive data points), or the total number of data points is below 15 (i.e., N$_{RVs} < 15$). We included these targets in the analysis for completeness and also because they are relevant targets with respect to expanding the sample in the near future. Nonetheless, we did not classify them in the groups mentioned above and we do not consider them for the discussion. The resulting $\alpha$ for all the tested models can be found in Table\,\ref{tab:alpha_res} and are displayed in Fig.\,\ref{fig:alp_gr} for the grouped targets, and Fig.\,\ref{fig:alp_ss} for the SS.
   
   Only one planet stands out as SC: GJ~3470\,b. This is a hot Neptune of around 14\,$M_{\oplus}$ in a 3.3-days orbit that has been intensively explored. There is no evidence for the existence of a second planet orbiting the M-dwarf, including no significant TTVs (below 500\,s, \citealt{2016MNRAS.463.2574A}). The outcome of the analysis corresponds with a co-orbital detection at a 3-$\sigma$ level ($\alpha=-0.16~\pm~0.05$). GJ~3470\,b is a very strong candidates not only by its significance, but also because it is not degenerated with the orbital architecture since the secondary eclipse of this planet has been observed with \textit{Spitzer}.  For more information on the analysis, we refer to Appendix\,\ref{sec:rv_target}.
   
   Remarkably, both transiting planets orbiting TOI-1130 have not null $\alpha$ values with a higher significance ($\sim$ 4- and 12-$\sigma$) than GJ~3470. Nonetheless, we downgrade them to WCs, since \citet{2023A&A...675A.115K} reported strong TTVs in both planets (more than 2 hours of amplitude for planet b) by using TESS and ground-based photometry. As \citet{2017A&A...599L...7L} warn, planets with coupled orbital periods in 2:1 MMR (as in the case of TOI-1130, with 4.1 and 8.4\,days) can induce false positives due to their possible TTVs. Nonetheless, the time span of the RV dataset we use (62\,days) covers around 15 orbits of the inner planet, which should mitigate the TTV impact in the inferred $\alpha$. For this reason, we either rule out this target as a candidate. Other systems near 2:1 MMR are present in the sample: HD~260655 (with periods of 2.8 and 5.7\,days), K2-199 (3.2 and 7.4\,days), TOI-776 (8.2 and 15.7\,days), and TOI-836 (3.8 and 8.6\,days). All of these targets have at least one not null $\alpha$, yet TOI-836\,c is the only planet that exhibits TTVs in the order of 20 minutes (\citealt{2021A&A...645A..41L, 2022A&A...664A.199L}; \citealt{2023MNRAS.520.3649H}). We recall, as mentioned above (Sect.~\ref{sec:sample}), that resonant chains are expected to help preserving the co-orbital stability.
   
   The cases of GJ~143\,b, and HIP~65\,A\,b are also noteworthy. The former planet orbits a presumably active star as its rotational period appears in the activity indicators. Nonetheless, when including a GP the uncertainty in $\alpha$ is greatly increased in almost a factor of five, taking this target from the WC group to the INC. For HIP~65\,A\,b, the result hardly depends on the eccentricity being a WC for the slightly eccentric model and ND for the circular. Since this planet has a grazing transit (b\,$\sim$\,1 according with \citealt{2020A&A...639A..76N}), breaking the degeneracy through the eclipse might not be possible.
   
   It is interesting to note that HAT-P-20\,b and WASP-43\,b have an extraordinary low uncertainty, namely: 0.2 and 0.5\,\% of the $\alpha$ parameter space. Both of them are favoured due to the fact that they are hot Jupiters that induce a big RV semi amplitude into their stars ($\sim$\,1240 and 555 m\,s$^{-1}$), by the determination of the eccentricity by the occultation and also by the high quality of their observations (using HARPS and ESPRESSO spectrographs). In the case of WASP~43\,b, the majority of the measurements were taken around the transit, which is also likely to favour an accurate mid-transit time being obtained through the RVs, as an advantage for our methodology. This was also found for WASP-80\,b (see Appendix \ref{sec:rv_target}).
   
   \section{Discussion}
   \label{sec:dis}

   \subsection{Co-orbital mass}
   \label{sec:dismass}
   
   In order to estimate the trojan mass we use Eq.~\ref{eq:alp} by fixing $\zeta$~=~$\pm60^\circ$. These resonant angles correspond to tadpole orbits in which the co-orbitals librate around L$_{\rm 4}$\,or L$_{\rm 5}$, the most common configurations based on numerical simulations (e.g., \citealt{2019A&A...631A...6L}). We take the percentiles at the 95.45\% confidence interval (2.3 and 97.7) of the $\alpha$ posterior distribution inferred for the best model per system (see Sect.\,\ref{sec:results}, and specified in Table\,\ref{tab:alpha_res}) to compute the upper limit on the trojan mass. We tabulate those masses in Table\,\ref{tab:alpha_res}. For the only SC in our sample (GJ~3470\,b), we obtain the predicted posterior distribution of the trojan mass in L$_{\rm 5}$\,by taking the whole distribution of $\alpha$, resulting in m$_{\rm t}$ = 2.6~$\pm$~0.7\,$M_{\oplus}$.
   
   \begin{figure}
     \begin{center}
    \subfigure{\includegraphics[width=86mm]{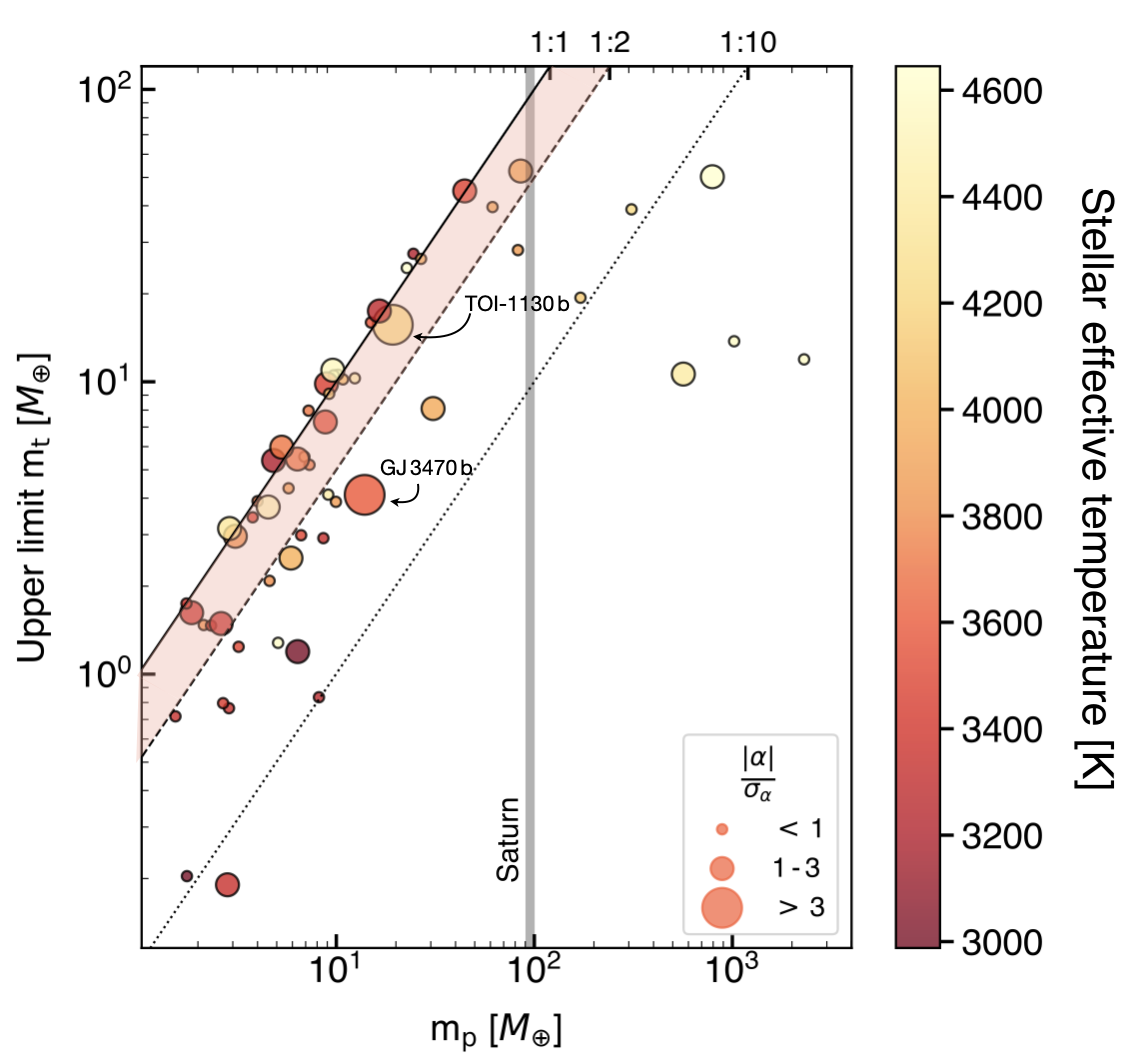}}
     \end{center}
    \caption{Upper limit of the trojan mass versus the mass of the confirmed planet. Black lines indicate equal mass ratios (solid line) along with the 1:2 (dashed) and 1:10 relations (dotted). The vertical grey solid line indicates the mass of Saturn. The light orange region corresponds to trojan masses higher than theoretically expected (between 1:1 and 1:2). {The size of the symbols informs about the significance of the co-orbital candidate as shown in the legend. Color code represents the stellar effective temperature as shown in the color bar.}}
    \label{fig:mt_mp}
   \end{figure}
   
   \begin{figure}
     \begin{center}
    \subfigure{\includegraphics[width=85mm]{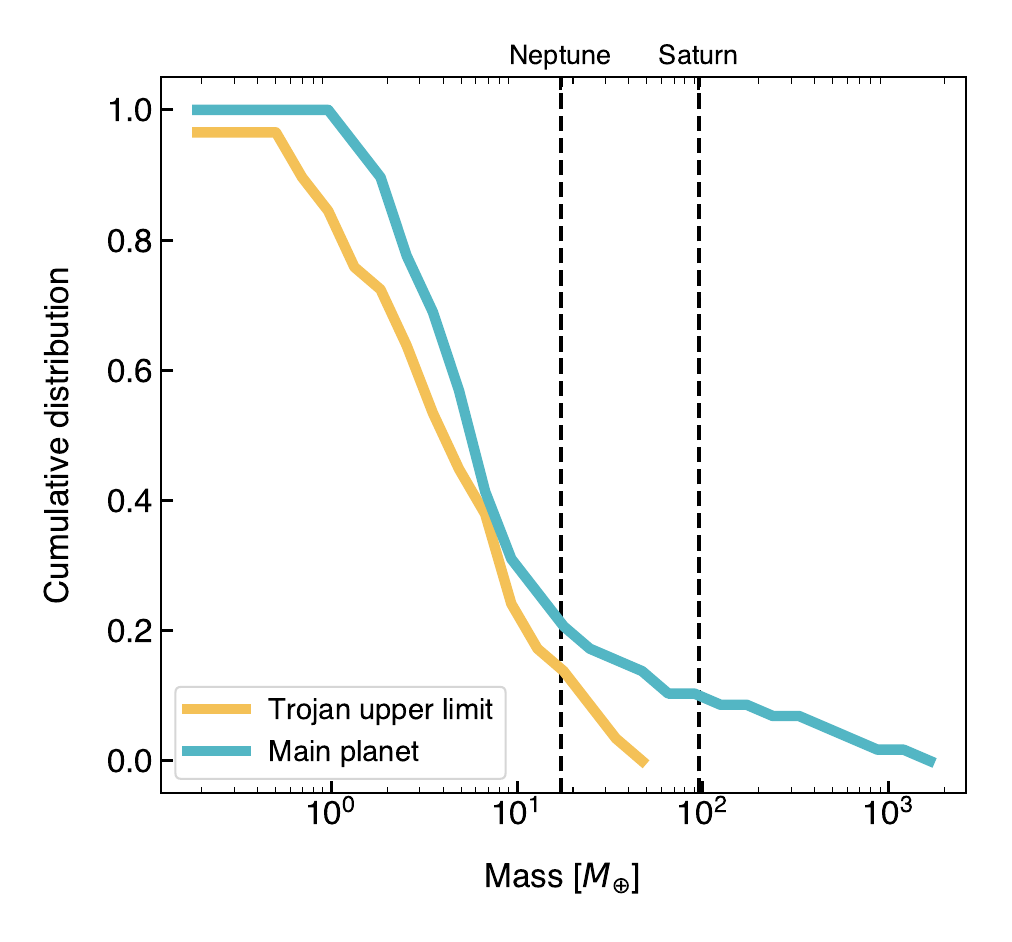}}
     \end{center}
    \caption{Occurrence rate of trojans and mass distribution of the main planets.}
    \label{fig:occurr_rate}
   \end{figure}
   
   Figure\,\ref{fig:mt_mp} graphs the maximum of the co-orbital mass upper limit (between L$_{\rm 4}$\,and L$_{\rm 5}$) as a function of the planetary mass. In this diagram, we find that for 34 of the planets (59\,\%), the upper limit of the hypothetical trojans is above half of the mass of the main planet (within the {light orange} region). We would expect this area to be empty based on theoretical studies (\citealt{2014MNRAS.442.2296P}); indeed, most of those targets correspond to the WC and INC groups, underscoring the fact that our current datasets do not constraint the co-orbital presence. On the other hand, only ten targets (17\,\%) have the mass ratio $\rm{m_t}/\rm{m_p}$ restricted to be below the 20\,\%. Interestingly, six of those are the only planets in our sample with planetary masses above Saturn ($\sim 100\,M_{\oplus}$). From this observational result based on the available time series and the $\alpha$-test method, we can infer two conclusions: (i) currently, for planets less massive than Saturn we cannot restrict the presence of trojans; and (ii) there is no observational evidence for the presence of trojans more massive than Saturn around low-mass stars. Notably, some of the less massive planets (e.g., GJ~486\,{b}, LHS~1140\,{c}, and TOI-244\,{b}) have their co-orbital companions restricted in more than the 50\,\%. The reason is that a more exhaustive RV monitoring with outstanding precision was needed for their detection, which shows that it is currently feasible to carry out dedicated searches to efficiently restrict the uncertainties in the $\alpha$ parameter.
   
   In Fig.\,\ref{fig:occurr_rate}, we show the cumulative lower limit occurrence rate of trojans as a function of their upper mass limits, compared with the distribution of the confirmed main planets. As already discussed, there are no Saturn-mass trojans expected accompanying the giant planets that orbit low-mass stars, which represent 15\,\% of our sample. Less than the 18\,\% of the planets are expected to be accompanied by Neptune-mass trojans, and roughly 80\,\% can have trojans less massive than the Earth. Comparing with the confirmed population of Neptunes (20\,\%) and with planets less massive than the Earth (100\,\%, since such planets remain undetectable to date), we notice again that current RV datasets are unable to give strong constraints to the presence of co-orbitals and, therefore, on the occurrence rate.
   
   \subsection{Search for Lagrangian point transits}
   \label{sec:lc}

   \begin{figure*}
   \begin{center}
     \rotatebox{0}{\subfigure{\includegraphics[width=160mm]{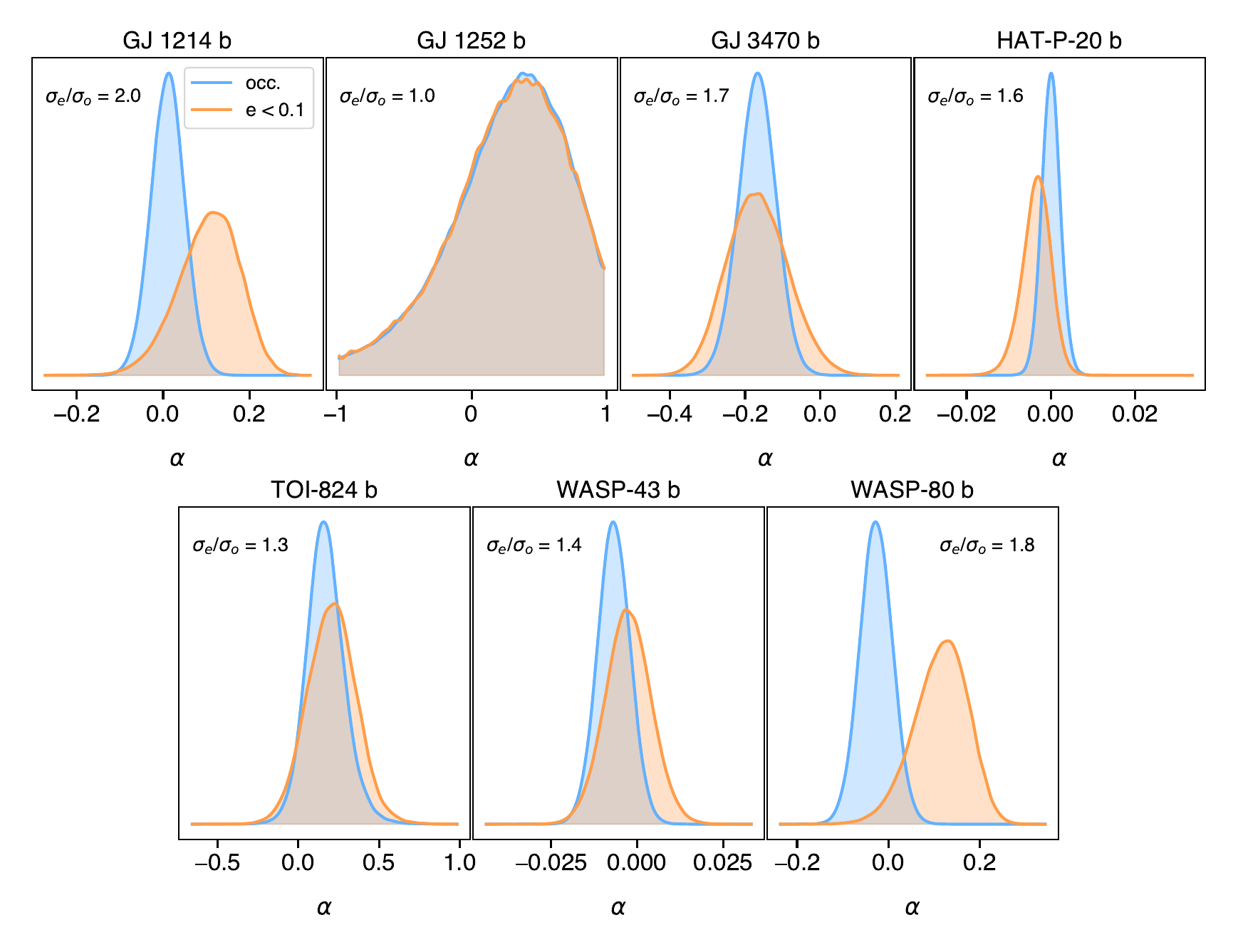}}}
   \end{center}
    \caption{Comparison of the $\alpha$ posterior distribution when using normal priors for the orbital architecture informed from the occultation (blue) and when using uniform priors with e < 0.1 (orange). The improvement in the precision as the ratio of their $\sigma$ is shown in the upper corner of each chart.}
    \label{fig:occ_vs_e01}
   \end{figure*}

   Under the assumption of (near) circular orbits and (near) coplanarity of the main planet and the potential co-orbital companion, if the main planet transits its host star from our line of sight, the co-orbital could also do so. Hence, for the cases where we find some hints for the presence of co-orbital bodies (WC sample), it is worth checking the available space-based time series photometry around the Lagrangian regions, seeking for a dimming induced by these (still weak) candidates. It is however relevant to mention that while having near-circular orbits is a requisite to apply our $\alpha$-test, coplanarity is neither required not necessary to keep the stability of a co-orbital pair (e.g., \citealt{2017A&A...599L...7L}). Consequently, non-detections within this exercise only constrain the parameter space for the existence of the trojan corresponding to the near-coplanar case, while larger mutual inclinations are still possible. 
   
   Our main light curve source is the Transiting Exoplanets Satellite Survey (TESS, \citealt{ricker14}), but when available, we also retrieved the K2 data (\citealt{borucki10,howell14}). We use the \texttt{lightkurve} \citep{lightkurve} package to retrieve the processed detrended light curves from these missions. For TESS, we use the SPOC pipeline \citep{jenkins16} with a 2-min cadence. For K2, we use the K2 pipeline with a 30-min cadence. For each of the planets in the WC list, we retrieved the light curve from all available sectors (TESS) or campaigns (K2). We then performed a detrending using \texttt{w{\={o}}tan} \citep{wotan}. To this end, all transits from any planet in the system are masked. The detrended and normalized light curves are then phase-folded with the planet ephemeris. The zoomed-in light curves around the Lagrangian points L$_{\rm 4}$\,and L$_{\rm 5}$\,for each planet are presented in Fig.~\ref{fig:LCs}, with bin sizes corresponding to 10\% (red symbols) and 20\% (blue symbols) of the planet’s transit duration. On each panel, we mark in color (orange for L$_{\rm 4}$\,and blue for L$_{\rm 5}$) the Lagrangian point where our $\alpha$-test has yielded the (weak) candidate. In most cases, the light curves around these regimes show a flat behaviour, with no detectable transits. 
   
   In some cases (specially in the K2 data) the light curves are very noisy, potentially due to stellar activity not appropriately removed with our detrending process. However, in some few cases, the light curves seem to show dimmings compatible with the duration and location expected for a co-orbital planet. In particular, this is the case for LHS~1140\,b, which shows very shallow dimmings at both L$_{\rm 4}$\,and L$_{\rm 5}$, although still compatible with the noise. Other two interesting cases are TOI-776\,b (L$_{\rm 5}$) and TOI-836\,b (L$_{\rm 4}$). They both show relatively clear dimmings at their corresponding Lagrangian points where the $\alpha$-test locates the co-orbital candidates. The dimmings would correspond to objects of around 1\,$\mathrm{R}_{\oplus}$. The estimated masses for the trojan candidates given the inferred $\alpha$ values are $2.8\pm1.6$~$\mathrm{M}_{\oplus}$ (TOI-776\,${\rm L_5\,b}$) and $1.9\pm0.9$~$\mathrm{M}_{\oplus}$ (TOI-836\,${\rm L_4\,b}$). The estimated planet radius for such masses using the empirical relations from \cite{chen17} are $1.44^{+0.78}_{-0.42}$~$\mathrm{R}_{\oplus}$ and $1.18^{+0.49}_{-0.25}$~$\mathrm{R}_{\oplus}$, respectively. Interestingly, these values match the detected dimmings. Further observations on these system will be critical to provide additional insights towards the confirmation of these candidates. {Meanwhile, there are no detectable dimmings for GJ~3470\,b. This means that there is no transiting counterpart for the strong trojan candidate with radius larger than $\sim$1\,$\mathrm{R}_{\oplus}$.}
    
   \subsection{Occultations}
   \label{sec:occ}
   
   \begin{figure*}
     \begin{center}
   \subfigure{\includegraphics[width=140mm]{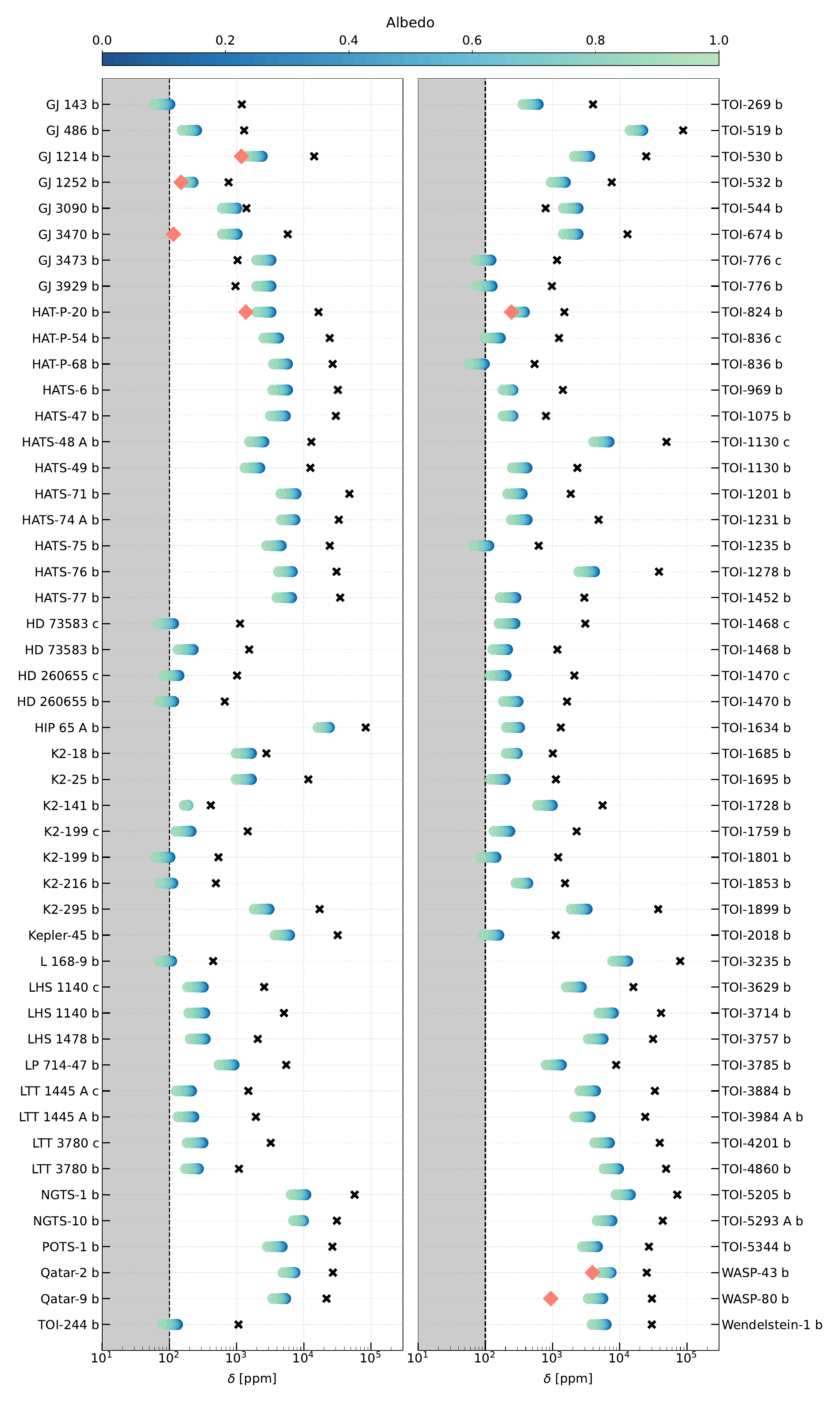}}
     \end{center}
    \caption{Depths of the transit (black cross), predicted occultation (color-coded as function of the albedo), and true occultation if measured (salmon diamond) per planet. Dashed line is at 100\,ppm, considering the grey area as the undetectable regime by current instrumentation.}
    \label{fig:occ}
   \end{figure*}

   Throughout this study, we have emphasized the critical importance of detecting occultations to reliably constrain the planetary eccentricities. This methodology stands as the sole means to resolve the degeneracy between eccentricity and $\alpha$. Only seven planets within our sample have had their secondary transits measured to our knowledge. In order to visualize the improvement, in Fig.~\ref{fig:occ_vs_e01}, we compare the posterior distribution of their $\alpha$ with the one obtained if no occultation were  to be observed (assuming e~<~0.1). For all the cases (except for GJ~1252 which is part of the SS targets), the precision in the inferred $\alpha$ is increased, being a factor two for GJ~1214. The magitude is also shifted in some of the cases, being a false candidate in GJ~1214 and WASP~80, when no occultation is measured. In this section, we focus on assessing the feasibility of detecting more occultations among our target planets.
   
   When the planet is eclipsed by the host star, the received flux is reduced the quantity corresponding to that of the emitted by the planet. Its emission is compound by the thermal radiation and the reflected starlight. The former is only not negligible as compared with the starlight when observing at long wavelengths. Hence, infrared instruments such as \textit{Spitzer}/IRAC were used to maximize the planet emission, which translates to a higher flux dimming. To estimate the eclipse depth we consider both the planet and the star emit as blackbodies, that in the mid-infrared can be approximated by Rayleigh-Jeans taking the form (e.g., \citealt{2010exop.book...55W}; \citealt{2014RSPTA.37230083E}):
   \begin{equation}
       \delta_{occ} = 0.01 \left( \frac{R_p/R_J}{R_{\star}/R_{\odot}} \right)^2 \frac{T_{eq}}{T_{\star}},
   \end{equation}
   where $R_p$ and $R_{\star}$ are the planet and stellar radius in Jovian and solar units respectively, and $T_{eq}/T_{\star}$ is the planet equilibrium and stellar temperature ratio. With the same assumption, we can also estimate $T_{eq}$ as:
   \begin{equation}
       T_{eq} = 279 \, (1 - a)^{0.25} \left( \frac{T_{\star}/K}{5770} \right) \left( \frac{R_{star}/R_{\odot}}{D/AU} \right)^{0.5}, 
   \end{equation}
   where $a$ is the geometric albedo and $D$ is the planet-star distance in au. At shorter wavelengths, the reflected emission from the star might be detectable and it also depends on $a$, $D$, and $R_p$. Nonetheless, the occultation depth is maximum at the mid-infrared where this component can be ignored. In Fig.\,\ref{fig:occ}, we show the measured transit depths and the estimated occultation depths as a function of the albedo for all the confirmed planets in the sample. For the few targets with this quantity measured, it is also represented here. We have found that they are compatible with the theoretical approximation (except for GJ~3470\,b and WASP-80\,b), but all of them are overestimated. This is probably because empirically the wavelenght used is around 4.5\,$\mu$m but the thermal emission would be increased at higher {wavelenghts} (e.g., using JWST/MIRI, which covers up to 28\,$\mu$m).
   
   Since this study is focused on the low-mass stellar domain, our sample enhances the occultation depth (i.e., $\delta_{occ} \propto R_{\star}^{-2}$). For this reason, most of the eclipses are expected to be detectable with current instrumentation ($\delta_{occ} > 100$\,ppm). We particularly encourage monitoring the secondary eclipse of the weak candidates resulted from this work, since it would translate in a reduction of the $\alpha$ uncertainty.
   
   \section{Conclusions}
   \label{sec:concl}
   We have carried out an extensive search for exotrojans around low-mass stars with spectral types later than K4\,V. We analyzed publicly available RV time series aiming at identifying mass imbalances within the orbital paths of their confirmed planets. We found a strong candidate for a trojan of $2.6 \pm 0.7 M_{\oplus}$ in the L$_{\rm 5}$\,point of the hot Neptune GJ~3470\,b. Additionally, for other two targets which show a weak evidence for the presence of co-orbitals, we also spot a dimming in the predicted Lagrangian point (L$_{\rm 5}$\,for TOI-776\,b, and L$_{\rm 4}$\,for TOI-836\,b) in their TESS light curves.
   
   The results of the analysis show that the current data permit constraints to be placed on the presence of co-orbitals accompanying planets more massive than Saturn. Interestingly, less massive planets ($< 3 M_{\oplus}$, such as GJ 486, LHS 1140, or TOI-244), which have been confirmed using the most precise instruments and a wealth of measurements, highlight the potential for significantly reducing the upper limits of co-orbital masses when using more demanding observational strategies (e.g., strategies devoted to the improvement in the precision of the planetary masses). This would lead to a more refined constraint on the presence of trojans and, consequently, would enhance our ability to infer their occurrence rate.
   
   The sample was divided into different groups based on the resulting $\alpha$ parameter. There is particular interest in the targets classified as WC and SS to continue the RV observations. Among the WC, those targets with fewer (or less precise) RV measurements are particularly promising for further investigation using this technique. These planets exhibit a high degree of uncertainty in $\alpha$, yet their 63\% confidence interval exclusively considers the possibility of a co-orbital scenario (non-null values). Some of these targets include K2-199, TOI-776, TOI-1452, TOI-3757, TOI-3884, and TOI-3984~A. Additionally, all targets classified as SS are highly desirable for continued observation via RV measurements. Special attention is warranted for GJ~1252 and TOI-824, since it is expected to derivate credible $\alpha$ values, given their well-known eccentricity as their secondary transit is measured.
   
   Conversely, for some other targets within the WC group, it is not worthwhile to continue the RV monitoring as the observational cost overweigh the potential for improvements. These candidates need to be studied in more detail through other techniques, such as photometrically inspecting the Lagrangian regions, combining the photometrical observations with dynamical models (e.g., \citealt{2018A&A...618A..42L}), and even directly imaging with future missions such as LIFE (\citealt{2022A&A...664A..21Q}). Priority should be given to the search for the occultation of the main planet in order to break the degeneracy of the trojan presence with the orbital architecture. Based on the estimation of the occultation depth (Sect.~\ref{sec:occ}), the most promising targets (with depths above $10^3$\,ppm) among the WC to measure the eclipse in infrared light are: GJ~3473\,b, K2-18\,b, LP~714-47\,b, Qatar~2\,b, TOI-544\,b, TOI-3757\,b, TOI-3884\,b, and TOI-3984\,b.
   
   It is important to note that a null detection in the context of our methodology does not definitively rule out the presence of co-orbital companions, yet it does significantly limit it. Adhering to the tadpole configurations assumption (Sect.~\ref{sec:dismass}), their masses are already tightly constrained, and further narrowing these limits is not expected. Conversely, it is plausible that co-orbitals could escape our detection approach. For instance, horseshoe configurations could have $\alpha$ values oscillating to zero at each instance that could result in a false negative provided the libration timescale is shorter than the RV time span. Such scenarios may be detectable through alternative detection methods meriting their exploration in future studies (e.g., RV modulations due to the trojan libration, e.g., \citealt{2015A&A...581A.128L}).
   
   This work serves to advocate for the RV monitoring of the stars devoted to the search for co-orbitals. It also provides guidelines that can aid in this objective, and a list of the most relevant targets. As demonstrated in previous works of this series (\citealt{2018A&A...609A..96L, 2018A&A...618A..42L}, increasing the number of measurements and their precision is crucial. In the present work, WASP-43\,b and WASP-80\,b suggest that not only an appropriate (i.e., homogeneous) coverage of the orbital phase is needed, but increasing the monitoring on phases close to the transit could benefit the determination of the spectroscopic mid-transit time, which directly translates into a lower uncertainty of the $\alpha$ parameter.

   \newpage
   \begin{acknowledgements}

   We thank to the anonymous referee for their effort reviewing this work. We also thank to all the authors who have helped us to access their data or who have provided relevant information on the targets: J.M.~Almenara, C.~Cadieux, D.~Dragomir, X.~Dumusque, E.~González-Álvarez, R.~Luque, and M.~Mallorquín. This research has made use of the NASA Exoplanet Archive, which is operated by the California Institute of Technology, under contract with the National Aeronautics and Space Administration under the Exoplanet Exploration Program. This Project has been funded by grant No.PID2019-107061GB-C61 by the Spanish Ministry of Science and Innovation/State Agency of Research  MCIN/AEI/10.13039/501100011033. O.\,B.~-~R. is supported by INTA grant PRE-MDM-07. J.L.-B. was partly funded by grants LCF/BQ/PI20/11760023 (La Caixa), Ram\'on y Cajal fellowship with code RYC2021-031640-I, and CNS2023-144309. P.~F. acknowledges the financial support of the SNSF, the work has been carried out within the framework of the National Centre of Competence in Research PlanetS supported by the Swiss National Science Foundation under grant 51NF40\_205606. AL acknowledges support from the Swiss NCCR PlanetS and the Swiss National Science Foundation. This work has been carried out within the framework of the NCCR PlanetS supported by the Swiss National Science Foundation under grants 51NF40$_{}$182901 and 51NF40$_{}$205606. AL acknowledges support of the Swiss National Science Foundation under grant number TMSGI2\_211697. Co-funded by the European Union (ERC, FIERCE, 101052347). Views and opinions expressed are however those of the author(s) only and do not necessarily reflect those of the European Union or the European Research Council. Neither the European Union nor the granting authority can be held responsible for them. This work was supported by FCT - Fundação para a Ciência e a Tecnologia
   through national funds and by FEDER through COMPETE2020 - Programa Operacional Competitividade e Internacionalização by these grants: UIDB/04434/2020; UIDP/04434/2020. E.H.-C. aknowledges support from grant PRE2020-094770 under project PID2019-109522GB-C51 funded by the Spanish Ministry of Science and Innovation / State Agency of Research, MCIN/AEI/10.13039/501100011033, and by ‘ERDF, A way of making Europe’. 
   
   \end{acknowledgements}

\newpage
\bibliography{references}

\begin{appendix}
\onecolumn

\section{Radial velocity analysis per system}
\label{sec:rv_target}
In this section, we give the details of the RV analysis for the individual targets.

\subsection{Strong candidate}
\subsubsection*{GJ 3470}
Located in the hot-Neptune desert, GJ~3470\,b has been object of many studies trying to bring light to the rare population to which it belongs. As a result, we account with multiple high-precision RV measurements. The RV analysis requires a linear trend probably caused by an undetected stellar companion, for which we inferred a slope of ($-0.0022~\pm~0.0011)\,{\rm m}\,{\rm s}^{-1}\,{\rm d}^{-1}$. The secondary eclipse of this planet has been observed, allowing us to set normal priors on the $c$ and $d$ parameters that were derived in this work using the eclipse published in \citet{2019NatAs...3..813B}, and obtaining compatible results with those from \citet{2019AJ....157...97K} (see Table~\ref{tab:occ}). Even there is no clear correlation between the RV measurements and the activity indicators, we included a GP informed with the $S_{\rm HK}$ index as it shows a peak in its GLS at $P_{\rm rot}\,=\,21.5~\pm~0.5$\,d. The resulting $\alpha$ places this target as a SC with a 3-$\sigma$ significance.

\subsection{Weak candidates}

\subsubsection*{GJ 486}
This single-planet target has an $\alpha$ parameter different from 0 in 1-$\sigma$ in both slightly eccentric ($e~<~0.1$) and circular orbit scenarios. We do not include a GP to the analysis since, according with \citet{2022A&A...665A.120C}, none of the periodograms of the activity indicators (e.g., H$_{\alpha}$, Ca~II, or Na~I) show any significant peak that could suggest the star is active.

\subsubsection*{GJ 3473}
Our RV analysis for the confirmed planet GJ~3473\,b resulted in an $\alpha$ parameter different from 0 within 1-$\sigma$ for both tested models, the slightly eccentric ($\alpha = -0.30^{+0.23}_{-0.24}$) and the circular orbit ($\alpha = -0.28^{+0.21}_{-0.22}$). Based on the $H_{\alpha}$ emission line, this target is considered inactive (\citealt{2018A&A...614A..76J}). 

\subsubsection*{HD 260655}
Both planets in this system transit the star with short periods, lying in a close 2:1~MMR ($P_b\,=\,2.8\,$d, and $P_c\,=\,5.7\,$d). However, these rocky worlds do not show significant TTVs according with \citet{2022A&A...664A.199L}. Based on the values obtained for the $H_{\alpha}$ emission and $R'_{\rm HK}$, this star is thought to be inactive with a rotation period of around $\sim$30\,d. Our analysis found that, contrary to the inner planet, the outer one has an $\alpha$ parameter different from zero within 1-$\sigma$ ($\alpha = -0.32^{+0.22}_{-0.25}$) indicating that the mass of its potential co-orbital companion could be higher than the one for the inner planet. This result is hold when considering circular orbits. As warned in \citet{2017A&A...599L...7L} (and discussed in Sect.~\ref{sec:results}), planets in 2:1~MMR are susceptible to mimic co-orbitals in the $\alpha$-test method. Nonetheless, as no TTVs are detected and the RV time span is huge (24\,years) in comparison with the orbital periods, a false positive caused by the planetary configuration is not expected. Detailed searches to constrain their presence is needed, such as photometrically inspect the Lagrangian regions in the search for dimmings (see Sect.~\ref{sec:lc}).

\subsubsection*{K2-18}
With an orbital period of 32.9\,d, K2-18\,b is a Super-Earth in the habitable zone (HZ) of an M-dwarf. We use the RVs extracted with the recent line-by-line (LBL) method {(\citealt{2022AJ....164...84A})} published in \citet{2022MNRAS.517.5050R}. They identified 14 outliers in CARMENES and three in the HARPS datasets, resulting in a total time series of 147 measurements after discarding those data points. This system has been claimed to host an additional not-transiting planet in a 9.2\,d orbit confidently detected with the LBL RVs after rejecting a particular night (December 25, 2016). We include a GP informed with the dLW, indicator showing a peak in the GLS at the estimated stellar rotational period ($\sim39\,$d). The analysis suggests the presence of a co-orbital for K2-18\,b with an $\alpha~=~0.48^{+0.29}_{-0.28}$. 

\subsubsection*{K2-141}
K2-141\,b is an ultra-short period planet orbiting at 0.28\,d. There is an additional validated transiting planet with a period of 7.7\,d that is not detected with the RV dataset. Therefore, we carried out the analysis only considering planet b. The RV dataset shows a considerably big scatter. Besides, all the GLS of the activity indicators present a peak at $P_{rot}~=~14\,$d, so we add a GP by using the FWHM as proxy. Our analysis was compatible with a configuration with an low-mass co-orbital within $L_4$ as $\alpha~=~0.07 \pm 0.07$.

\subsubsection*{K2-199}
The K2 mission detected two transiting planets orbiting around K2-199, at 3.2 and 7.4\,d orbital periods. Activity tracers suggest the star is moderately active. Nonetheless, \citet{2021AJ....162..294A} opted for a model not including a GP since the spot behaviour of the star changed between the campaigns and therefore they did not achieve the convergence of the hyperparameters. Not accounting for the activity either, we found that the outer planet has a similar $\alpha$ for both eccentric ($\alpha = -0.25^{+0.23}_{-0.24}$) and circular ($\alpha = -0.26^{+0.20}_{-0.21}$) models.

\subsubsection*{LHS 1140}
Two transiting planets orbit this M-dwarf star with periods of 3.8 (LHS~1140\,c) and 24.7\,d (LHS~1140\,b, lying within the HZ). LHS~1140 has already been target of co-orbital search through the $\alpha$-test in \citet{2020A&A...642A.121L}. There, {it is} discarded the presence of a co-orbital for LHS~1140\,b more massive than 1\,M$_{\oplus}$, but found a co-orbital candidate for LHS~1140\,c. Here, we re-analyze the system using the same observations but with a different RV extraction (using the LBL method, \citealt{2024ApJ...960L...3C}) that is expected to improve the uncertainty on the inferred $\alpha$ parameter, and with some changes in the analysis as mentioned in Sect.~\ref{sec:sample}. An additional external planet was proposed in \citet{2020A&A...642A.121L} based on the RV signal, but it was not supported by the new LBL dataset (\citealt{2024ApJ...960L...3C}). Therefore, we consider LHS~1140 as a two-planets system. A GP is clearly needed in the model, with a rotational period of around 131\,days appearing in the FWHM. Both transiting planets arise to be WC, with $\alpha_b = -0.11^{+0.07}_{-0.02}$, and $\alpha_c = -0.04~\pm~0.02$, assuming circular orbits.

\subsubsection*{LP 714-47}
LP~714-47\,b is a planet located in the hot-Neptune desert ($P\,=\,4.1\,$d, $R\,=\,4.7\,R_{\oplus}$). The scientific interest of this target motivated its RV follow-up using multiple instruments, such as ESPRESSO and CARMENES. In the original publication, \citealt{2020A&A...644A.127D} found that the rotational period of the star is present in the GLS of the RVs ($\sim$\,33\,d). Though, according with the periodogram of the activity tracers presented in their work, that signal is not obvious in any of them. For this reason, and since such indicators are not available and our residuals do not show much scatter, we do not include a GP in our model.

\subsubsection*{LTT 3780}
A pair of TESS-candidate planets were confirmed using spectroscopic data around LTT~3780 with orbital periods of 0.8\,d and 12.3\,d. The star is known to be rather inactive, as indicated by the $R'_{\rm HK}$ value of $-5.59$ (\citealt{2020AJ....160....3C}). Additionally, no modulation appears in the photometric data. Given that no strong activity impact is expected, we chose not to include a GP in this test.

\subsubsection*{Qatar-2}
Qatar-2\,b is a hot Jupiter in a 1.3\,d orbit around a K-dwarf star. Its strong photometric variations reveal the active nature of the star with a rotational period of $\sim$\,18.5\,d,  which corresponds with a gyrochronological young age of around 1.4\,Gyr (\citealt{2017A&A...601A..53E}; \citealt{2017MNRAS.471..394M}). From the RV dataset in hand, only HARPS-N data have activity tracers available (nine out of the 80 out of transit data points). Such few FWHM measurements are not enough to account for the activity contribution on the signal and thus we do not include a GP to our analysis. We obtain an $\alpha$ different from zero in {1}-$\sigma$ ($\alpha = -0.02~\pm~0.02$, in the circular case). 

\subsubsection*{TOI-269}
The sub-Neptune (2.8\,R$_{\oplus}$, 8.8\,M$_{\odot}$) orbiting this M2-dwarf star with a 3.7\,d period has been reported to be highly eccentric ($\sim$\,0.43). This solution is obtained by \citet{2021A&A...650A.145C} when fitting together transits and RV datasets. Based on the mature stage of the system ($\sim$2-4\,Gyr) and the short orbital period of the planet, it would be expected to have circularized the orbit. To explain the puzzling result, they invoked the inward migration of the planet through a planet-planet interaction as an hypothesis. As explained in {Sect.\,\ref{sec:meth}}, the eccentricity is highly degenerated with $\alpha$, and thus it was not surprising to obtain an $\alpha$ different from zero ($\alpha = 0.29^{+0.20}_{-0.18}$ for a circular orbit) in our analysis. Since the occultation of this planet has not been measured, the eccentricity is not constrained and the co-orbital scenario is also a plausible alternative to the uncommon high-eccentricity. As the authors explain in their work, the host star is though to be quiet and probably has a long rotation period ($\sim$\,69\,d). As in the original work, we did include a linear trend to the analysis with a resulting slope significantly different from zero (0.021\,$\pm$\,0.008$\,\rm{m}\,\rm{s}^{-1}\,\rm{d}^{-1}$).

\subsubsection*{TOI-544}
The star hosts a small planet discovered by TESS in a short period of 1.5\,days. The RV dataset is composed of HARPS and HARPS-N campaigns taken over nearly three years, which enabled the detection of a non-transiting Neptune-mass planet at a 50.1\,day period. The S-index shows a very clear peak at 19\,days, which we interpret as the rotational period. When we include a GP the scatter of the RV residuals are greatly reduced, yet it is still larger than in \citet{2024MNRAS.52711138O} since they use a two-dimensional GP. The result of the analysis is compatible with the existence of a co-orbital companion within the $L_4$ region ($\alpha = 0.57^{+0.19}_{-0.22}$).

\subsubsection*{TOI-776}
Three signals are required in order to model the RVs of this system: two Keplerians to account for the two transiting planets, and a sinusoidal function describing the activity. The sinusoidal function has the periodicity of the rotation of the star (34.4\,days), and an amplitude of 2.7\,m\,s$^{-1}$. It was shown by \citet{2021A&A...645A..41L} that this model explains well the observations and it is not necessary to appeal to a GP. Both planets are mini Neptunes with orbital periods of P$_b$\,=\,8.3\,d and P$_c$\,=\,15.7\,d. The signal for the outer planet is inadequately sampled based on the criterion adopted at the beginning of this section (i.e., there is a phase gap greater than 15\% for P$_c$), and hence, we only put attention on the $\alpha$ parameter inferred for the inner planet. The inferred aforementioned parameter is  different from zero ($\alpha = -0.62^{+0.29}_{-0.24}$ for the circular case) and thus it could point to a mass imbalance towards the $L_5$ region. We warn that, even the period of planet b is well sampled according with our criterion, there exists a scarcity of data around the orbital phase $\phi$\,=\,0.

\subsubsection*{TOI-836}
Two transiting planets were detected using two TESS Sectors (11 and 38): a super-Earth (P$_b$ = 3.8\,d), and a mini-Neptune (P$_c$ = 8.6\,d). The K\,V star is very active being imperative the use of a GP to detect the planetary signals in the HARPS RV dataset. We use the FWHM, which informs on the stellar rotational period being around 22\,days. Based on the discovery paper (\citealt{2023MNRAS.520.3649H}), the existence of TTVs with an amplitude of $\sim$20 \,min for TOI-836\,c may suggests the existence of an outer planet not yet detected either by RVs or transits. As discussed in Sect.~\ref{sec:results}, TTVs can induce false positive co-orbitals since the $T_0$ varies. However, the TTVs amplitude is very low in comparison with the RV time-span so we do not expect a big impact on our analysis. The results locate both targets as interesting candidates with not-null $\alpha$ at a 2.4 and 2.8-$\sigma$ level for the inner and outer planets respectively. Furthermore, the light curve inspection comes as a suprise, with an interesting dimming in the L$_{\rm 4}$ region of TOI-836\,b that is compatible with the location and the radius inferred from its $\alpha$ value (see Sect.~\ref{sec:lc}).

\subsubsection*{TOI-1130}
This system is composed of a K-dwarf orbited by an inner Neptune-sized planet (TOI-1130\,b) and an outer hot Jupiter (TOI-1130\,{c}). TESS light curves show that the time of mid-transit of both planets varies at each orbit, with an amplitude of 2.5 h for TOI-1130\,b, and 0.25 h for TOI-1130\,c. This is cause by the resonant configuration in 2:1 MMR (P$_b$\,=\,4.1\,d, P$_c$\,=\,8.4\,d). The HARPS RV time series present a linear trend with a slope of $\sim0.5\,{\rm m}\,{\rm s}^{-1}\,{\rm d}^{-1}$ likely induced by an undetected outer companion. Both planets are clearly detected with these RVs, and the result of our analysis return not-null $\alpha$ values for both planets with very high significance ($\alpha_b = -0.44^{+0.12}_{-0.11}$, and $\alpha_c = -0.09 \pm 0.01$). However, there is a high probability that these results are affected by the known TTVs, even the time span of the RV dataset is higher than the TTVs period. For this reason, we do not consider them as SC but as WC (see Sect. \ref{sec:results}).

\subsubsection*{TOI-1235}
This M0.5\,dwarf is orbited by the transiting planet TOI-1235\,b, which is located in the radius gap ($\sim$1.7\,$R_{\oplus}$, $\sim$6-7\,$M_{\oplus}$) in a 3.44\,days orbit (\citealt{2020A&A...639A.132B}; \citealt{2020AJ....160...22C}). The RV data set comprises CARMENES, HARPS, and HIRES measurements. Its GLS shows an additional signal around 21 days, which is also present in the dLW GLS. Therefore, we include a GP informed with this activity indicator and considering a normal prior for $\eta_3$ of $P_{\rm rot} = 20.93 \pm 0.6$\,d. The result barely places the target as a WC, with a not-null $\alpha$ just at 1-$\sigma$ level ($\alpha = 0.12^{+0.12}_{0.11}$).

\subsubsection*{TOI-1452}
TOI-1452\,b is a super-Earth in a 11.1-days orbital period first detected by its transit with TESS, and later confirmed with SPIRou and IRD RVs (\citealt{2022AJ....164...96C}). For the analysis, we opt for not including the IRD data since they are few measurements in comparison with the SPIRou not including additional information in our case but increasing in one the number of free parameters. For most of the SPIRou nights around three to four measurements were taken. For this reason, we use the night-binned measurement to reduce the measurement uncertainty. As detailed in the discovery article, the host star does not show any obvious rotational period, so we consider it as a quiet star. As the orbital period is higher than 10\,days, we do not test the circular model (see Sect.\,\ref{sec:meth}). The inferred $\alpha$ suggest the presence of a co-orbital companion {($-0.51^{+0.33}_{-0.30}$)}. 

\subsubsection*{TOI-1695}
A wealth of SPIRou and HARPS-N measurements enabled confirming the planetary nature of the transiting super-Earth TOI-1695\,b identified by TESS ($P = 3.13$\,d, $\sim 2.0 R_{\oplus}$, $\sim 6 M_{\oplus}$, \citealt{2023A&A...670A.136K}, \citealt{2023AJ....165..167C}). The photometric estimation of the rotational period of the star ($P_{\rm rot} \sim 48$\,d) corresponds with an inactive early M-dwarf, which is consistent with the spectroscopic indices showing very weak indicators of activity. The circular model of our analysis is compatible with the presence of a co-orbital within 2.5-$\sigma$ ($\alpha = -0.40^{+0.14}_{-0.16}$).

\subsubsection*{TOI-3757}
TOI-3757\,b is one of the few gas giants orbiting an M-dwarf ($P = 3.4$\,d, around 12\,$R_{\oplus}$, and $\sim 85 M_{\oplus}$, \citealt{2022AJ....164...81K}). The available RV dataset is composed of HPF and NEID {measurements}. We do not include a GP since it does not show hints of activity. The slightly eccentric and the circular model provide compatible results, with an $\alpha_{\rm c} = -0.25 \pm 0.13$. We note that this target labeled as a WC is a good opportunity to keep its RV monitoring in order to improve the data set (higher precision and more measurements).

\subsubsection*{TOI-3884}
The TESS photometry of this M4-dwarf shows the transit of a super-Neptune of 6.4\,$R_{\oplus}$ in a 4.5\,days orbit affected by star-spots in several orbits. An upper limit to its mass has been confirmed with a dataset of HPF (\citealt{2023AJ....165..249L}) and a couple of measurements from ESPRESSO (\citealt{2022A&A...667L..11A}). Interestingly, even the star is known to be active, a GP does not improve the RV analysis as tested by \citet{2023AJ....165..249L}. In both our eccentric ($\alpha = -0.42^{+0.27}_{-0.28}$) and circular ($\alpha = -0.41^{+0.25}_{-0.27}$) models, this target is a WC with a 1-$\sigma$ significance.

\subsubsection*{TOI-3984 A}
The RV analysis for this M4-dwarf star is based on a dataset of 41 NEID and HPF measurements. The star is orbited by a transiting giant of 4.4\,day period (0.14\,$M_J$, 0.71\,$R_J$), and is the primary of a multiple system with a wide binary. With the available data, this target results a WC at a 1.3-$\sigma$ level ($\alpha = 0.34^{+0.27}_{-0.26}$). 

\subsubsection*{WASP-43}
The case of the hot Jupiter WASP-43\,b ($\sim 2 M_{J}$ with a period of 0.8\,days) around a K7V star has been widely studied since its discovery (\citealt{2011A&A...535L...7H}). Despite what would be expected, considering its location and mass, this planet does not experience orbital decay (\citealt{2016AJ....151..137H}). For our analysis, we opted for using only the HARPS and ESPRESSO data available in the ESO Archive with Program IDs 0102.C-0820, 60.A-9128, 089.C-0151, 096.C-0331, and 0104.C-0849, since there are a high quantity of them and are of great quality. We note that the observations were mainly taken around the transit as a cause of the original scientific goal. Such observational strategy probably favours obtaining a more accurate measurement of the mid-transit time through RVs, which favours our methodology. Moreover, the occultation of the planet has been observed breaking the $\alpha$-eccentricity degeneracy (\citealt{2014ApJ...781..116B}). The host star is though to be very active based on the high value of the log($R'_{\rm HK}$), which could be enhanced by the close Jupiter and its youth (\citealt{2017A&A...601A..53E}). However, the rotational period of the star (around 16\,d according with \citet{2011A&A...535L...7H}) is not present in the GLS of the FWHM, therefore we do not include a GP. The result of the analysis suggest the presence of a co-orbital up to 11\,$M_{\oplus}$ within L$_{\rm 5}$\,($\alpha = -0.007 \pm 0.005$).

\subsection{Inconclusive}

\subsubsection*{GJ 143}
Apart from the confirmed planet (GJ~143\,b), this star hosts an additional inner transiting candidate at 7.8\,d (GJ~143\,c). However, the available RVs are not sensitive enough to detect that candidate (\citealt{2019ApJ...875L...7D}). Consequently, we only include GJ~143\,b in our model and a linear trend to account for a long-term. Among the GLS periodogram of the available activity indicators for the HARPS data subset, we find a peak at 37.1\,d (close but not overlapping with the planet orbital period, $P_b\,=\,35.6\,$d) in the {FWHM} of the cross correlation function (CCF), which is compatible with the rotational period proposed in \citet{2019ApJ...875L...7D}. Therefore, we add a GP informed with the FWHM as activity proxy. The uncertainty in the $\alpha$ parameter is greatly increased when including the GP in contrast with the not-null $\alpha$ value (2-$\sigma$) inferred for the model without the GP. Therefore, although it falls within the group of INC targets, it is convenient to study this planet by alternative methods such as transits (see Sect.~\ref{sec:lc}).

\subsubsection*{GJ 3929}
The RV dataset shows the presence of the confirmed transiting hot Earth-sized GJ~3929\,b ($P_{b}\,=\,2.6\,$d), and a candidate not transiting sub-Neptune ($P_{c}\,=\,15.0\,$d) around the M3.5V star (\citealt{2022A&A...659A..17K}; \citealt{2022ApJ...936...55B}). This star is thought to be inactive based on different indicators (e.g., H$\alpha$, flat light curve). Even in the discovery article they include a GP, based on the quiet behavior and  since the $\alpha$-test do not show hints for the presence of a Trojan candidate, we opt for not including a GP.

\subsubsection*{HD 73583}
This system is composed of two transiting mini-Neptunes discovered by TESS (TOI-560) which are close to a 3:1~MMR ($P_b\,=\,6.4\,$d, and $P_c\,=\,18.9\,$d). This is a clear case of a young ($\sim 500\,$Myr) active star affecting the RVs. As in \citet{2022MNRAS.514.1606B}, we used the $S_{\rm HK}$ time-series as activity indicator to feed the GP, which has a peak in the GLS periodogram at $P~\sim~$12\,d, as the periodicity found in the TESS light curves. The posterior distributions found are compatible with those from their work. We note that they use a more sophisticated method using a multidimensional GP by adding the derivate of the QP kernel as a component of the RV model, resulting in cleaner residuals of the model. None of the two planets resulted in an $\alpha$ parameter different from zero.

\subsubsection*{K2-25}
Young planetary system ($730\,\pm\,50\,$Myr, \citealt{2016ApJ...818...46M}) hosting a Neptune-sized planet at an orbital period of 3.5\,d. As expected due to its youth, our preliminary RV analysis show large residuals and thus we decide to test a GP based on the dLW as activity tracer. We set a normal prior for $\eta_3$ as the rotational period estimated by \citet{2020AJ....160..192S}, $P_{rot}\,=\,(1.878\,\pm\,0.005\,)$d. The result did not support the presence of a co-orbital. Nonetheless, we note that obtaining more precise RVs is currently possible for this target and it would likely improve this test. Thus, this analysis should be revised in the future.

\subsubsection*{L 168-9}
This M1V dwarf hosts a transiting super-Earth discovered by TESS in a close orbit (1.4\,days). Looking at the GLS of the activity indicators, it is not evident whether there is an effect of stellar activity in the RVs: HARPS indices show a forest of peaks near $P_{rot}$ and its harmonics ($\sim$20\,days based on \citealt{2020A&A...636A..58A}), meanwhile for PFS, there are no peaks related. Thus, we opt for not including a GP. Our analysis resulted in an $\alpha$ compatible with no co-orbital mass.

\subsubsection*{LHS 1478}
TESS detected a hot super-Earth in a 1.9\,days orbital period that was later confirmed through CARMENES RVs (\citealt{2021A&A...649A.144S}). In our analysis, we decide to not include the IRD data since they are sparse and do not add significant information to the $\alpha$-test. This M-dwarf star is inactive based on the $H_{\alpha}$ emission and therefore, we do not include any GP. We do not find hints for the presence of co-orbitals, but improvement in the analysis could be made by acquiring more RV measurements.

\subsubsection*{TOI-532}
This M-dwarf hosts a super Neptune ($\sim$\,62\,M$_{\oplus}$) orbiting with a 2.3 day period. The flat light curve detected by TESS suggests that this star is probably inactive (\citealt{2021AJ....162..135K}). There is no suspect so far that the planet could harbor a co-orbital based on the RVs dataset. Although we note that this target have not been extensively monitored and the uncertainty in $\alpha$ is potentially narrowable.

\subsubsection*{TOI-969}
The late K-dwarf star TOI-969 is transited by a close mini-Neptune in a 1.8\,day period (TOI-969\,b, located at the lower boundary of the hot-Neptune desert). The combination of 94 RV measurements with HARPS, PFS and CORALIE revealed the presence of a second companion in the system, the candidate TOI-969\,c ($\sim$\,11\,$M_{J}$). We do not include the CAROLIE dataset in our analysis since the associated uncertainties are too high to detect the transiting planet (tens of m/s). As in the original article (\citealt{2023A&A...669A.109L}), we model the RV long-term with a Keplerian of $P_c = 1700 \pm 290$\,d, which is the signal of the outer candidate. To account for the activity, we include a GP informed with the $\log{R'_{\rm HK}}$ computed for HARPS. The GLS of such indicator has its strongest peak at $\sim$24\,d, with aliases at 12 and 8\,d. The same signal appears in the WASP photometry according with \citet{2023A&A...669A.109L}, which confirms that it corresponds with the rotational period of the K-dwarf. The result does not favor the presence of a co-orbital companion ($\alpha = 0.03^{+0.16}_{-0.14}$).

\subsubsection*{TOI-1468}
Two transiting low-mass planets were detected and confirmed around this M3\,V star using TESS photometry, and a dataset of 81 CARMENES and MAROON-X RVs (\citealt{2022A&A...666A.155C}). The inner planet orbits the star with a 1.9\,d period (TOI-1468\,b), while the outer orbits every 15.5\,d. The H$\alpha$ index is a good proxy for the rotational period as its GLS shows a clear signal close to 41\,d. The preferred model of our analysis (including the GP and with the inner planet in a circular orbit) is compatible with no co-orbitals ($\alpha_b = 0.12^{+0.17}_{-0.10}$, and $\alpha_c = 0.01^{+0.11}_{-0.21}$).

\subsubsection*{TOI-1470}
The RV monitoring of the M1.5\,V TOI-1470 with CARMENES confirmed the TESS transiting sub-Neptune TOI-1470\,b ($P_b = 2.5$\,d), as well as other transiting planet in a 18.1\,d orbit first reported by \citet{2023A&A...675A.177G}, TOI-1470\,c. We use H$\alpha$ as spectral {activity} indicator of the rotational period ($\sim$\,29\,d) to perform a GP analysis. The results are compatible with no co-orbitals but with wide uncertainties ($\alpha_b = -0.18^{+0.14}_{-0.19}$, $\alpha_c = 0.44^{+0.39}_{-0.51}$).

\subsubsection*{TOI-1685}
The CARMENES (\citealt{2021A&A...650A..78B}) and IRD (\citealt{2021AJ....162..161H}) RV time series allowed us to confirm the planetary nature of the transiting super-Earth TOI-1685\,b, which orbits its M3\,V star with an ultra-short period of just 0.7\,d. \citet{2021A&A...650A..78B} reported a moderated evidence for the presence of a second not-transiting planet in the system (TOI-1685\,c) with an orbital period of 9.0\,d based on the CARMENES dataset. In our IRD + CARMENES combined analysis, we find an increase in the evidence for its presence, with a logarithm of the Bayes Factor of {10} when comparing the two-planets model with the single transiting planet. We also include a GP informed with the CARMENES dLW indicator, with a normal prior in the rotational period of ($18.7\pm0.7$)\,d. The result is affected of high uncertainty in the inferred $\alpha$ with the available dataset: $\alpha = 0.18^{+0.27}_{-0.16}$.

\subsubsection*{TOI-1728}
With an orbital period of  3.5\,d, the super-Neptune TOI-1728\,b transits an old and inactive M0-dwarf star (\citealt{2020ApJ...899...29K}). We include a linear trend to our RV model to account for a long-term with an inferred slope of ($-0.14 \pm 0.17$)\,$\rm{m}\,\rm{s}^{-1}\,\rm{d}^{-1}$. The results of the $\alpha$-test (compatible with no trojans with $\alpha = -0.22^{+0.23}_{-0.29}$) are here limited due to the relatively low-precision of the uncertainties associated with its HPF RV measurements, which are in the order of tens of m/s.

\subsubsection*{TOI-1759}

The sub-Neptune TOI-1759\,b was first detected through three transits from different TESS sectors. Later, independent works using CARMENES (\citealt{2022AJ....163..133E}) and SPIRou (\citealt{2022A&A...660A..86M}) RVs confirmed the planet orbiting the M0-dwarf star with a 18.8\,d period. Our analysis is only based on the CARMENES dataset since it has a better presicion compared with the SPIRou measurements, and we checked that the latter do not improve the inferred $\alpha$. We introduce a GP informed with the dLW, which shows the rotational periodicity at $\sim 36$\,d and is consistent with the inferred from the SPIRou dataset. The inferred $\alpha$ is $-0.23^{+0.24}_{-0.27}$.

\subsubsection*{TOI-1801}
TOI-1801\,b was identified by TESS photometry and confirmed by CARMENES and HIRES RVs. It is a transiting young mini-Neptune orbiting with a 10.6\,d period a M dwarf younger than 1\,Gyr (\citealt{2023A&A...680A..76M}). We include to the MCMC analysis a GP informed with the H$\alpha$ time series (peak at P$_{\rm rot} \sim 16$\,d in its GLS). The high scatter in the RVs even after the GP correction results in high uncertainties for the inferred $\alpha$ ($-0.05 ^{+0.33}_{-0.25}$).

\subsubsection*{TOI-2018}

The metal poor K-dwarf TOI-2018 is transited by a mini-Neptune with an orbital period of 7.4\,d. We use the available HIRES measurements correcting them from the stellar activity informing a GP through the $S_{\rm HK}$ index ($P_{\rm rot} \sim 21$\,d). A vague constraint can be made for the presence of co-orbitals given the high uncertainty of the result ($\alpha = 0.19^{+0.33}_{-0.25}$).

\subsubsection*{TOI-3785}

The M2-dwarf star TOI-3785 hosts a low-density transiting Neptune of 4.7\,d orbital period. The parent star is thought to be inactive (\citealt{2023AJ....166...44P}) thus no GP is introduced even there is a high scatter in the RV measurements. Note that six of the RV data points have a bigger uncertainty since they were taken with a shorter exposure time. The quality of the data do not allow the $\alpha$ parameter to be constrained, resulting in a non-informative $\alpha = -0.33 \pm 0.33$.

\subsection{Null detections}

\subsubsection*{GJ 1214}
This is a extensively studied planetary system in the literature. Only one planet has been detected (the sub-Neptune GJ~1214\,b) and the star is known to be inactive. The recent detection of its secondary eclipse (\citealt{2023Natur.620...67K}) confirmed that the planet is almost in a circular orbit. Our analysis was very well constrained not requiring a GP and with informative priors on $c$, and $d$. For this reason and in light of the result, this target is perfectly compatible with an scenario with no Trojan, with a very low upper limit to the Trojan mass (0.8\,$M_{\oplus}$, see Sect.~\ref{sec:dismass}). 

\subsubsection*{HAT-P-20}
HAT-P-20 has been thoroughly studied since its discovery (\citealt{2011ApJ...742..116B}). Multitude of transits and RV measurements have been gathered, including observations of the R-M (\citealt{2017A&A...601A..53E}), the absence of significant TTVs have been confirmed (e.g., \citealt{2017AJ....153...28S}; \citealt{2018A&A...618A..42L}), and the secondary eclipse has been observed twice in two \textit{Spitzer} bands (\citealt{2015ApJ...805..132D}) confirming a slightly eccentric orbit. As first noticed by \citet{2014ApJ...785..126K}, the RVs have a linear trend which is potentially caused by the visual companion of HAT-P-20. Thanks to these previous efforts, $\alpha$ is estimated with an small uncertainty being able to reject this object as a co-orbital candidate based on the $\alpha$-test ($\alpha = 0.000 \pm 0.002$). The result is compatible with the one already presented in \citealt{2018A&A...618A..42L}.

\subsubsection*{HIP 65 A}
This star (\object{TOI-129}) hosts a hot-Jupiter of $\sim3\,M_{J}$ in a $0.98\,$d orbital period. Its light curve reveals that the star is active, showing modulations with a periodicity of $\sim13\,$d. As argued by \citet{2020A&A...639A..76N}, the available RV dataset is unaffected by the activity since the impact is of the order of $\sim10\,\rm{m}\,\rm{s}^{-1}$, which is comparable to the RV uncertainty of the FEROS measurements, and 2-3 times smaller than those from CORALIE and CHIRON. Additionally, none of the available activity indicators show any relevant periodicity. The result places this target as a co-orbital candidate. Interestingly, the result differs when assuming a circular orbit, where the $\alpha$ drops to a value compatible with zero. This is a good example to show the importance of breaking the $\alpha$-eccentricity degeneracy through the secondary eclipse. In Sect.~\ref{sec:dis} we predict the expected occultation depths for all the targets, being HIP\,65\,A one of the most promising ($\delta~>~10^4\,$ppm). Our solution where $\alpha$ differs from zero corresponds with a slightly eccentric orbit that would imply that the occultation would happen around $18\,$min before of the expected time, at an orbital phase of $0.487\,\pm\,0.003$. Nonetheless, since this planet has a grazing transit, this exercise might not be possible.

\subsubsection*{LTT 1445 A}
LTT~1445~A is the primary star of a hierarchical triple system. It is orbited by the transiting planet LTT~1445~A\,b, which has an orbital period of 5.4\,d. \citealt{2022AJ....163..168W} found the RV signal of a second planetary body in a closest orbit, LTT~1445~A\,c with 3.1\,d. The result of the analysis reject this target as a candidate for both planets ({$\alpha_b$} = -0.07 $\pm$ 0.08, and $\alpha_c = 0.10^{+14}_{-0.13}$, for the circular case). 

\subsubsection*{TOI-244}
The bright M2V dwarf hosts a transiting super-Earth in a 7.4\,days orbital period with a characteristic low density. The spectral activity indicators (FWHM, $H_{\alpha}$ and S-index) clearly show the stellar rotational period in their periodograms at $P_{rot} \sim 56$\,days, indicating that this is an active star. As in the discovery article (\citealt{2023A&A...675A..52C}), we include a GP to mitigate this effect from the RVs by using the FWHM as proxy. Since after subtracting the stellar contribution the HARPS data still showing a big scatter, we opt for only including the ESPRESSO dataset to our analysis. As a result of the wealth of high-precision RV time-series, we can constrain the presence of co-orbitals to be below 0.8\,{$M_{\oplus}$} ($\alpha = -0.01 \pm 0.12$).

\subsubsection*{TOI-1075}
The Super-Earth TOI-1075\,b orbits its M0-type parent star with an ultra-short period of just 0.6\,d. The star does not show any sign of activity. The RV dataset requires a linear trend with an slope that we infer to be of 0.14\,$\rm{m}\,\rm{s}^{-1}\,\rm{d}^{-1}$, which might be attributed to an outer planetary-mass companion. The PSF RV time-series do not show any shift in the mid-transit time of TOI-1075\,b ($\alpha = 0.07 \pm 0.13$).

\subsubsection*{TOI-3629}
TESS discovered a Jupiter orbiting close to this M1-dwarf star with a 3.9 day period. The subsequent gathering of 28 RV measurements with NEID and HPF confirmed this candidate and enables rejecting the possibility of having a co-orbital object more massive than 28.2\,$M_{\oplus}$. Even this target is classified according with Table~\ref{tab:groups} as a \textit{Null Detection} ($\alpha = -0.01 \pm 0.13$), it would be desirable to decrease the Trojan upper limit by extending the sample and even increasing the precision of the {RVs} ($< 10$\,m\,s$^{-1}$).

\subsubsection*{WASP-80}
This interesting confirmed transiting planet, halfway between warm Neptunes and hot Jupiters (171\,$M_{\oplus}$, P = 3\,days), was already studied through the $\alpha$-test in \citet{2018A&A...609A..96L}. As its occultation was already known in this previous article, the target has a low uncertainty with an $\alpha = 0.038^{+0.083}_{-0.085}$. Six years later, the RV sample has been increased from 47 to 73 measurements, mainly grouped around the transit since its atmosphere has been studied with HARPS. Thanks to that observational effort, we now get $\alpha = -0.03\pm0.04$, reducing the uncertainty in a factor two and therefore, greatly decreasing the upper limit of the Trojan mass (from 44.7\,$M_{\oplus}$ to 8.7\,$M_{\oplus}$ in L$_{\rm 4}$, and from 28.5\,$M_{\oplus}$ to 19.4\,$M_{\oplus}$ in L$_{\rm 5}$). This example shows that gathering numerous RVs with orbital phases close to 0 might favor the search for co-orbitals by this method as the mid-transit time could be better constrained.

\subsection{Sparsely sampled}

\subsubsection*{GJ 1252}
GJ~1252\,b stands out as a Super-Earth of particular interest for characterization due to its notably short orbital period ($P\,=\,0.5\,$d), low stellar activity, and the observation of a day-side eclipse. Even though, the RVs for GJ~1252\,b have been poorly monitored, with a limited time span of less than 11\,days. Consequently, continued tracking of this target is imperative to accumulate more comprehensive data and derive a conclusive $\alpha$ parameter. 

\subsubsection*{TOI-824}
The K4-dwarf star is orbited every 1.4\,days by one of the few known hot Neptunes (with a radius of 2.9\,R$_{\oplus}$, and a mass around 18.5\,M$_{\oplus}$). The exotic of its nature and its elevated surface temperature makes it a very interesting target for atmospheric characterization. Therefore, its occultation has been measured with \textit{Spitzer} which has constrained the planet to be in a circular orbit. This under-sampled target is perfect to gather additional data to better constrain the $\alpha$ parameter and thus its possibility of harboring a co-orbital companion.

\subsection{Rejected case}

\subsubsection*{GJ 3090}
This star hosts a transiting mini-Neptune on a 2.9\,d orbit. \citet{2022A&A...665A..91A} reported the RV signal of a candidate second planet with an orbital period of 12.7\,d. Their RV analysis resulted in a solution where both planets are in very eccentric orbits ($e_b\,\sim\,0.2$, and $e_c\,\sim\,0.3$). Our priors do not allow for eccentric orbits for GJ~3090\,b since Eq.~\ref{eq:rv} is only valid when e~<~0.1. Despite the degeneracy between $e$ and $\alpha$, our solution do not favor the co-orbital scenario. The model found is quite noisy, which translates into a big uncertainty for the $\alpha$ parameter, probably caused by the presence of stellar activity. Indeed, the GLS periodogram of the RVs show some signal at the rotational period ($\sim\,18\,$d) and its first harmonic ($\sim\,9\,$d). As argued in \citet{2022A&A...665A..91A}, the quasi-periodic kernel in the GP does not converge, but adding a cosine component (QPC kernel) it improves. We repeat the same procedure using the QPC informed with the $H_{\alpha}$ indicator. Nonetheless, we still found problems in the convergence since GJ~3090\,b is restricted to not very eccentric orbits. This target is therefore outside the validity domain of the $\alpha$-test. For this reason, we do not include this target in the sample for the discussion.

\section{Additional tables}

\begin{landscape}
{\fontsize{7}{10}\selectfont
\begin{longtable}{lcccccccc}
\caption{Summary of the archival radial velocities and parameters adopted in this work.\label{tab:sumrvs_pt0}} \\

\hline 
\hline 
System & Planet &  N$_{\rm RVs}$ & N$_{\rm ins}^a$ & Timespan {[}d{]} & Transits & P {[}d{]} & $T_0$ -- 2 450 000 {[}d{]} & Ref.$^b$ RV / param. \\

\hline 

\endfirsthead

\caption{Continuation.} \\

\hline
\hline

System & Planet &  N$_{\rm RVs}$ & N$_{\rm ins}^a$ & Timespan {[}d{]} & Transits & P {[}d{]} & $T_0$ -- 2 450 000 {[}d{]} & Ref.$^b$ RV / param. \\
\hline

\endhead
\hline
\endfoot

GJ 143 & b & 283 & 4 & 5942.8 &  Yes  & 35.61253$^{+0.00060}_{-0.00062}$ & 8385.92502$^{+0.00054}_{-0.00055}$ & Dra19 \\
GJ 486 & b & 243 & 7 & 1967.1 &  Yes  & 1.467119$^{+0.000031}_{-0.000030}$ & 8931.15935~$\pm$~0.00042 & Cab22 / Tri21  \\
GJ 1214 & b & 163 & 2 & 3734.8 &  Yes  & 1.58040433~$\pm$~0.00000013 & 5701.413328$^{+0.000066}_{-0.000059}$ & Clo21a \\
GJ 1252 & b & 20 & 1 & 10.9 &  Yes  & 0.51824160~$\pm$~0.00000069 & 8668.09748~$\pm$~0.00032 & Shp20 / Luq22b  \\
GJ 3090 & b & 55 & 1 & 326.0 &  Yes  & 2.853136$^{+0.000064}_{-0.00038}$ & 8370.41849~$\pm$~0.00034 & Alm22a  \\
GJ 3470 & b & 193 & 5 & 4709.1 &  Yes  & 3.3366496$^{+0.0000039}_{-0.0000033}$ & 5983.70421$\pm$0.00010 & Lil18, Ste22 / Awi16  \\
GJ 3473 & b & 148 & 3 & 358.2 &  Yes  & 1.1980035$^{+0.0000018}_{-0.0000019}$ & 8492.20408$^{+0.00043}_{-0.00042}$ & Kem20 \\
 & c &  &  &  & No & 15.509~$\pm$~0.033 & 8575.6~$\pm$~2.2 &  \\
GJ 3929 & b & 97 & 2 & 528.6 &  Yes  & 2.616235~$\pm$~0.000005 & 8956.39620~$\pm$~ 0.00050 & Bea22, Kem22 / Bea22 \\
 & c &  &  &  &  No  & 15.040~$\pm$~0.030 & 9070.9~$\pm$~2.0 &  \\
HAT-P-20 & b & 57 & 4 & 2857.8 &  Yes  & 2.8753172~$\pm$~0.0000003 & 5917.64480461~$\pm$~0.00000010 & Bak11, Knu14, Esp17, Lil18 / Sun17  \\
HAT-P-54 & b & 17 & 2 & 410.0 &  Yes  & 3.799847~$\pm$~0.000014 & 6299.3037~$\pm$~0.00024 & Bak15  \\
HAT-P-68 & b & 13 & 3 & 1100.0 &  Yes  & 2.29840551~$\pm$~0.00000052 & 6614.20355~$\pm$~0.00014 & Lin21   \\
HATS-6 & b & 15 & 3 & 259.2 &  Yes  & 3.3252725~$\pm$~0.0000021 & 6643.74058~$\pm$~0.000084 & Har15 \\
HATS-47 & b & 12 & 1 & 151.7 &  Yes  & 3.9228038~$\pm$~0.0000022 & 7365.35804~$\pm$~0.00029 & Har20  \\
HATS-48 A & b & 11 & 1 & 125.8 &  Yes  & 3.1316666~$\pm$~0.0000037 & 7100.55022~$\pm$~0.00045 & Har20 \\
HATS-49 & b & 8 & 1 & 536.3 &  Yes  & 4.1480467~$\pm$~0.0000037 & 7105.1648~$\pm$~0.00054 & Har20 \\
HATS-71 & b & 7 & 1 & 744.0 &  Yes  & 3.7955202~$\pm$~0.0000010 & 7858.80134~$\pm$~0.00023 & Bak20 \\
HATS-74 A & b & 5 & 1 & 5.0 &  Yes  & 1.73185606~$\pm$~0.00000055 & 8392.02654~$\pm$~0.00024 & Jor22 \\
HATS-75 & b & 5 & 1 & 6.1 &  Yes  & 2.7886556~$\pm$~0.0000011 & 8611.05487~$\pm$~0.00027 & Jor22 \\
HATS-76 & b & 3 & 1 & 6.1 &  Yes  & 1.9416423~$\pm$~0.0000014 & 8424.55556~$\pm$~0.00053 & Jor22 \\
HATS-77 & b & 5 & 1 & 25.9 &  Yes  & 3.0876262~$\pm$~0.0000016 & 9136.69378~$\pm$~0.00020 & Jor22 \\
HD 73583 & b & 87 & 1 & 334.1 &  Yes  & 6.398042$^{+0.0000067}_{-0.0000062}$ & 8517.69013$^{+0.00056}_{-0.00059}$ & Tes21, Bar22 / Odd23  \\
 & c &  &  &  &  Yes  & 18.87974$^{+0.00086}_{-0.00074}$ & 9232.1682$^{+0.0019}_{-0.0024}$ & Tes21, Bar22 / Bar22   \\
HD 260655 & b & 180 & 2 & 8794.5 &  Yes  & 2.76953~$\pm$~0.000030 & 9497.91020~$\pm$~0.00030 & Luq22a \\
 & c &  &  &  &  Yes  & 5.70588~$\pm$~0.000070 & 9490.36460~$\pm$~0.00040 &  \\
HIP 65 A & b & 90 & 3 & 491.9 &  Yes  & 0.9809734~$\pm$~0.0000031 & 8326.102581$\pm$ 0.000050 & Nie20, Par21 / Nie20  \\
K2-18 & b & 146 & 2 & 1189.9 &  Yes  & 32.9396~$\pm$~0.0004 & 7264.39140~$\pm$~0.00070 & Rad22 \\
 & c &  &  &  &  No   & 9.207$^{+0.007}_{-0.006}$ & 7267.581$^{+0.480}_{-0.467}$ &  \\
K2-25 & b & 31 & 1 & 427.8 &  Yes  & 3.484545$^{+0.000042}_{-0.000043}$ & 7062.57958$^{+0.00054}_{-0.00056}$ & Ste20 / Tha20  \\
K2-141 & b & 74 & 2 & 1500.9 &  Yes  & 0.2803244~$\pm$~0.0000015 & 7744.07160~$\pm$~0.00022 & Bar18, Mal18, Bon23 / Bon23  \\
 & c &  &  &  &  Yes  & 7.7485~$\pm$~0.00022 & 7751.1546~$\pm$~0.001 &  \\
K2-199 & b & 45 & 1 & 1071.0 &  Yes      & 3.2253993~$\pm$~0.0000024 & 7218.73733$^{+0.00051}_{-0.00055}$ & Aka21 \\
 & c &  &  &  &  Yes       & 7.3744897~$\pm$~0.0000037 & 7222.93034~$\pm$~0.00035 & Aka21 \\
K2-216 & b & 26 & 4 & 488.7 &  Yes        & 2.1748~$\pm$~0.000050 & 7394.04170~$\pm$~0.00090 & Per18 \\
K2-295 & b & 6 & 1 & 108.7 &  Yes        & 4.024867~$\pm$~0.000015 & 7395.4140500~$\pm$~0.0000010 & Smi19 \\
Kepler-45 & b & 14 & 1 & 92.7 &  Yes        & 2.455239~$\pm$~0.0000040 & 5003.822000~$\pm$~0.000036 & Joh12 \\
L 168-9 & b & 123 & 2 & 106.0 &  Yes        & 1.4015~$\pm$~0.00018 & 8340.04781$^{+0.00088}_{-0.00122}$ & Ast20 \\
LHS 1140 & b & 254 & 3 & 1470.9 &  Yes        & 24.737230~$\pm$~0.000020 & 8399.93000~$\pm$~+0.00030 & Men19, Cad24 / Cad24    \\
 & c &  &  &  &  Yes  & 3.777940~$\pm$~0.000020 & 8389.29390~$\pm$~+0.00020 & Men19, Cad24 / Cad24   \\
LHS 1478 & b & 57 & 1 & 261.3 &  Yes        & 1.9495378$^{+0.0000040}_{-0.0000041}$ & 8786.75425~$\pm$~0.00042 & Sot21 \\
LP 714-47 & b & 81 & 5 & 314.5 &  Yes        & 4.052037~$\pm$~0.0000040 & 9196.11490 $^{+0.00031}_{-0.00030}$ & Dre20 \\
LTT 1445 A & b & 135 & 5 & 643.4 &  Yes       & 5.3587657$^{+0.0000043}_{-0.0000042}$ & 8412.70851$^{+0.00040}_{-0.00039}$ & Win22 / Odd23    \\
 & c &  &  &  &  Yes        & 3.1239035$^{+0.0000034}_{-0.0000036}$ & 8412.58159$^{+0.00059}_{-0.00057}$ & Win22 \\
LTT 3780 & b & 119 & 4 & 267.1 &  Yes      & 0.768448$^{+0.000055}_{-0.000053}$ & 8543.9115~$\pm$~0.0011 & Now20, Clo20a / Luq22b   \\
 & c &  &  &  &  Yes      & 12.2519$^{+0.0028}_{-0.0030}$ & 8546.8484~$\pm$~0.0014 & Now20, Clo20a / Luq22b   \\
NGTS-1 & b & 7 & 1 & 10.9 &  Yes      & 2.647298~$\pm$~0.000020 & 7720.659396~$\pm$~0.00062 & Bay18 \\
NGTS-10 & b & 10 & 1 & 413.8 &  Yes      & 0.7668944~$\pm$~0.00000030 & 7518.84377~$\pm$~0.00017 & McC20 \\
POTS-1 & b & 13 & 2 & 462.9 &  Yes      & 3.1606296~$\pm$~0.0000016 & 4231.65488~$\pm$~0.00044 & Kop13 \\
Qatar-2  & b & 80 & 4 & 2365.6 &  Yes      & 1.33711647~$\pm$~0.00000026 & 5624.267096~$\pm$~0.000087 & Bon17, Esp17 / Man14   \\
Qatar-9  & b & 9 & 1 & 107.9 &  Yes      & 1.540731~$\pm$~0.000038 & 8227.75643~$\pm$~0.00027 & Als19 \\
TOI-244  & b & 57 & 1 & 357.9 &  Yes       & 7.397225~$\pm$~0.000026 & 8357.3627~$\pm$~0.0020 & Cas23 \\
TOI-269  & b & 81 & 1 & 275.2 &  Yes       & 3.6977104~$\pm$~0.0000037 & 2458381.84668~$\pm$~0.00033 & Coi21 \\
TOI-519  & b & 17 & 1 & 265.2 &  Yes      & 1.2652328~$\pm$~0.00000050 & 8491.877117~$\pm$~0.00013 & Kag23 \\
TOI-530  & b & 15 & 1 & 9.0 &  Yes      & 6.387597$^{+0.000019}_{-0.000018}$ & 8470.1998$^{+0.0016}_{-0.0017}$ & Gan22 \\
TOI-532  & b & 18 & 1 & 134.8 &  Yes      & 2.3266508~$\pm$~0.0000030 & 8470.57678$^{+0.00086}_{-0.00090}$ & Kan21 \\
TOI-544  & b & 122 & 2 & 1051.2 &  Yes      & 1.5483510~$\pm$~0.0000015 & 8469.75700~$\pm$~0.00050 & Osb23 / Gia22   \\
 & c &  &  &  &   No  & 50.089~$\pm$~0.24 & 9212.0~$\pm$~1.9 & Osb23  \\
TOI-674 & b & 17 & 1 & 54.9 &  Yes     & 1.9771430~$\pm$~0.0000030 & 8641.404552~$\pm$~0.00010 & Mur21 \\
TOI-776 & b & 64 & 2 & 545.7 &  Yes      & 8.246620$^{+0.000024}_{-0.000031}$ & 9288.8713$^{+0.0010}_{-0.0011}$ & Fri24 \\
 & c &  &  &  &  Yes      & 15.665323$^{+0.000075}_{-0.000070}$ & 9324.53478$^{+0.00080}_{-0.00077}$ & Fri24 \\
TOI-824 & b & 17 & 2 & 39.9 &  Yes    & 1.392978$^{+0.000018}_{-0.000017}$ & 8639.60354~$\pm$~0.00035 & Bur20 \\
TOI-836 & b & 52 & 1 & 351.1 &  Yes  & 3.816730~$\pm$~0.000010 & 8599.9953~$\pm$~0.0019 & Haw23 \\
 & c &  &  &  &  Yes  & 8.595450~$\pm$~0.000010 & 8599.76230~$\pm$~0.00080 &  \\
TOI-969 & b & 73 & 2 & 141.7 &  Yes  & 1.8237305$^{+0.0000020}_{-0.0000021}$ & 9248.37709$^{+0.00036}_{-0.00039}$ & Lil23 \\
 & c &  &  &  &  No  & 1700.0$^{+290.0}_{-280.0}$ & 10640.0~$\pm$~260.0 & Lil23 \\
TOI-1075 & b & 54 & 1 & 176.7 &  Yes      & 0.6047328~$\pm$~0.0000032 & 8654.25100$^{+0.00040}_{-0.00050}$ & Ess23  \\
TOI-1130 & b & 48 & 1 & 62.0 &  Yes  &  4.07445~$\pm$~0.00046  &  8658.7405~$\pm$~0.0013  & Kor23 \\
 & c &  &  &  &  Yes      & 8.350381$^{+0.000032}_{-0.000033}$ & 8657.90322~$\pm$~0.00030 & Kor23   \\
TOI-1201 & b & 33 & 1 & 100.8 &  Yes      & 2.4919863$^{+0.0000030}_{-0.0000031}$ & 9169.23222$^{+0.00052}_{-0.00054}$ & Kos21 \\
TOI-1231 & b & 14 & 1 & 272.3 &  Yes       & 24.245586$^{+0.000064}_{-0.000066}$ & 8685.11630$^{+0.00048}_{-0.00049}$ & Bur21 \\
TOI-1235 & b & 97 & 3 & 128.9 &  Yes        & 3.444717$^{+0.000040}_{-0.000042}$ & 8683.6155$^{+0.0017}_{-0.0015}$ & Blu20, Clo20b / Luq22b    \\
TOI-1278 & b & 10 & 1 & 154.7 &  Yes        & 14.47567~$\pm$~0.00021 & 8711.9595~$\pm$~0.0013 & Art21 \\
TOI-1452 & b & 52 & 1 & 124.8 &  Yes        & 11.06201~$\pm$~0.000020 & 8691.5321~$\pm$~0.0015 & Cad22 \\
TOI-1468 & b & 81 & 2 & 686.4 &  Yes        & 1.8805136$^{+0.0000024}_{-0.0000026}$ & 8765.68079$^{+0.00007}_{-0.00069}$ & Cha22 \\
 & c &  &  &  &  Yes      & 15.532482$^{+0.000034}_{-0.000033}$ & 8766.9269~$\pm$~0.0012 & Cha22 \\
TOI-1470 & b & 43 & 1 & 228.7 &  Yes      & 2.527093~$\pm$~0.000040 & 8766.47020~$\pm$~0.00060 & Gon23 \\
TOI-1634 & b & 32 & 1 & 209.7 &  Yes      & 0.989343~$\pm$~0.000015 & 8791.51473~$\pm$~0.00061 & Clo21b / Luq22b   \\
TOI-1685  & b & 91 & 2 & 206.1 &  Yes       & 0.6691403$^{+0.0000023}_{-0.0000021}$ & 8816.22615$^{+0.00050}_{-0.0006}$ & Blu21, Luq22b / Luq22b   \\
TOI-1695  & b & 220 & 2 & 417.4 &  Yes       & 3.1342791$^{+0.0000071}_{-0.0000063}$ & 8791.5206$^{+0.0010}_{-0.0011}$ & Che23, Kie23 / Che23   \\
TOI-1728  & b & 30 & 1 & 52.0 &  Yes      & 3.49151$^{+0.000062}_{-0.000057}$ & 8843.27427~$\pm$~0.00043 & Kan20 \\
TOI-1759  & b & 57 & 1 & 177.7 &  Yes      & 18.85019~$\pm$~0.00013 & 8745.4654~$\pm$~0.0011 & Esp22  \\
TOI-1801  & b & 109 & 2 & 585.7 &  Yes  & 10.64386$^{+0.00005}_{-0.00006}$ & 2 458 903.54333$^{+0.00337}_{-0.00340}$ & Mal23 \\
TOI-1899  & b & 15 & 1 & 56.9 &  Yes       & 29.02$^{+0.36}_{-0.23}$ & 8711.9578~$\pm$~0.0012 & Cañ20 \\
TOI-2018  & b & 38 & 1 & 3725.9 &  Yes      & 7.435583~$\pm$~0.000022 & 8958.258~$\pm$~0.0013 & Dai23 \\
TOI-3235  & b & 7 & 1 & 12.1 &  Yes        & 2.59261842~$\pm$~0.00000041 & 9690.00173~$\pm$~0.000045 & Hob23 \\
TOI-3629  & b & 28 & 2 & 361.0 &  Yes        & 3.936551$^{+0.0000050}_{-0.0000060}$ & 8784.256~$\pm$~0.0010 & Cañ22 \\
TOI-3714  & b & 26 & 3 & 136.0 &  Yes        & 2.154849~$\pm$~0.0000010 & 8840.5093~$\pm$~0.00040 & Cañ22, Har23 / Cañ22     \\
TOI-3757  & b & 27 & 2 & 130.9 &  Yes        & 3.438753~$\pm$~0.0000040 & 8838.77148$^{+0.00062}_{-0.00061}$ & Kan22 \\
TOI-3785  & b & 39 & 2 & 557.7 &  Yes        & 4.674737~$\pm$~0.0000038 & 8861.49553~$\pm$~0.00060 & Pow23 \\
TOI-3884  & b & 19 & 2 & 222.5 &  Yes       & 4.5445697~$\pm$~0.0000094 & 9642.86314~$\pm$~0.00012 & Alm22b, Lib23 / Alm22b  \\
TOI-3984 A  & b & 41 & 2 & 260.1 &  Yes      & 4.353326~$\pm$~0.0000050 & 9715.022680~$\pm$~0.000095 & Cañ23 \\
TOI-4201  & b & 12 & 1 & 122.9 &  Yes      & 3.5819134~$\pm$~0.0000017 & 8470.96190~$\pm$~0.00040 & Har23 \\
TOI-4860  & b & 7 & 1 & 26.1 &  Yes       & 1.52275959~$\pm$~0.00000035 & 9832.641439~$\pm$~0.000032 & Tri23 \\
TOI-5205  & b & 7 & 1 & 28.9 &  Yes      & 1.630757~$\pm$~0.0000010 & 9443.47179~$\pm$~0.00019 & Kan23 \\
TOI-5293 A  & b & 16 & 1 & 60.8 &  Yes      & 2.930289~$\pm$~0.0000040 & 9448.91480~$\pm$~0.00040 & Cañ23 \\
TOI-5344  & b & 13 & 1 & 130.9 &  Yes      & 3.7926220~$\pm$~0.0000062 & 9848.99030~$\pm$~0.00019 & Har23 \\
WASP-43  & b & 246 & 2 & 2898.3 &  Yes      & 0.8134750~$\pm$~0.0000010 & 5528.86774~$\pm$~0.00014 &  ESO Archive$^{c}$ / Hoy16  \\
WASP-80 & b & 54 & 2 & 776.8 &  Yes      & 3.06785234$^{+0.00000083}_{-0.00000079}$ & 6487.425006$^{+0.000023}_{-0.000025}$ & Tri13 / Tri15   \\
Wendelstein-1  & b & 9 & 2 & 1443.9 &  Yes      & 2.6634160~$\pm$~0.0000010 & 5367.738464~$\pm$~0.000014 & Obe20 \\

\end{longtable}
\tablefoot{\tablefoottext{a}{We consider as two different instruments a single instrument with a change in the offset at a given epoch (e.g., caused by an instrumental upgrade).}
\tablefoottext{b}{Aka21: \citet{2021AJ....162..294A};
Alm22a: \citet{2022A&A...665A..91A};
Alm22b: \citet{2022A&A...667L..11A};
Als19: \citet{2019AJ....157..224A};
Art21: \citet{2021AJ....162..144A};
Ast20: \citet{2020A&A...636A..58A};
Awi16: \citet{2016MNRAS.463.2574A};
Bak11: \citet{2011ApJ...742..116B};
Bak15: \citet{2015AJ....149..149B};
Bak20: \citet{2020AJ....159..267B};
Bay18: \citet{2018MNRAS.475.4467B};
Bar18: \citet{2018A&A...612A..95B};
Bar22: \citet{2022MNRAS.514.1606B};
Bea22: \citet{2022ApJ...936...55B};
Blu20: \citet{2020A&A...639A.132B};
Blu21: \citet{2021A&A...650A..78B};
Bon17: \citet{2017A&A...602A.107B};
Bon23: \citet{2023A&A...677A..33B};
Bur20: \citet{2020AJ....160..153B};
Bur21: \citet{2021AJ....162...87B};
Cab22: \citet{2022A&A...665A.120C};
Cad22: \citet{2022AJ....164...96C};
Cad24: \citet{2024ApJ...960L...3C};
Cas23: \citet{2023A&A...675A..52C};
Cañ20: \citet{2020AJ....160..147C};
Cañ22: \citet{2022AJ....164...50C};
Cañ23: \citet{2023AJ....166...30C};
Cha22: \citet{2022A&A...666A.155C};
Che23: \citet{2023AJ....165..167C};
Clo20a: \citet{2020AJ....160....3C};
Clo20b: \citet{2020AJ....160...22C};
Clo21a: \citet{2021AJ....162..174C};
Clo21b: \citet{2021AJ....162...79C};
Coi21: \citet{2021A&A...650A.145C};
Dai23: \citet{2023AJ....166...49D};
Dra19: \citet{2019ApJ...875L...7D};
Dre20: \citet{2020A&A...644A.127D};
Ess23: \citet{2023AJ....165...47E};
Esp17: \citet{2017A&A...601A..53E};
Esp22: \citet{2022AJ....163..133E};
Fri24: \citet{2024A&A...684A..12F}; 
Gan22: \citet{2022MNRAS.511...83G};
Gia22: \citet{2022AJ....163...99G};
Gon23: \citet{2023A&A...675A.177G};
Har15: \citet{2015AJ....149..166H};
Har20: \citet{2020AJ....159..173H};
Har23: \citet{2023AJ....166..163H};
Haw23: \citet{2023MNRAS.520.3649H};
Hob23: \citet{2023ApJ...946L...4H};
Hoy16: \citet{2016AJ....151..137H};
Joh12: \citet{2012AJ....143..111J};
Jor22: \citet{2022AJ....163..125J};
Kag23: \citet{2023PASJ...75..713K};
Kan20: \citet{2020ApJ...899...29K};
Kan21: \citet{2021AJ....162..135K};
Kan22: \citet{2022AJ....164...81K};
Kan23: \citet{2023AJ....165..120K};
Kem20: \citet{2020A&A...642A.236K};
Kem22: \citet{2022A&A...659A..17K};
Kie23: \citet{2023A&A...670A.136K};
Knu14: \citet{2014ApJ...785..126K};
Kop13: \citet{2013MNRAS.435.3133K};
Kor23: \citet{2023A&A...675A.115K};
Kos21: \citet{2021A&A...656A.124K};
Lib23: \citet{2023arXiv230204757L};
Lil18: \citet{2018A&A...618A..42L};
Lil23: \citet{2023A&A...669A.109L};
Lin21: \citet{2021AJ....161...64L};
Luq22a: \citet{2022A&A...664A.199L};
Luq22b: \citet{2022Sci...377.1211L};
Mal18: \citet{2018AJ....155..107M};
Mal23: \citet{2023A&A...680A..76M};
Man14: \citet{2014MNRAS.443.2391M};
McC20: \citet{2020MNRAS.493..126M};
Men19: \citealt{2019AJ....157...32M};
Mur21: \citet{2021A&A...653A..60M};
Nie20: \citet{2020A&A...639A..76N};
Now20: \citet{2020A&A...642A.173N};
Obe20: \citet{2020A&A...639A.130O};
Odd23: \citet{2023AJ....165..134O};
Osb23: \citet{2023arXiv231014908O};
Par21: \citet{2021AJ....162..176P};
Per18: \citet{2018A&A...618A..33P};
Pow23: \citet{2023AJ....166...44P};
Rad22: \citet{2022MNRAS.517.5050R};
Smi19: \citet{2019AcA....69..135S};
Sot21: \citet{2021A&A...649A.144S};
Sph20: \citet{2020ApJ...890L...7S};
Ste20: \citet{2020AJ....160..192S};
Ste22: \citet{2022ApJ...931L..15S};
Sun17: \citet{2017AJ....153...28S};
Tes21: \citet{2021ApJS..256...33T};
Tha20: \citet{2020AJ....159...32T};
Tri13: \citet{2013A&A...551A..80T};
Tri15: \citet{2015MNRAS.450.2279T};
Tri21: \citet{2021Sci...371.1038T};
Tri23: \citet{2023MNRAS.525L..98T};
Win22: \citet{2022AJ....163..168W}.
}
\tablefoottext{c}{\url{http://archive.eso.org/eso/eso_archive_main.html}}
}
}
\end{landscape}

\setlength{\tabcolsep}{10pt}
\begin{table*}[h!]
\centering
\caption[]{Radial velocity datasets per system.}
\label{tab:rvs}
\begin{tabular}{ccccc}
\hline
\hline 
System & JD & RV [m/s] & $\Delta$RV[m/s] & Instrument \\ 
\hline 
GJ 1214 & 2454993.76749	& -- 8.239 & 2.275 & HARPS pre-upgrade \\ 
GJ 1214	& 2455036.57373	& -- 18.365 & 2.503 & HARPS pre-upgrade \\ 
GJ 1214	& 2455036.65153	& -- 13.028 & 2.385 & HARPS pre-upgrade \\ 
GJ 1214	& 2455037.58578	& 4.096	& 2.272	& HARPS pre-upgrade \\ 
GJ 1214	& 2455037.65309	& -- 9.916 & 2.048 & ... \\ 
\hline 
\end{tabular}
\tablefoot{Full table is available at the CDS.}
\end{table*}

{\footnotesize
\begin{table}[h!]
\centering
\caption[]{Eccentricity constrained by secondary eclipses. Orbital parameters referred to the planet frame (transit at a true anomaly $\nu$ = -90$^\circ$). \label{tab:occ}}
\label{tab:occ}
\begin{tabular}{lcccccc}
\hline
\hline

Planet & $e\cos{\omega}$ & $e\sin{\omega}$ & $e$ & Ref. occ.$^a$\\

\hline

GJ 1214 b & --\,0.00094~$\pm$~0.00018    & --\,0.0005~$\pm$~0.0029    & 0.0011~$\pm$~0.0014 & Kem23 \\ 
GJ 1252 b & --\,0.0027~$\pm$~0.0058 & --\,0.012~$\pm$~0.012  & 0.012~$\pm$~0.012  & Cro22 \\ 
GJ 3470 b & --\,0.019~$\pm$~0.007    & 0.001~$\pm$~0.178 & 0.019~$\pm$~0.014 & Ben19 \\ 
HAT-P-20 b & --\,0.01352~$\pm$~0.00060    & 0.0170~$\pm$~0.0034     & 0.0217~$\pm$~0.0027 &  Dem15 \\ 
TOI-824 b  & 0.0001~$\pm$~0.0083     & --\,0.0003~$\pm$~0.0083 & 0.0003~$\pm$~0.0042& Roy22 \\ 
WASP-43 b  & --\,0.049~$\pm$~0.011 & --\,0.03~$\pm$~0.14 & 0.057~$\pm$~0.076 & Ble14 \\ 
WASP-80 b  & --\,0.023~$\pm$~0.001 & --\,0.007~$\pm$~0.014 & 0.024~$\pm$~0.004 & Tri15 \\ 
\hline
\end{tabular}
\end{table}
\tablefoot{
\tablefoottext{a}{
Ben19: \citet{2019NatAs...3..813B};
Ble14: \citet{2014ApJ...781..116B};
Cro22: \citet{2022ApJ...937L..17C};
Dem15: \citet{2015ApJ...805..132D};
Kem23: \citet{2023Natur.620...67K};
Roy22: \citet{2022ApJ...941...89R};
Tri15: \citet{2015MNRAS.450.2279T}
}
}
}

\begin{landscape}
{\fontsize{8}{10}\selectfont
\begin{longtable}{lccccccccc}
\caption{Results of the $\alpha$ parameter per model and upper limit to the trojan mass.\label{tab:alpha_res}} \\

\hline
\hline
& \multicolumn{4}{c}{$\alpha$} & & & & \multicolumn{2}{c}{Upper limit Trojan mass [$M_{\oplus}$]} \\  
  \cline{2-5} \cline{9-10}
  \multirow{-2}{*}{Planet} & e & c & GPe & GPc & \multirow{-2}{*}{Best model} & \multirow{-2}{*}{$|\alpha|$~/~$\sigma_{\alpha}$} & \multirow{-2}{*}{$m_p$ [$M_{\oplus}$]} & L$_{\mathrm{4}}$ & L$_{\mathrm{5}}$ \\

\hline

\endfirsthead

\caption{Continuation.} \\

\hline
\hline
& \multicolumn{4}{c}{$\alpha$} & & & & \multicolumn{2}{c}{Upper limit Trojan mass [$M_{\oplus}$]} \\  
  \cline{2-5} \cline{9-10}
  \multirow{-2}{*}{Planet} & e & c & GPe & GPc & \multirow{-2}{*}{Best model} & \multirow{-2}{*}{$|\alpha|$~/~$\sigma_{\alpha}$} & \multirow{-2}{*}{$m_p$ [$M_{\oplus}$]} & L$_{\mathrm{4}}$ & L$_{\mathrm{5}}$ \\
\hline
\endhead
\hline
\endfoot
\hline 

& & & & Strong Candidates & & & & & \\
\hline
GJ~3470\,b  & --\,0.17~$\pm$~0.05 &    & --\,0.16~$\pm$~0.05 &   &  GPe &  3.2 & 13.9 &  0.0 & 4.1 \\
\hline
& & & & Weak Candidates & & & & & \\
\hline
GJ~486\,b  & --\,0.05~$\pm$~0.05 & -0.02~$\pm$~0.02 &     &      &  c &  1.2 & 2.8 & 0.0 & 0.2 \\
GJ~3473\,b  & --\,0.30$_{-0.24}^{+0.23}$ & --\,0.28$_{-0.22}^{+0.21}$ &      &      &  c &  1.2 & 1.9 & 0.3 & 1.6 \\
HD 260655 c &  --\,0.32$_{-0.25}^{+0.22}$ & --\,0.34$_{-0.23}^{+0.20}$ &    &    &  c &  1.5 & 3.1 & 0.2 & 3.0 \\
K2-18 b & 0.38$_{-0.26}^{+0.28}$ & 0.39$_{-0.26}^{+0.28}$ & 0.31$_{-0.28}^{+0.30}$ & 0.48$_{-0.28}^{+0.29}$ &  GPc &  1.7 & 8.9 & 9.8 & 0.9 \\
K2-141 b & --\,0.13$_{-0.25}^{+0.26}$ & --\,0.13$_{-0.23}^{+0.24}$ & --\,0.04$_{-0.09}^{+0.11}$ & 0.07~$\pm$~0.07 &  GPc &  1.0 & 5.1 & 1.3 & 0.4 \\
K2-199 c &  --\,0.25$_{-0.24}^{+0.23}$ & --\,0.26$_{-0.21}^{+0.20}$ &     &     &  c &  1.2 & 12.4 & 2.3 & 10.3 \\
LHS 1140 b & 0.10~$\pm$~0.11 & 0.14$_{-0.11}^{+0.10}$ & --\,0.06$_{-0.06}^{+0.02}$ & --\,0.11$_{-0.02}^{+0.07}$ &  GPc &  1.5 & 6.4 & 0.2 & 1.2 \\
LHS 1140 c &  0.06$_{-0.17}^{+0.16}$ & 0.01~$\pm$~0.12 & 0.08$_{-0.07}^{+0.03}$ & --\,0.04~$\pm$~0.02 &  GPc &  1.8 & 1.8 & 0.0 & 0.2 \\
LP 714-47 b & --\,0.15~$\pm$~0.09 & --\,0.12~$\pm$~0.05 &      &      &  c &  2.2 & 30.8 & 0.0 & 8.1 \\
LTT 3780 b & --\,0.20$_{-0.17}^{+0.16}$ & --\,0.22~$\pm$~0.13 &      &      &  c &  1.7 & 2.6 & 0.1 & 1.5 \\
Qatar-2 b & --\,0.05~$\pm$~0.03 & --\,0.02~$\pm$~0.02 &      &      &  c &  1.0 & 792.6 & 14.2 & 50.3 \\
TOI-269 b & 0.27$_{-0.21}^{+0.22}$ & 0.29$_{-0.18}^{+0.20}$ &     &     &  c &  1.5 & 8.8 & 7.3 & 0.8 \\
TOI-544 b & 0.43$_{-0.36}^{+0.33}$ & 0.43$_{-0.35}^{+0.32}$ & 0.57$_{-0.24}^{+0.19}$ & 0.57$_{-0.22}^{+0.19}$ &  GPc &  2.6 & 2.9 & 3.1 & 0.0 \\
TOI-776 b & --\,0.62$_{-0.25}^{+0.31}$ & --\,0.62$_{-0.24}^{+0.29}$ &     &     & c &  2.1 & 4.0 & 0.1 & 4.5 \\
TOI-836 b & 0.32$_{-0.43}^{+0.40}$ & 0.33$_{-0.39}^{+0.37}$ & 0.38$_{-0.18}^{+0.19}$ & 0.38$_{-0.14}^{+0.16}$ &  GPc &  2.4 & 4.5 & 3.7 & 0.0 \\
TOI-836 c &  --\,0.45$_{-0.34}^{+0.39}$ & --\,0.48$_{-0.33}^{+0.38}$ & --\,0.71$_{-0.20}^{+0.28}$ & --\,0.74$_{-0.18}^{+0.27}$ &  GPc &  2.8 & 9.6 & 0.0 & 11.0 \\
TOI-1130 b & --\,0.11$_{-0.18}^{+0.17}$ & --\,0.44$_{-0.11}^{+0.12}$ &     &     &  c &  3.6 & 19.3 & 0.0 & 15.7 \\
TOI-1130 c &  --\,0.00~$\pm$~0.02 & --\,0.09~$\pm$~0.01 &     &     &  c &  9.0 & 309.6 & 0.0 & 38.9 \\
TOI-1235 b & 0.23$_{-0.23}^{+0.25}$ & 0.22$_{-0.20}^{+0.23}$ & 0.10~$\pm$~0.15 & 0.12$_{-0.11}^{+0.12}$ &  GPc &  1.0 & 5.9 & 2.5 & 0.7 \\
TOI-1452 b & --\,0.51$_{-0.30}^{+0.33}$ &     &     &     & e &  1.5 & 4.8 & 1.2 & 5.4 \\
TOI-1695 b & --\,0.34$_{-0.19}^{+0.40}$ & --\,0.40$_{-0.16}^{+0.14}$ &      &      &  c &  2.5 & 6.4 & 0.0 & 5.4 \\
TOI-3757 b & --\,0.27~$\pm$~0.17 & --\,0.25~$\pm$~0.13 &     &     &  c &  1.9 & 85.3 & 0.6 & 52.6 \\
TOI-3884 b & --\,0.42$_{-0.28}^{+0.27}$ & --\,0.41$_{-0.27}^{+0.25}$ &      &      &  c &  1.5 & 16.5 & 2.8 & 17.5 \\
TOI-3984 A b & 0.34~$\pm$~0.28 & 0.34$_{-0.26}^{+0.27}$ &      &      &  c &  1.3 & 44.5 & 44.9 & 11.5 \\
WASP-43 b & --\,0.007~$\pm$~0.005 &      &    &      & e &  1.5 & 565.7 & 1.7 & 10.6 \\
\hline
& & & &  Inconclusive & & & & & \\
\hline
GJ~143\,b & --\,0.17$_{-0.09}^{+0.08}$ &      & --\,0.27$_{-0.40}^{+0.44}$ &     &  GPe &  0.6 & 22.7 & 19.1 & 24.5 \\
GJ 3929 b & --\,0.13$_{-0.38}^{+0.35}$ & --\,0.14$_{-0.36}^{+0.33}$ &     &     &  c &  0.4 & 1.8 & 1.2 & 1.7 \\
HD 73583 b & --\,0.29$_{-0.41}^{+0.46}$ & --\,0.29$_{-0.41}^{+0.39}$ & --\,0.11~$\pm$~0.41 & --\,0.02$_{-0.05}^{+0.06}$ &  GPc &  0.3 & 10.2 & 9.4 & 10.6 \\
HD 73583 c &  0.41$_{-0.39}^{+0.36}$ & 0.42$_{-0.39}^{+0.35}$ & 0.25$_{-0.28}^{+0.33}$ & --\,0.02$_{-0.05}^{+0.06}$ &  GPc &  0.7 & 9.7 & 10.5 & 5.1 \\
HD 260655 b & 0.17~$\pm$~0.23 & 0.14$_{-0.20}^{+0.21}$ &    &    &  c &  0.7 & 2.1 & 1.5 & 0.7 \\
K2-25 b & --\,0.22$_{-0.46}^{+0.50}$ & --\,0.22$_{-0.46}^{+0.50}$ & --\,0.18$_{-0.52}^{+0.61}$ & --\,0.16$_{-0.51}^{+0.59}$ &  GPc &  0.7 & 24.5 & 22.8 & 27.4 \\
K2-199 b & 0.02~$\pm$~0.32 & 0.01$_{-0.29}^{+0.30}$ &      &      &  c &  0.0 & 6.9 & 5.5 & 5.2 \\
L 168-9 b & 0.08~$\pm$~0.19 & --\,0.05~$\pm$~0.16 &     &     &  c &  0.3 & 4.6 & 2.1 & 1.6 \\
LHS 1478 b & 0.15$_{-0.23}^{+0.22}$ & 0.10~$\pm$~0.20 &      &      &  c &  0.5 & 2.3 & 1.5 & 0.9 \\
TOI-532 b & 0.07$_{-0.27}^{+0.25}$ & 0.02~$\pm$~0.25 &     &     &  c &  0.1 & 61.5 & 39.6 & 39.4 \\
TOI-969 b & 0.21$_{-0.28}^{+0.31}$ & 0.20$_{-0.2}^{+0.29}$ & 0.06$_{-0.18}^{+0.19}$ & 0.03$_{-0.14}^{+0.16}$ &  GPc &  0.2 & 9.1 & 4.1 & 2.7 \\
TOI-1468 b & 0.00~$\pm$~0.21 & 0.04~$\pm$~0.19 & --\,0.05$_{-0.10}^{+0.22}$ & 0.13$_{-0.10}^{+0.17}$ &  GPc &  0.8 & 3.2 & 1.2 & 0.7 \\
TOI-1468 c &  --\,0.19$_{-0.27}^{+0.26}$ & --\,0.20~$\pm$~0.25 & 0.03$_{-0.22}^{+0.11}$ & 0.01$_{-0.21}^{+0.11}$ &  GPc &  0.0 & 6.6 & 1.8 & 3.0 \\
TOI-1470 b & --\,0.12$_{-0.28}^{+0.29}$ & --\,0.13$_{-0.25}^{+0.28}$ & --\,0.20~$\pm$~0.20 & --\,0.18$_{-0.19}^{+0.14}$ &  GPc &  0.9 & 7.3 & 1.3 & 5.2 \\
TOI-1470 c &  --\,0.14$_{-0.29}^{+0.28}$ & --\,0.13$_{-0.29}^{+0.28}$ & 0.30$_{-0.47}^{+0.43}$ & 0.44$_{-0.51}^{+0.39}$ &  GPc &  0.9 & 7.2 & 8.0 & 4.9 \\
TOI-1685 b & 0.16$_{-0.20}^{+0.21}$ & 0.25$_{-0.21}^{+0.24}$ & 0.22~$\pm$~0.17 & 0.18$^{+0.27}_{-0.26}$ &  GPc &  0.7 & 3.8 & 3.4 & 1.8 \\
TOI-1728 b & --\,0.23$_{-0.29}^{+0.25}$ & --\,0.22$_{-0.29}^{+0.23}$ &      &      &  c &  0.8 & 26.8 & 7.3 & 26.3 \\
TOI-1759 b & --\,0.17$_{-0.38}^{+0.37}$ &      & --\,0.23$_{-0.27}^{+0.24}$ &     &  GPe &  0.9 & 10.8 & 3.4 & 10.2 \\
TOI-1801 b & --\,0.10~$\pm$~0.30 &      & --\,0.05$_{-0.25}^{+0.33}$ &     &  GPe &  0.1 & 5.7 & 4.3 & 4.3 \\
TOI-2018 b & 0.27$_{-0.31}^{+0.34}$ & 0.28$_{-0.29}^{+0.32}$ & 0.24$_{-0.16}^{+0.21}$ & 0.19$_{-0.25}^{+0.33}$ &  GPc &  0.6 & 9.2 & 9.1 & 4.4 \\
TOI-3785 b & --\,0.33$_{-0.34}^{+0.36}$ & --\,0.33~$\pm$~0.33 &     &     &  c &  1.0 & 15.0 & 7.4 & 16.0 \\
\hline
& & & & Null Detections & & & & & \\
\hline
GJ 1214 b & 0.01~$\pm$~0.04 &     &     &     & e &  0.3 & 8.2 & 0.8 & 0.6 \\
HAT-P 20 b & 0.000~$\pm$~0.002 &     &     &     & e &  0.0 & 2302.9 & 11.9 & 11.7 \\
HIP 65 A b$^a$ & 0.01~$\pm$~0.01 & 0.002~$\pm$~0.004 &    &    & c &  0.5 & 1021.2 & 13.8 & 8.4 \\
LTT 1445 A b & --\,0.02$_{-0.13}^{+0.12}$ & --\,0.07~$\pm$~0.08 &      &      & c &  0.8 & 2.9 & 0.3 & 0.8 \\
LTT 1445 A c & 0.06$_{-0.16}^{+0.18}$ & 0.10$_{-0.13}^{+0.14}$ &      &     & c &  0.7 & 1.5 & 0.7 & 0.3 \\
LTT 3780 c & 0.04$_{-0.14}^{+0.13}$ & 0.04$_{-0.14}^{+0.13}$ &      &      &  c &  0.3 & 8.6 & 2.9 & 2.4 \\
TOI-244 b & --\,0.32$_{-0.35}^{+0.36}$ & --\,0.29~$\pm$~0.34 & --\,0.03~$\pm$~0.16 & --\,0.01~$\pm$~0.12 &  GPc &  0.1 & 2.7 & 0.7 & 0.8 \\
TOI-1075 b & 0.12$_{-0.17}^{+0.15}$ & 0.07~$\pm$~0.13 &      &      &  c &  0.5 & 10.0 & 3.9 & 2.5 \\
TOI-3629 b & --\,0.04$_{-0.16}^{+0.17}$ & --\,0.01~$\pm$~0.13 &     &     &  c &  0.1 & 82.6 & 22.6 & 28.2 \\
WASP-80 b & --\,0.03~$\pm$~0.04 &  &      &      & e & 0.8 & 171.0 & 8.7 & 19.4 \\
\hline 
& & & & Sparsely Sampled & & & & & \\
\hline 
GJ 1252 b & 0.32$_{-0.48}^{+0.38}$ &     &    &     &       &    &  &   \\
HAT-P-54 b & 0.07$_{-0.23}^{+0.22}$ & 0.04~$\pm$~0.07 &    &     &       &    &  &   \\
HAT-P-68 b & 0.12~$\pm$~0.15 & 0.07$_{-0.10}^{+0.11}$ &    &     &       &    &  &   \\
HATS-6 b & 0.27$_{-0.29}^{+0.31}$ & 0.28$_{-0.26}^{+0.29}$ &    &     &       &    &  &   \\
HATS-47 b & --\,0.22$_{-0.42}^{+0.48}$ & --\,0.23$_{-0.41}^{+0.48}$ &    &     &       &    &   &  \\
HATS-48 A b & --\,0.48$_{-0.23}^{+0.24}$ & --\,0.46$_{-0.23}^{+0.21}$ &    &     &       &    &   &  \\
HATS-49 b & --\,0.35$_{-0.31}^{+0.40}$ & --\,0.38$_{-0.30}^{+0.40}$ &    &     &       &    &  &   \\
HATS-71 b & 0.56$^{+0.34}_{-0.69}$ & 0.58$^{+0.32}_{-0.69}$ &    &    &       &    &   &  \\
HATS-74 A b & 0.05$^{+0.12}_{-0.14}$ & 0.01~$\pm$~0.03 &    &     &       &    &   &  \\
HATS-75 b & --\,0.01$^{+0.11}_{-0.10}$ & --\,0.01~$\pm$~0.07 &    &    &       &    &  &   \\
HATS-76 b & 0.08~$\pm$~0.07 & 0.08$^{+0.05}_{-0.06}$ &    &     &       &    &   &  \\
HATS-77 b & --\,0.03$^{+0.12}_{-0.09}$ & 0.06$^{+0.04}_{-0.05}$ &    &    &       &    &  &   \\
K2-216 b & 0.40$^{+0.30}_{-0.34}$ & 0.40$^{+0.28}_{-0.33}$ &    &     &       &    &   &  \\
K2-295 b & --\,0.16$^{+0.44}_{-0.34}$ & --\,0.12$^{+0.44}_{-0.37}$ &    &    &       &   &   &   \\
Kepler-45 b & --\,0.25$^{+0.17}_{-0.19}$ & --\,0.29$^{+0.14}_{-0.17}$ &    &    &       &   &   &   \\
NGTS-1 b & 0.05$^{+0.12}_{-0.13}$ & 0.01$^{+0.06}_{-0.05}$ &    &    &       &    &   &  \\
NGTS-10 b & 0.01~$\pm$~0.04 & 0.00$^{+0.01}_{-0.02}$ &    &    &       &    &   &  \\
POTS-1 b & --\,0.03$^{+0.68}_{-0.65}$ & --\,0.06$^{+0.68}_{-0.63}$ &    &     &      &   &    &   \\
Qatar-9 b & --\,0.36$^{+0.30}_{-0.28}$ & --\,0.39$^{+0.26}_{-0.23}$ &    &     &      &   &    &   \\
TOI-519 b & --\,0.03~$\pm$~0.14 & --\,0.04$_{-0.07}^{+0.08}$ &    &     &       &    &  &   \\
TOI-530 b & 0.43$^{+0.23}_{-0.22}$ & 0.52$^{+0.22}_{-0.23}$ &    &    &       &    &   &  \\
TOI-674 b & 0.17$^{+0.18}_{-0.19}$ & 0.19$^{+0.13}_{-0.15}$ &    &    &       &    &   &  \\
TOI-776 c & 0.16$_{-0.29}^{+0.27}$ & 0.15$_{-0.29}^{+0.27}$ &    &    &       &    &   &  \\
TOI-824 b & 0.16$^{+0.12}_{-0.11}$ &     &    &    &       &    &    &  \\
TOI-1201 b & --\,0.14~$\pm$~0.32 & --\,0.12$^{+0.29}_{-0.30}$ &    &     &       &    &    &  \\
TOI-1231 b & --\,0.23~$\pm$~0.27 &      &    &    &       &    &    &  \\
TOI-1278 b & --\,0.01~$\pm$~0.02 &      &    &    &       &    &   &  \\
TOI-1634 b & 0.08$^{+0.27}_{-0.28}$ & 0.01~$\pm$~0.25 &    &    &       &    &   &  \\
TOI-1899 b & 0.23$^{+0.24}_{-0.22}$ &     &    &     &       &    &   &  \\
TOI-3235 b & --\,0.05~$\pm$~0.09 & --\,0.03~$\pm$~0.02 &    &    &       &    &   &  \\
TOI-3714 b & 0.08$^{+0.08}_{-0.10}$ & 0.02$^{+0.06}_{-0.05}$ &    &    &       &    &   &  \\
TOI-4201 b & --\,0.03$^{+0.07}_{-0.06}$ & 0.09$^{+0.03}_{-0.04}$ &    &    &       &    &   &  \\
TOI-4860 b & 0.27$^{+0.28}_{-0.33}$ & 0.23$^{+0.26}_{-0.30}$ &    &     &       &    &   &  \\
TOI-5205 b & --\,0.16$^{+0.13}_{-0.11}$ & --\,0.07~$\pm$~0.05 &    &     &       &    &   &  \\
TOI-5293 b & --\,0.02~$\pm$~0.18 & --\,0.03~$\pm$~0.17 &    &     &       &    &    &  \\
TOI-5344 b & 0.10$^{+0.13}_{-0.12}$ & 0.14$^{+0.10}_{-0.09}$ &    &     &       &    &    &  \\
Wendelstein-1 b & --\,0.05$^{+0.27}_{-0.28}$ & --\,0.04$^{+0.23}_{-0.27}$ &    &     &       &    &   &  \\
\hline 
\end{longtable}
}
\end{landscape}

\section{Additional figures}

\begin{figure*}[h!]
  \begin{center}
  \rotatebox{0}{\subfigure{\includegraphics[width=155mm]{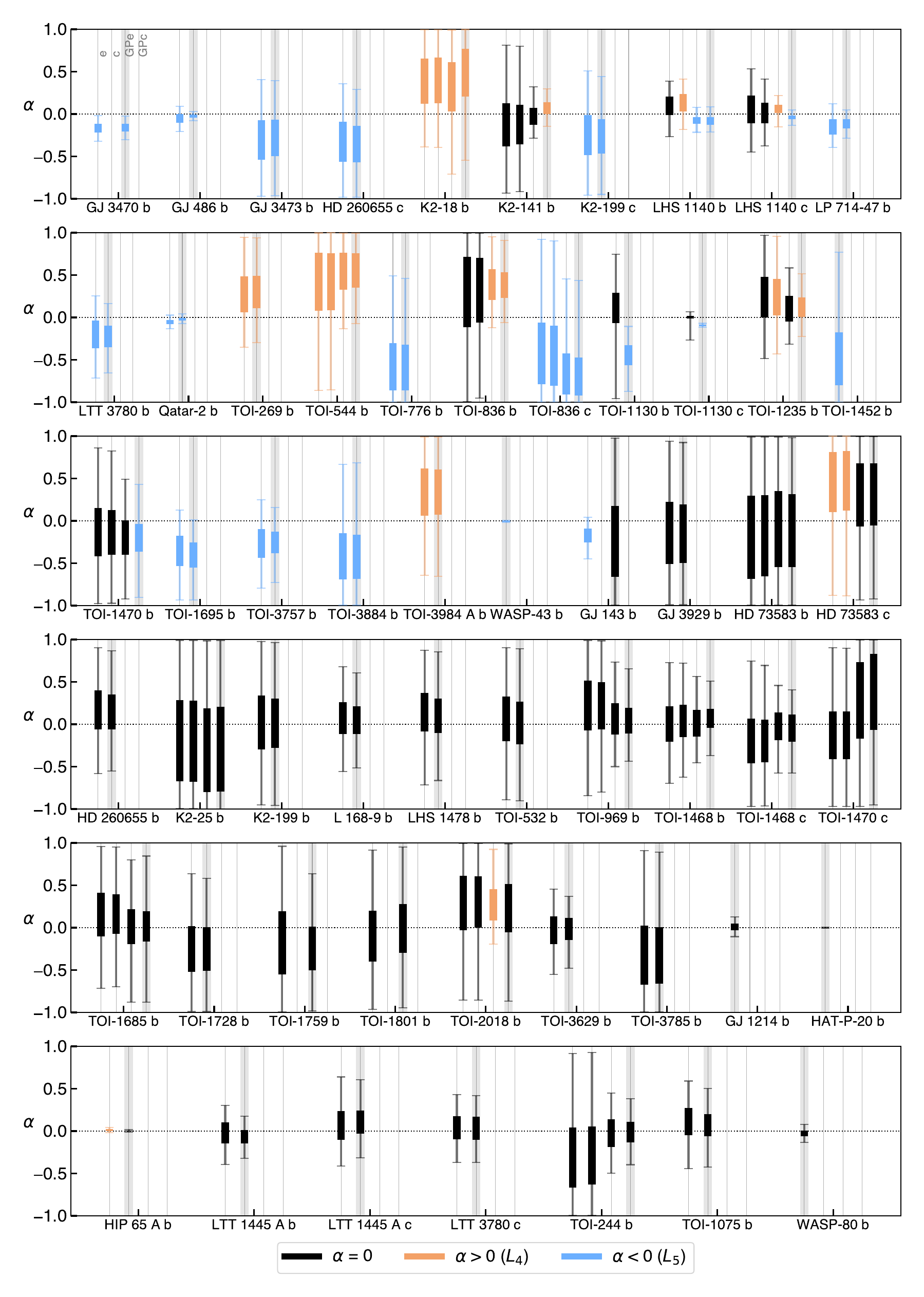}}}
  \end{center}
 \caption{Inferred $\alpha$ parameter for the transiting planets ordered by groups as in Table\,\ref{tab:alpha_res}. The analysis of each planet is divided in four different models (dotted vertical lines) corresponding, from left to right, with the slightly eccentric orbit (e), the circular orbit when the period of the planet is under 10\,d (c), and both of these adding a GP if required (GPe and GPc). The model used for the co-orbital mass estimation in Sect.\ref{sec:dismass} is indicated with a vertical grey region. The null value is indicated with an horizontal dashed line. Color-code informs on the result of $\alpha$ within 1-$\sigma$ as indicated in the legend. Errorbars show the 68.3\% (wide bar) and 99.7\% (narrow bar) confidence intervals (1 and 3-$\sigma$).}
 \label{fig:alp_gr}
\end{figure*}

\begin{figure*}
\begin{center}
  \rotatebox{0}{\subfigure{\includegraphics[width=160mm]{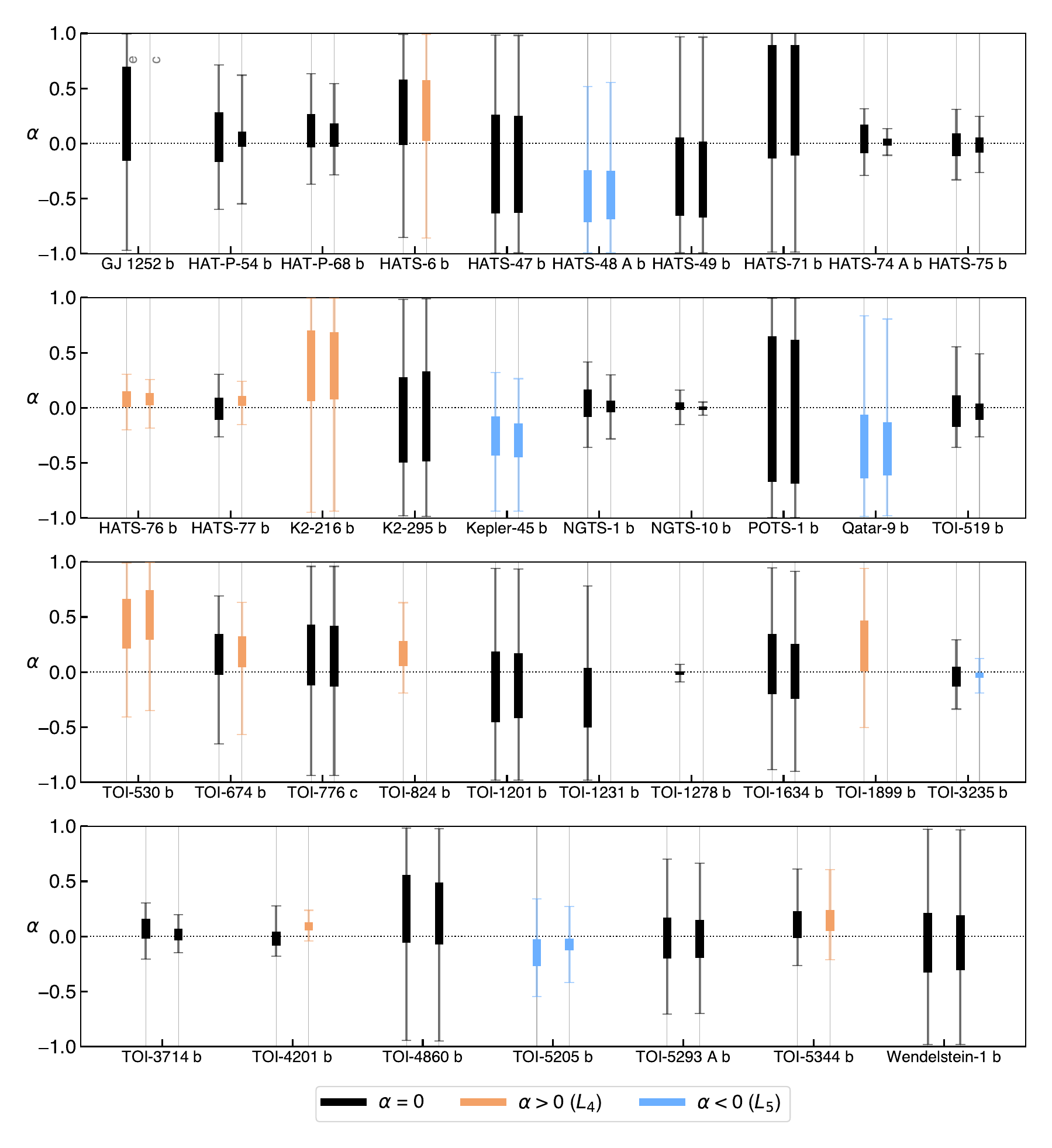}}}
\end{center}
 \caption{Inferred $\alpha$ parameter for the targets sparsely sampled. See details in the caption of Fig.\,\ref{fig:alp_gr}.}
 \label{fig:alp_ss}
\end{figure*}

\begin{figure*}[h!]
 \begin{center}
 \subfigure{\includegraphics[width=160mm]{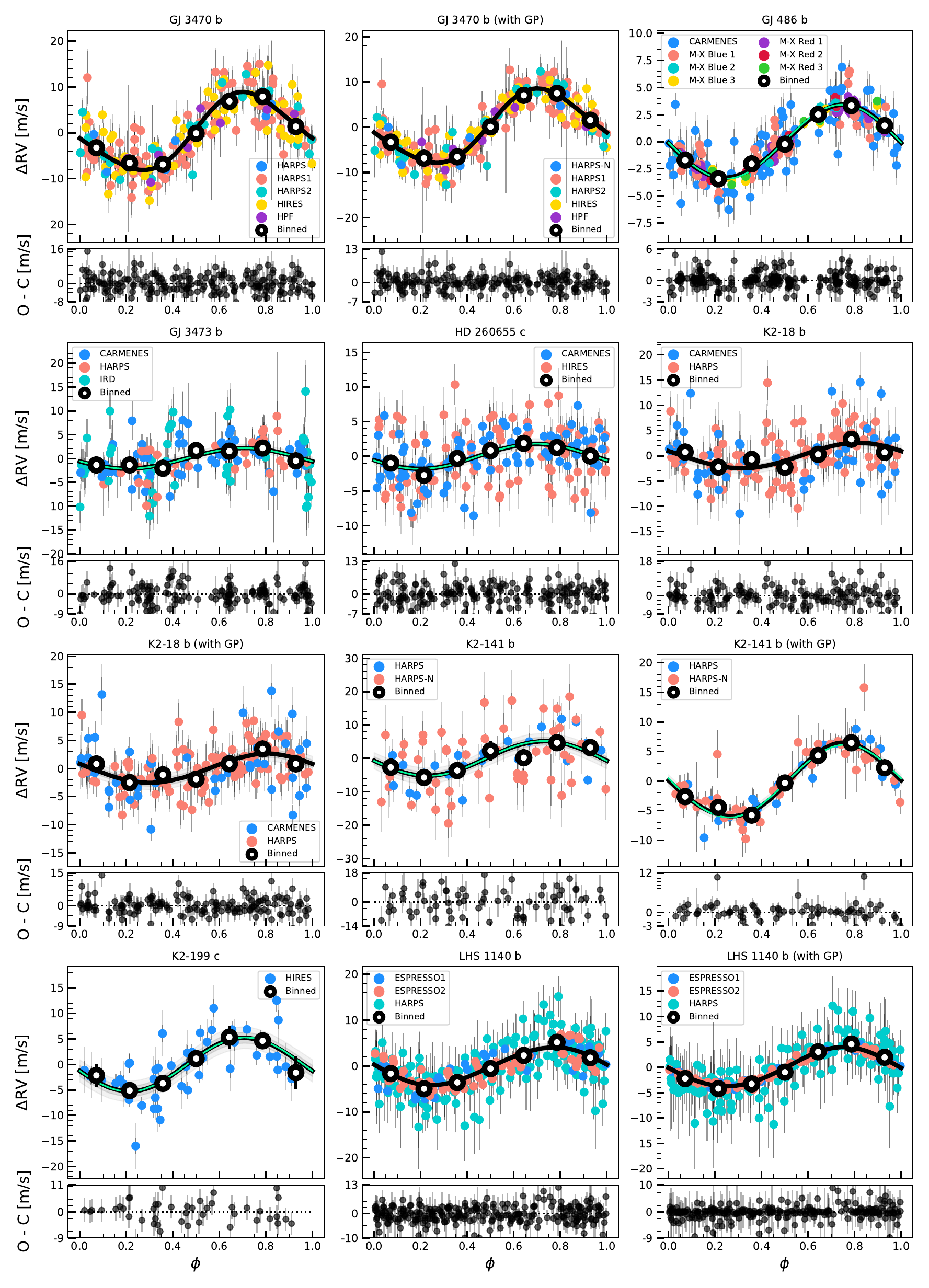}}
\end{center}
 \caption{Phase-folded radial velocity curves for the transiting planets ordered by groups as in Table\,\ref{tab:alpha_res}. The colors of the measurements identify the instrument as shown in each legend. Error bars show the measurement uncertainty (dark grey) and its quadratic sum with the instrument jitter (light grey). Instrument offsets, trends, and GPs have been substracted when required. Black solid line corresponds with the median of the predicted posterior distribution of the model, and the dark and light grey shaded regions are its corresponding 1-$\sigma$ and 2-$\sigma$ intervals. Green line shows the model for the circular case. We assume the trends and GPs of the circular models to be the same as in the eccentric except for TOI-1130\,b, for which we separate both cases in different charts.}
 \label{fig:rv_inph0}
\end{figure*}

\begin{figure*}[h!]
 \begin{center}
 \addtocounter{figure}{-1}
 \subfigure{\includegraphics[width=170mm]{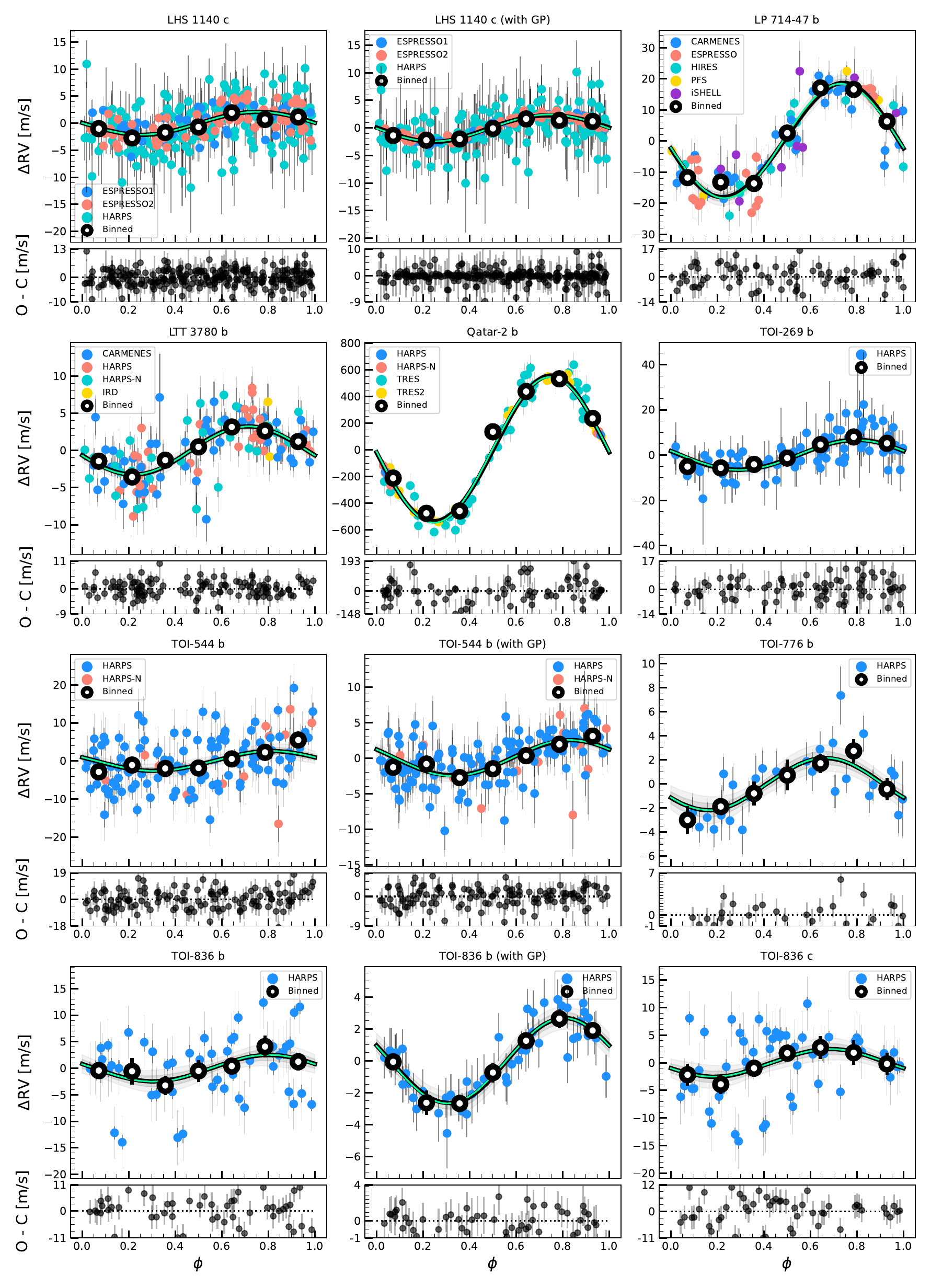}}
\end{center}
 \caption{Continuation of the phase-folded radial velocity curves.}
 \label{fig:rv_inph1}
\end{figure*}

\begin{figure*}[h!]
 \begin{center}
 \addtocounter{figure}{-1}
 \subfigure{\includegraphics[width=170mm]{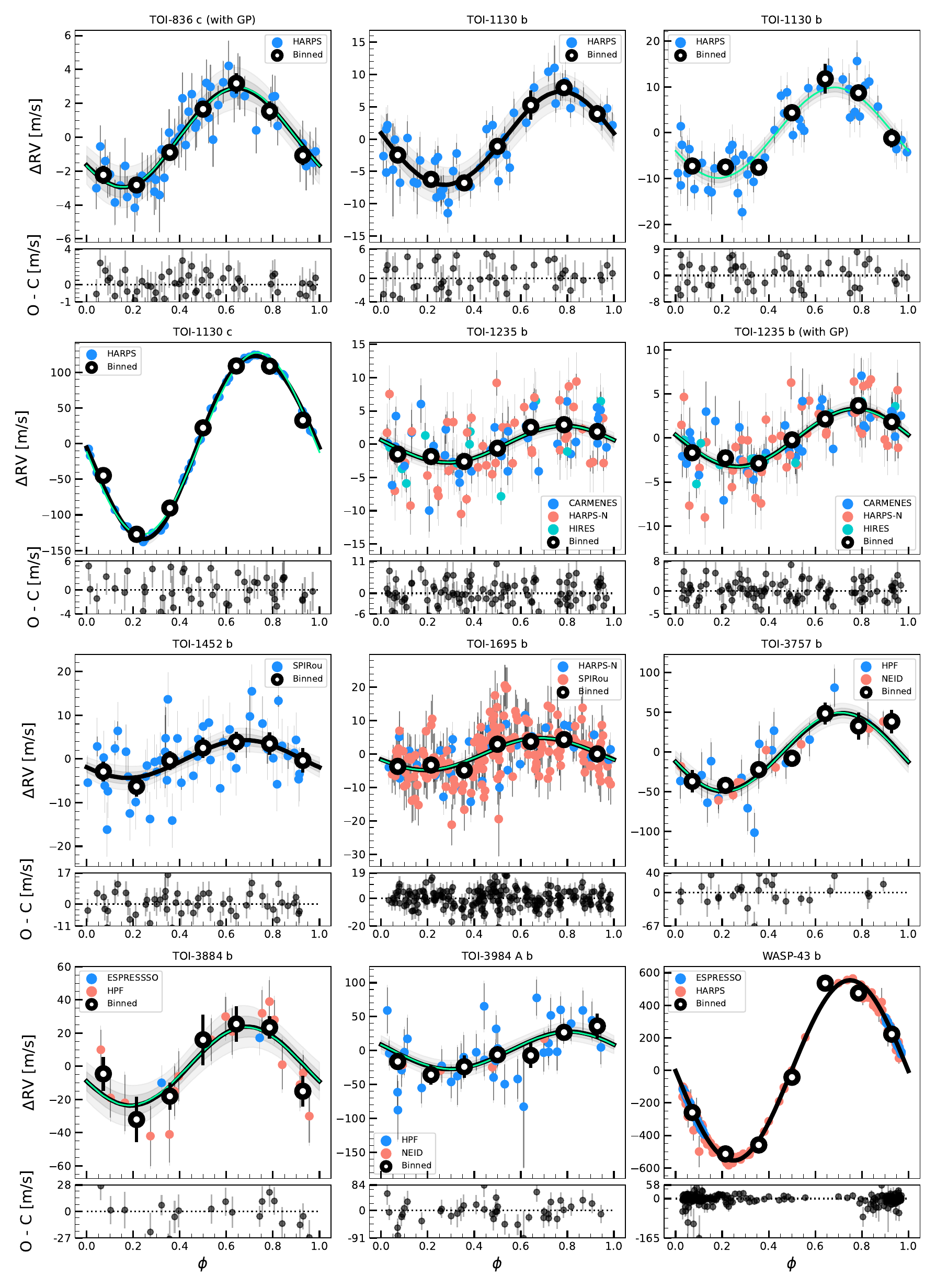}}
\end{center}
 \caption{Continuation of the phase-folded radial velocity curves.}
 \label{fig:rv_inph2}
\end{figure*}

\begin{figure*}[h!]
 \begin{center}
 \addtocounter{figure}{-1}
 \subfigure{\includegraphics[width=170mm]{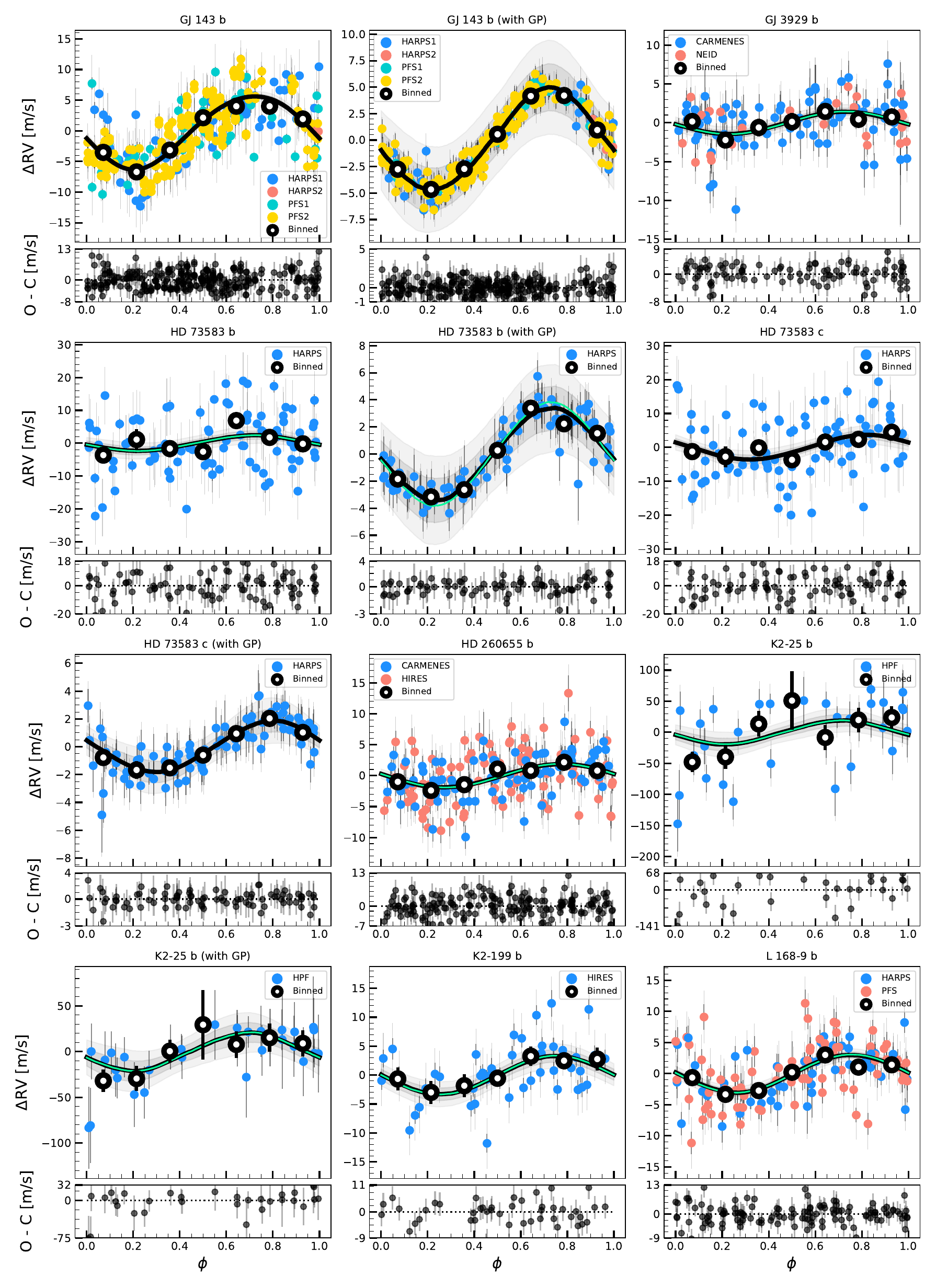}}
\end{center}
 \caption{Continuation of the phase-folded radial velocity curves.}
 \label{fig:rv_inph3}
\end{figure*}

\begin{figure*}[h!]
 \begin{center}
 \addtocounter{figure}{-1}
 \subfigure{\includegraphics[width=170mm]{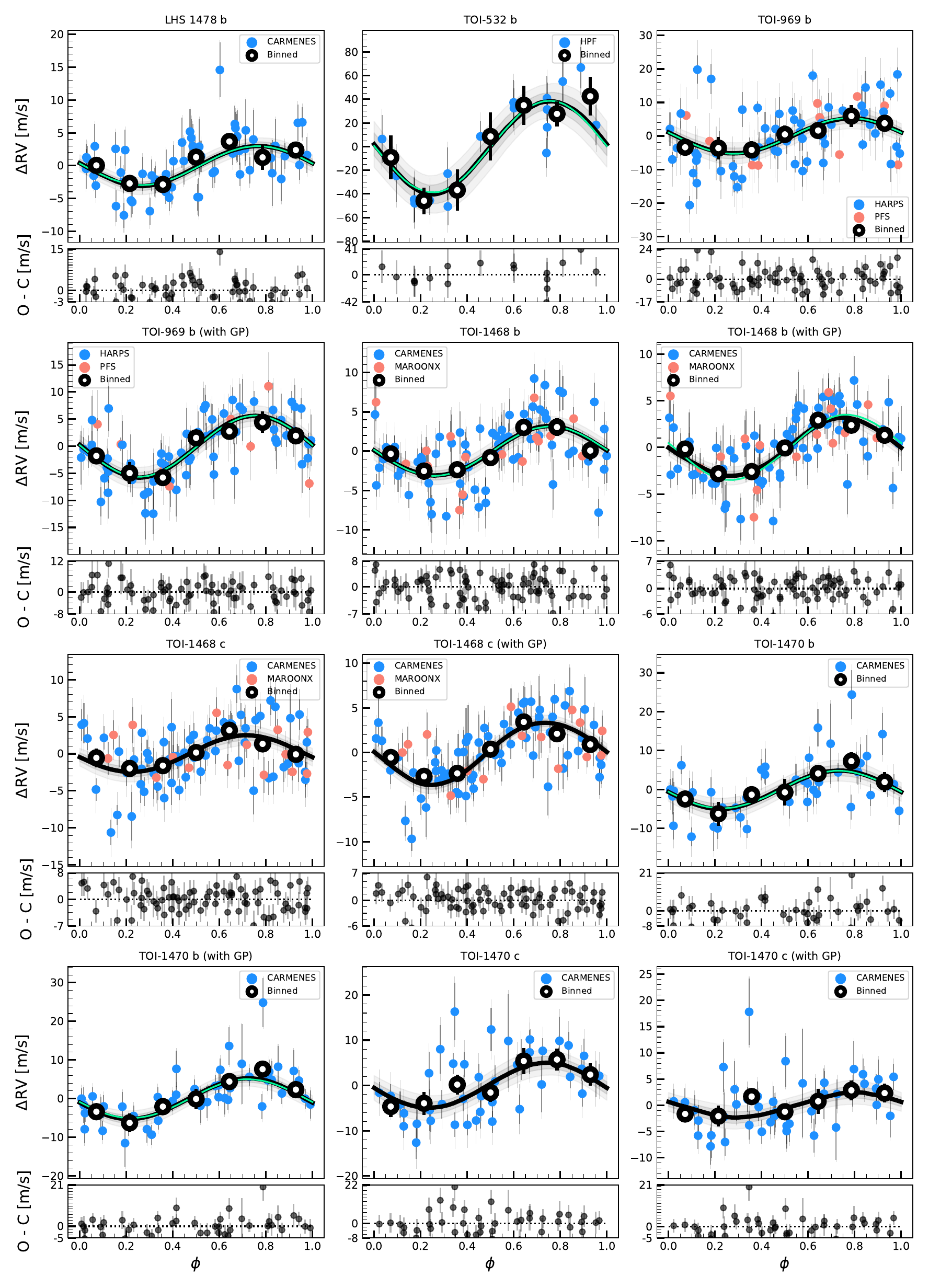}}
\end{center}
 \caption{Continuation of the phase-folded radial velocity curves.}
 \label{fig:rv_inph4}
\end{figure*}

\begin{figure*}[h!]
 \begin{center}
 \addtocounter{figure}{-1}
 \subfigure{\includegraphics[width=170mm]{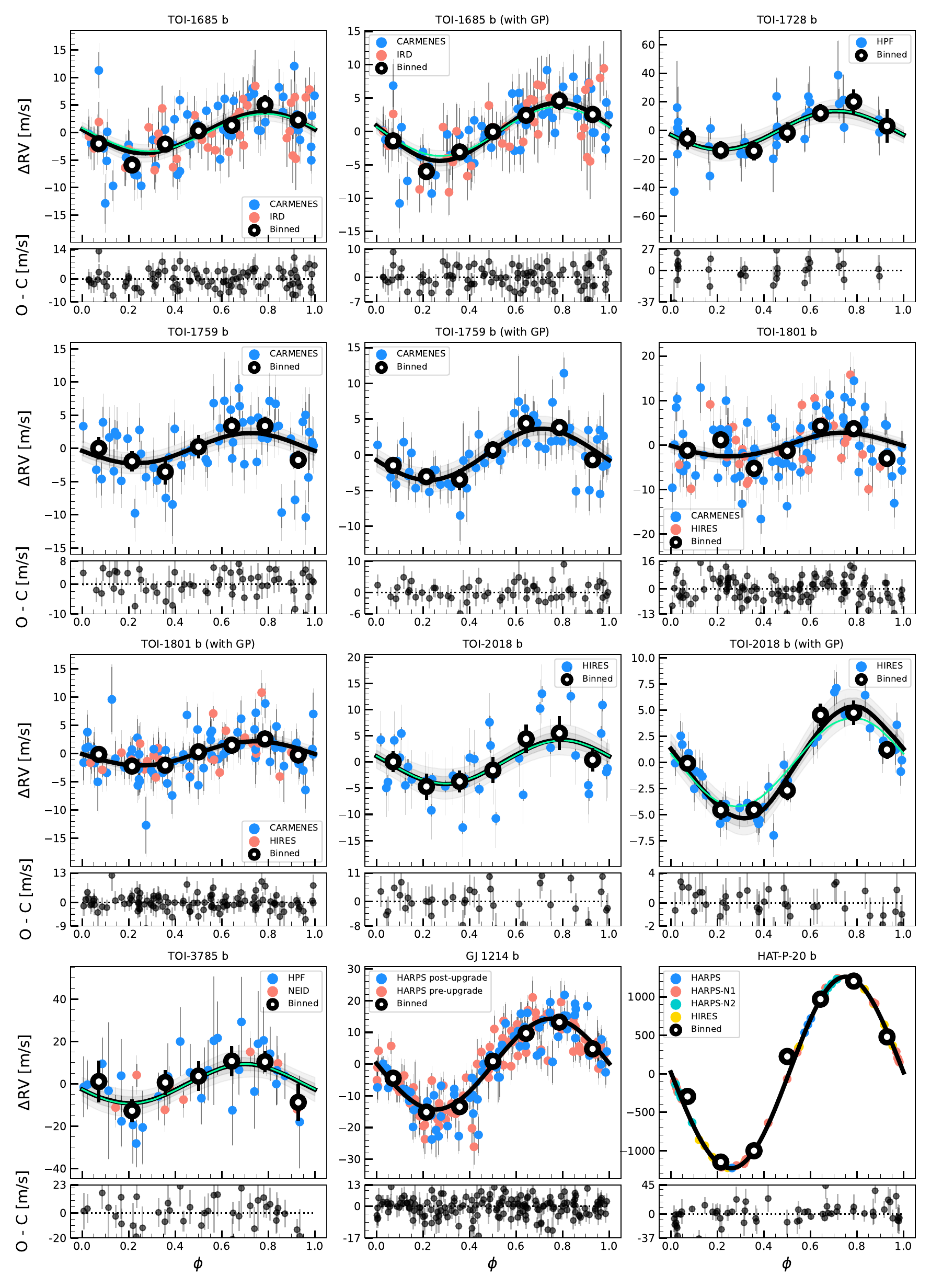}}
\end{center}
 \caption{Continuation of the phase-folded radial velocity curves.}
 \label{fig:rv_inph5}
\end{figure*}

\begin{figure*}[h!]
 \begin{center}
 \addtocounter{figure}{-1}
 \subfigure{\includegraphics[width=170mm]{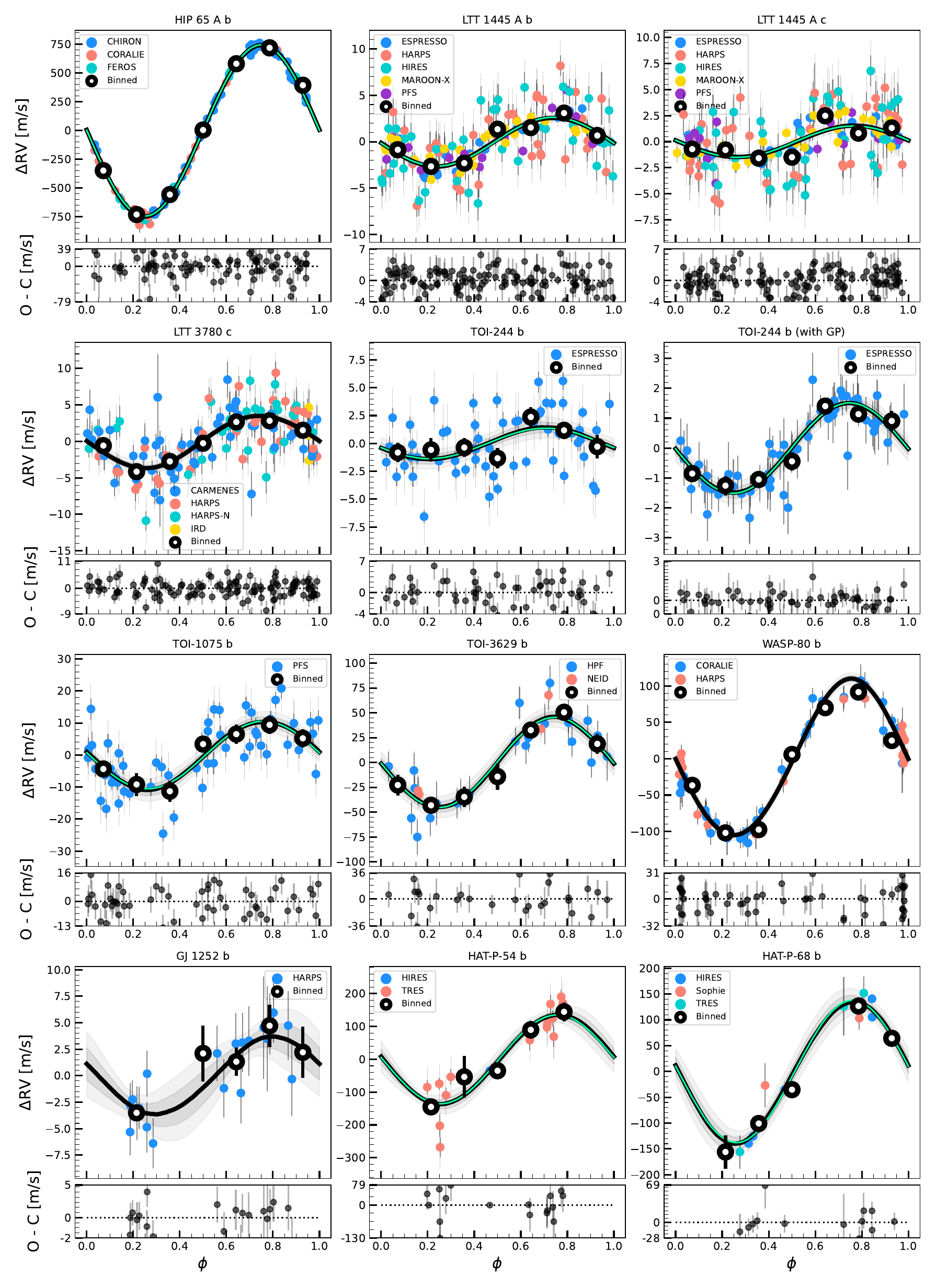}}
\end{center}
 \caption{Continuation of the phase-folded radial velocity curves.}
 \label{fig:rv_inph6}
\end{figure*}

\begin{figure*}[h!]
  \begin{center}
   \addtocounter{figure}{-1}
 \subfigure{\includegraphics[width=170mm]{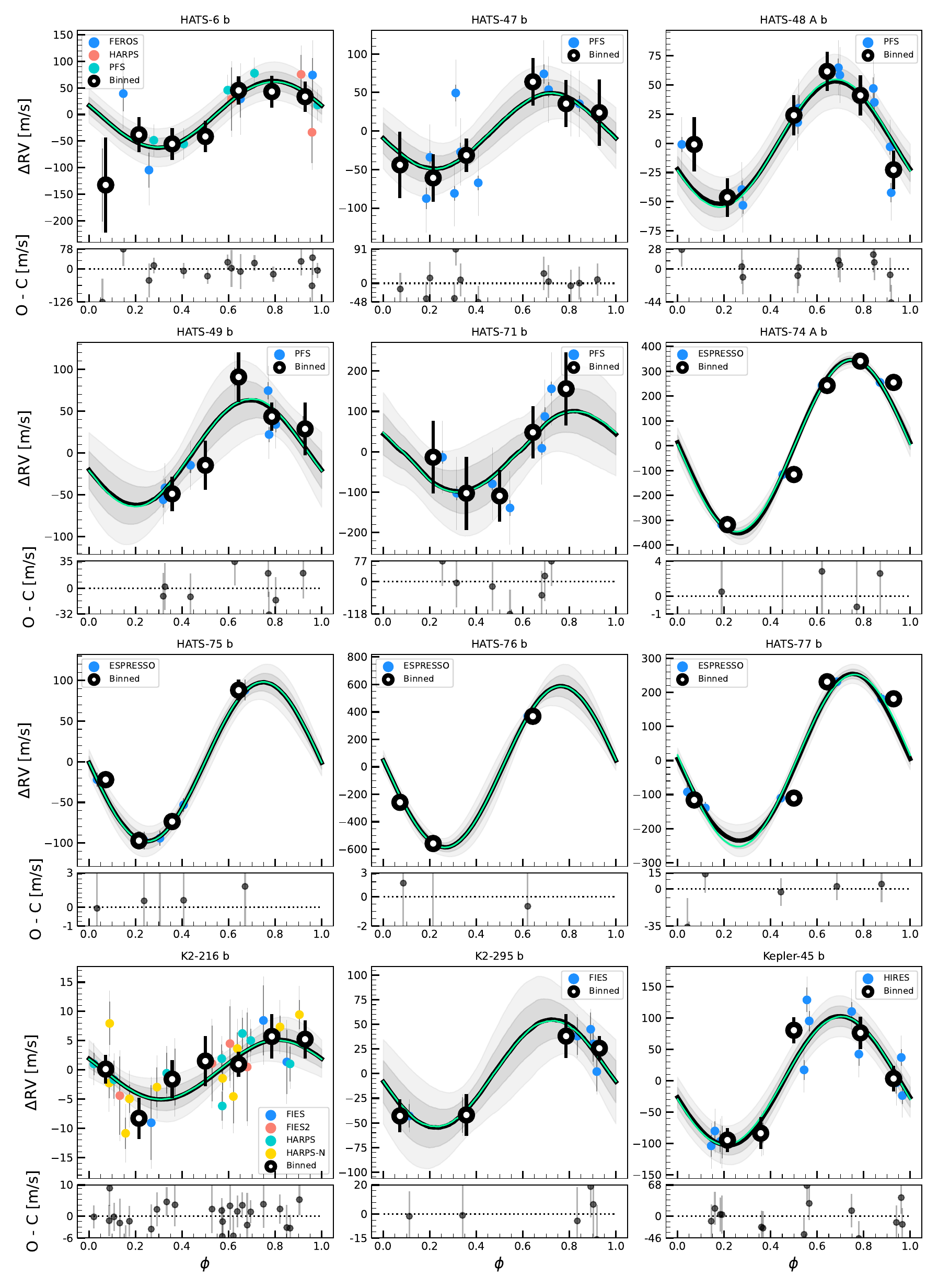}}
  \end{center}
 \caption{Continuation of the phase-folded radial velocity curves.}
 \label{fig:ainc1}
\end{figure*}

\begin{figure*}[h!]
  \begin{center}
   \addtocounter{figure}{-1}
 \subfigure{\includegraphics[width=170mm]{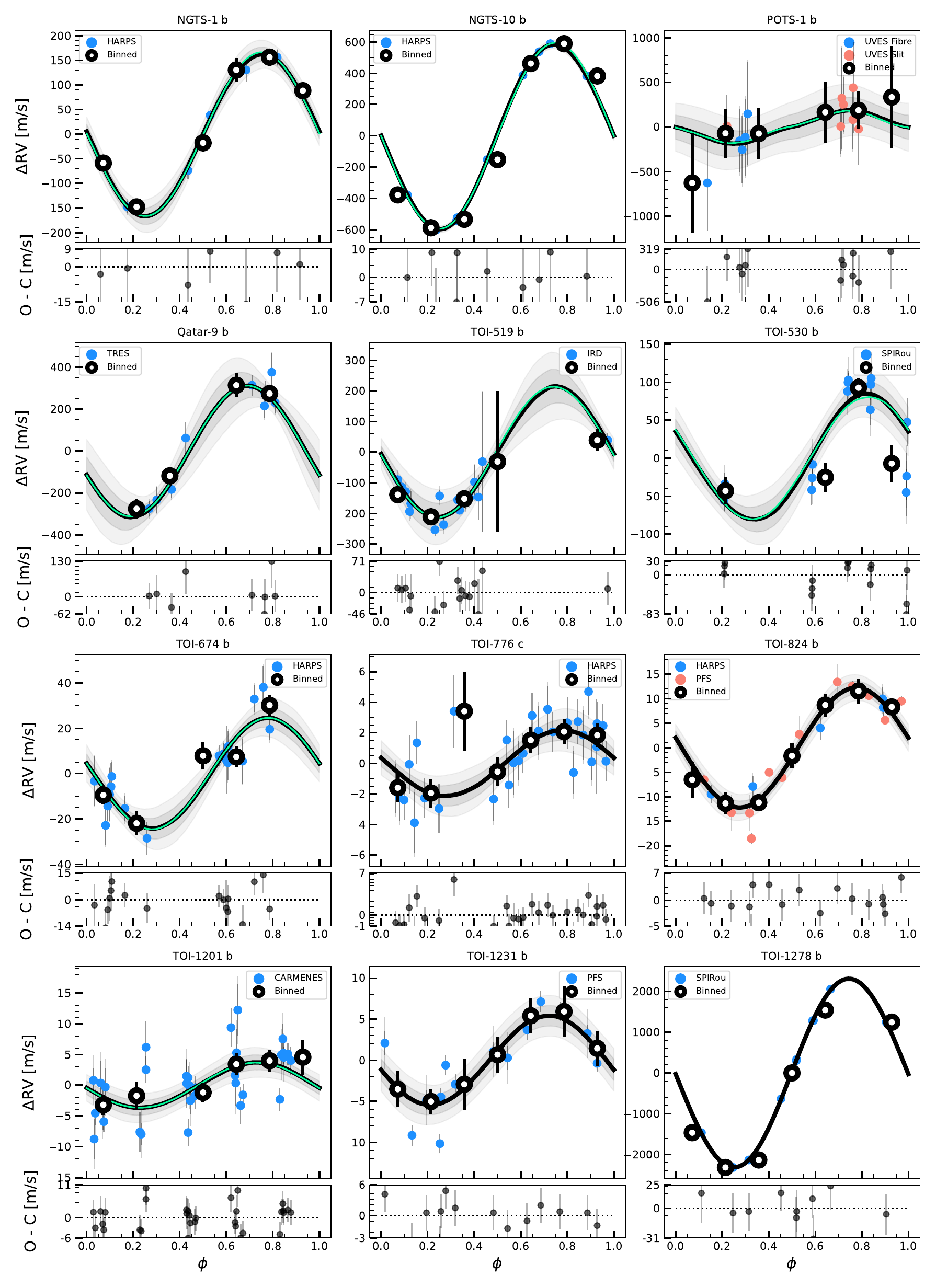}}
  \end{center}
 \caption{Continuation of the phase-folded radial velocity curves.}
 \label{fig:ainc2}
\end{figure*}

\begin{figure*}[h!]
  \begin{center}
   \addtocounter{figure}{-1}
 \subfigure{\includegraphics[width=170mm]{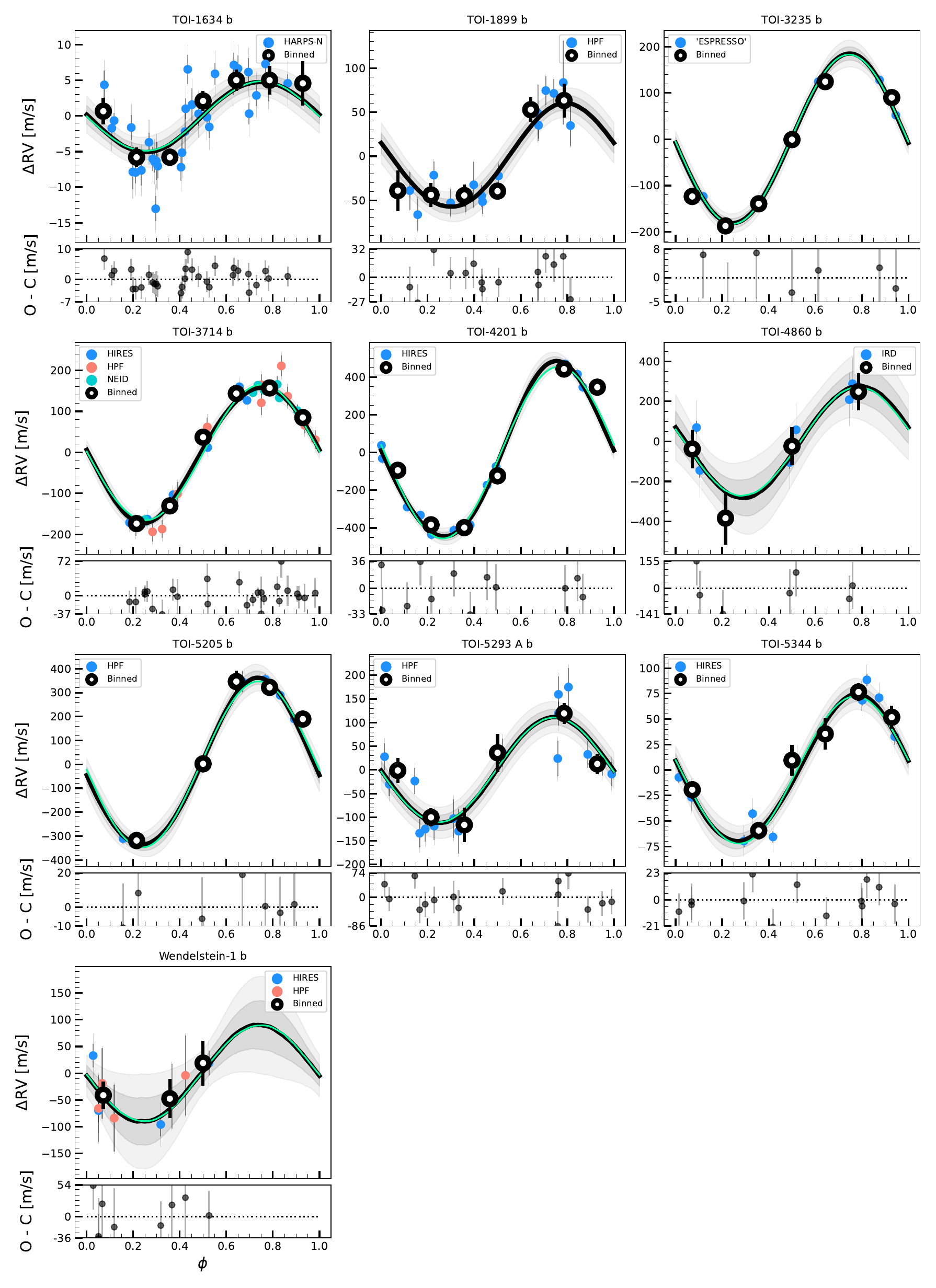}}
  \end{center}
 \caption{Continuation of the phase-folded radial velocity curves.}
 \label{fig:ainc3}
\end{figure*}

\begin{figure*}[h!]
 \begin{center}
\includegraphics[width=0.48\textwidth]{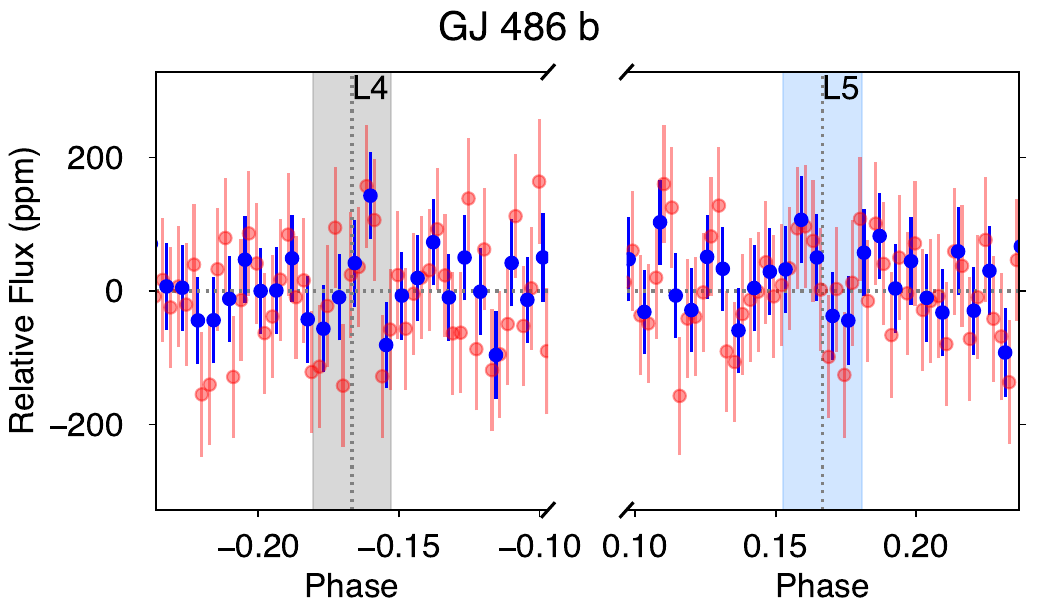}
\includegraphics[width=0.48\textwidth]{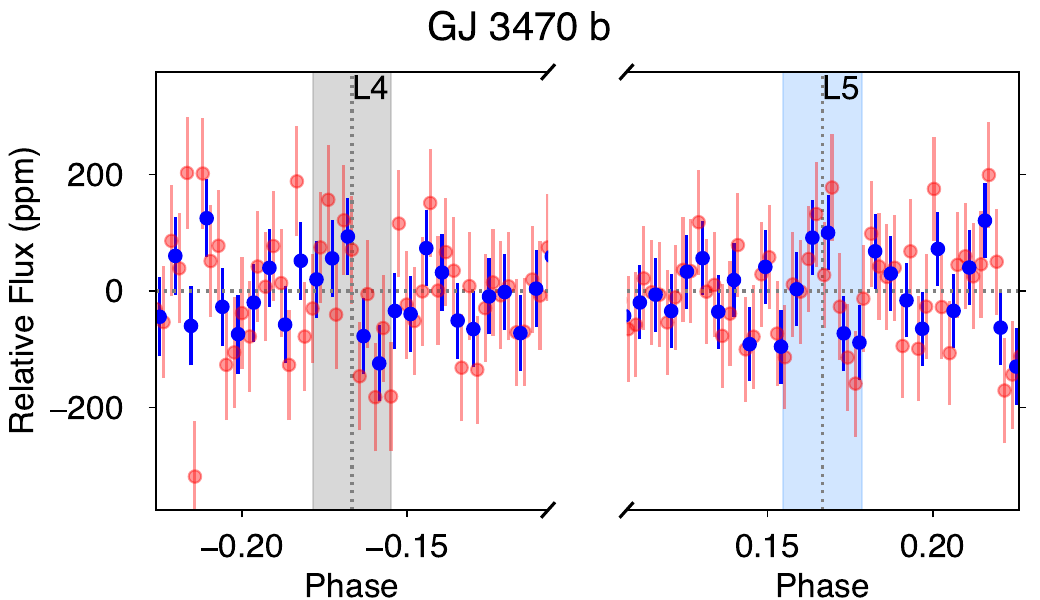}
\includegraphics[width=0.48\textwidth]{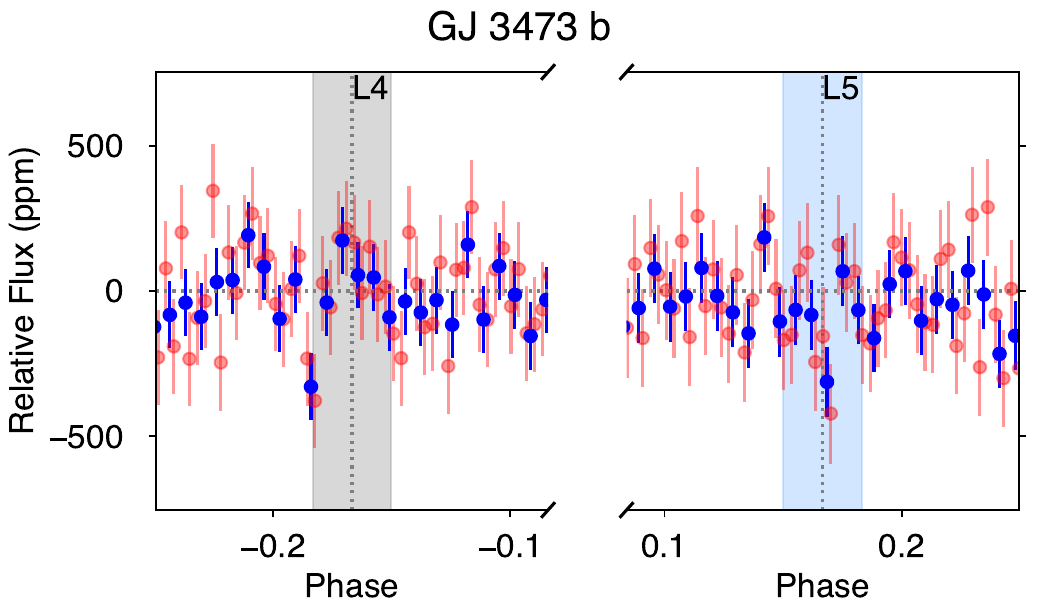}
\includegraphics[width=0.48\textwidth]{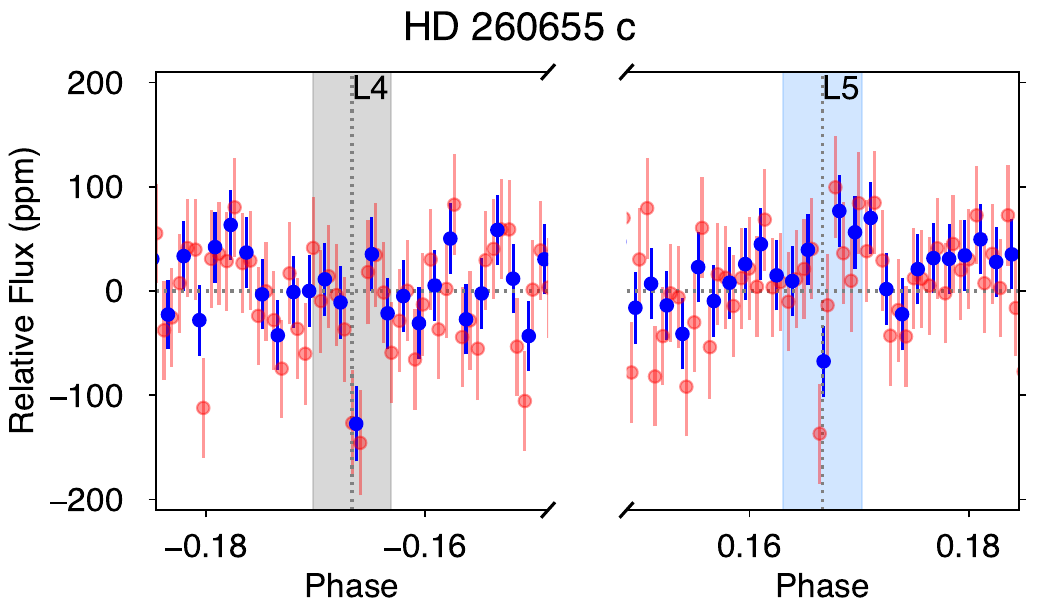}
\includegraphics[width=0.48\textwidth]{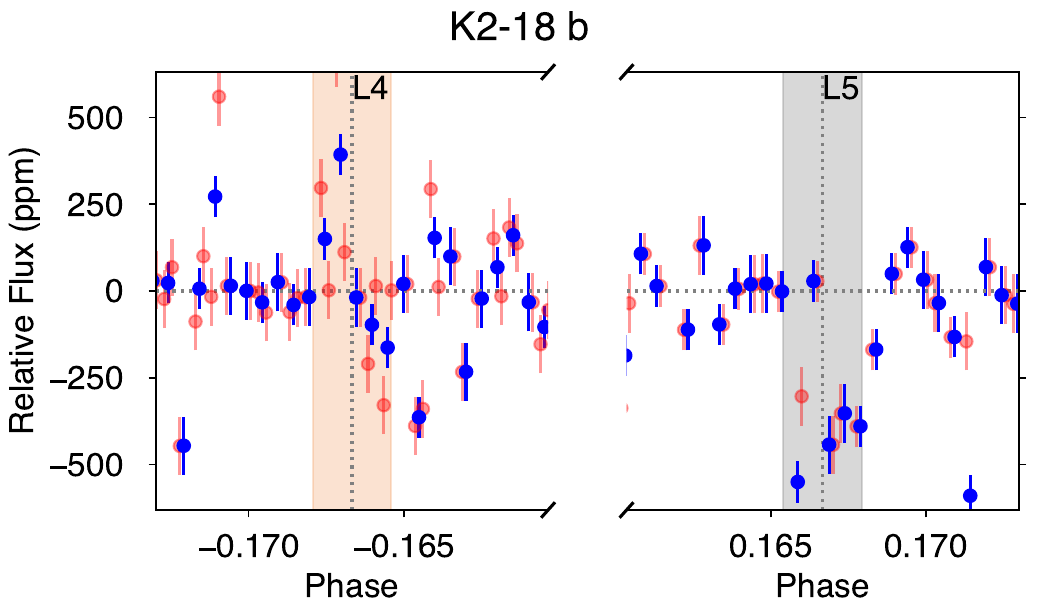}
\includegraphics[width=0.48\textwidth]{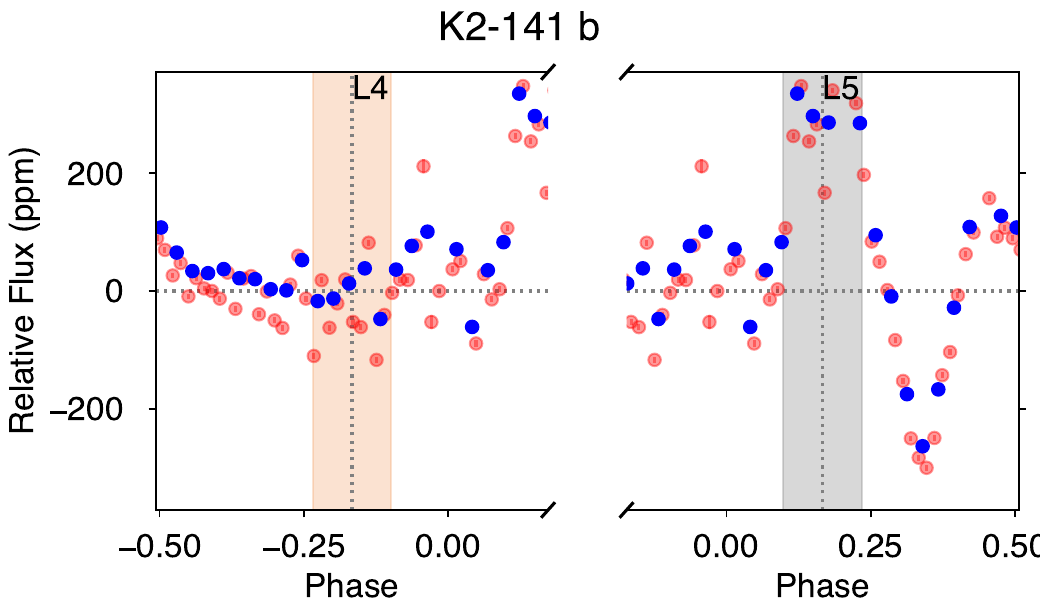}
\includegraphics[width=0.48\textwidth]{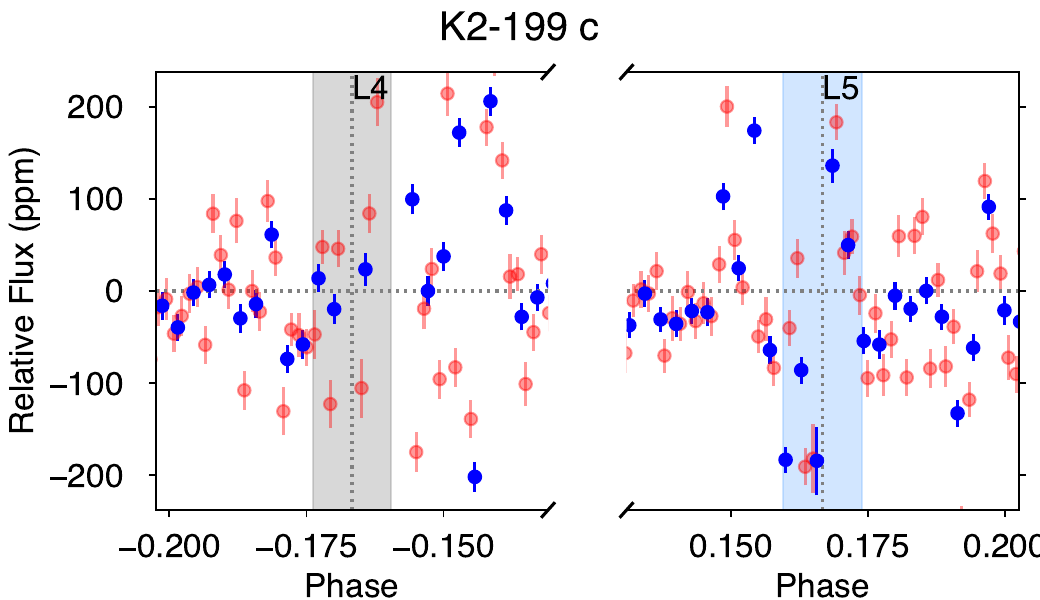}
\includegraphics[width=0.48\textwidth]{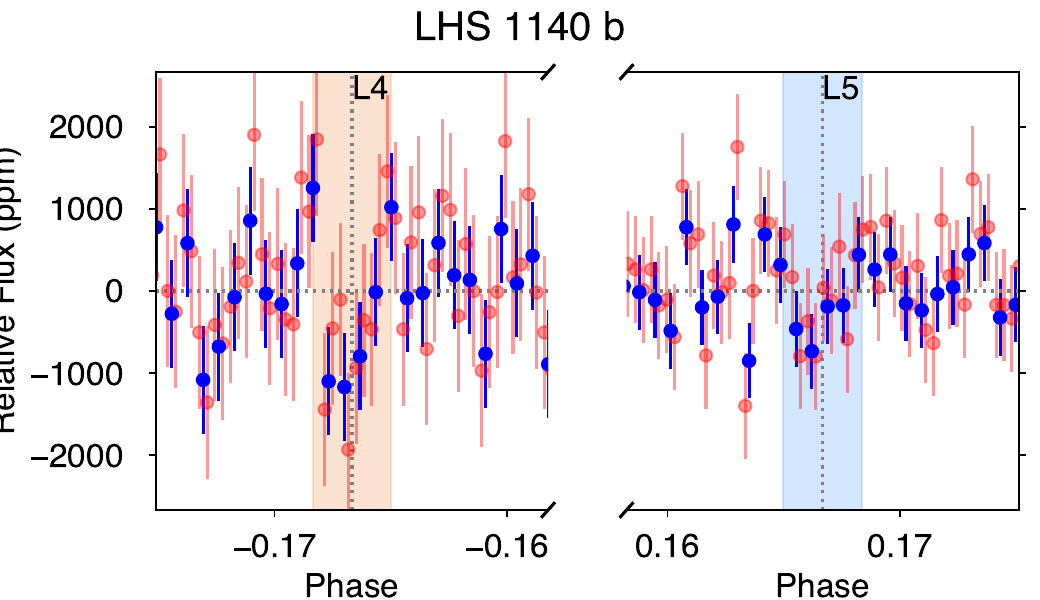}
\end{center}
 \caption{Phase-folded light curves (from the TESS mission, except for K2-18\,b, K2-141\,b, and K2-199\,c where we used the K2 data) around the $L_4$ and $L_5$ regions for the weak candidate (WC) sample. The phase-folded light curves are binned with a bin size corresponding to 10\% (red symbols) and 20\% (blue symbols) of the main-planet transit duration. The Lagrangian point phase of conjunction is marked by a dotted vertical line and the shaded region indicates the duration of the main-planet transit. We highlight in color (blue for $L_5$ and orange for $L_4$) the region where our model indicates the potential location of the co-orbital, while in gray the opposite region. Cases where both regions are colored indicate systems where some of the models (circular versus eccentric) showed candidates at either Lagrangian point.}
 \label{fig:LCs}
\end{figure*}

\begin{figure*}[h!]
 \begin{center}
 \addtocounter{figure}{-1}
\includegraphics[width=0.48\textwidth]{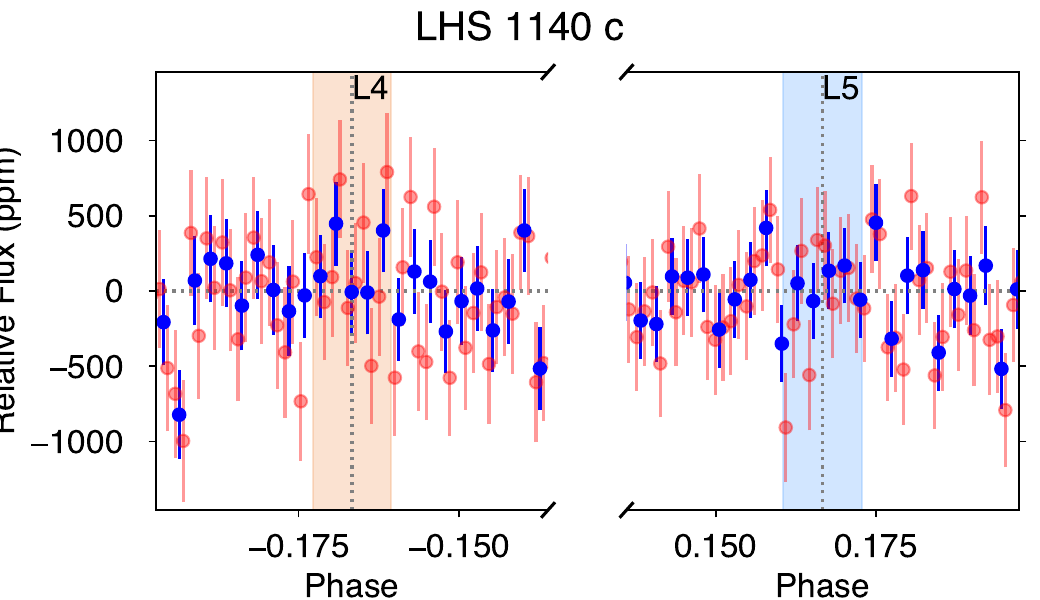}
\includegraphics[width=0.48\textwidth]{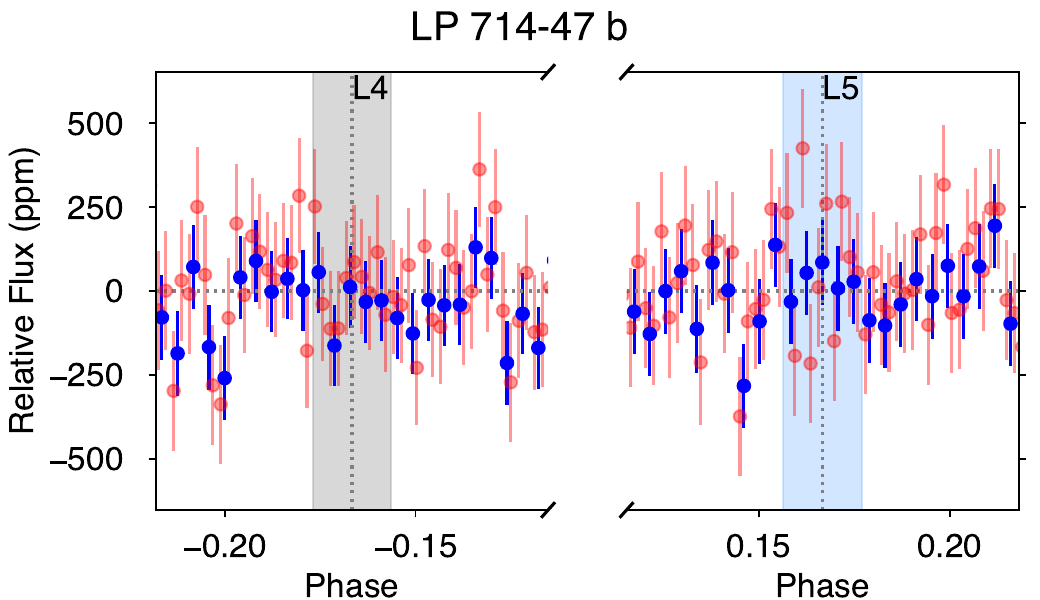}
\includegraphics[width=0.48\textwidth]{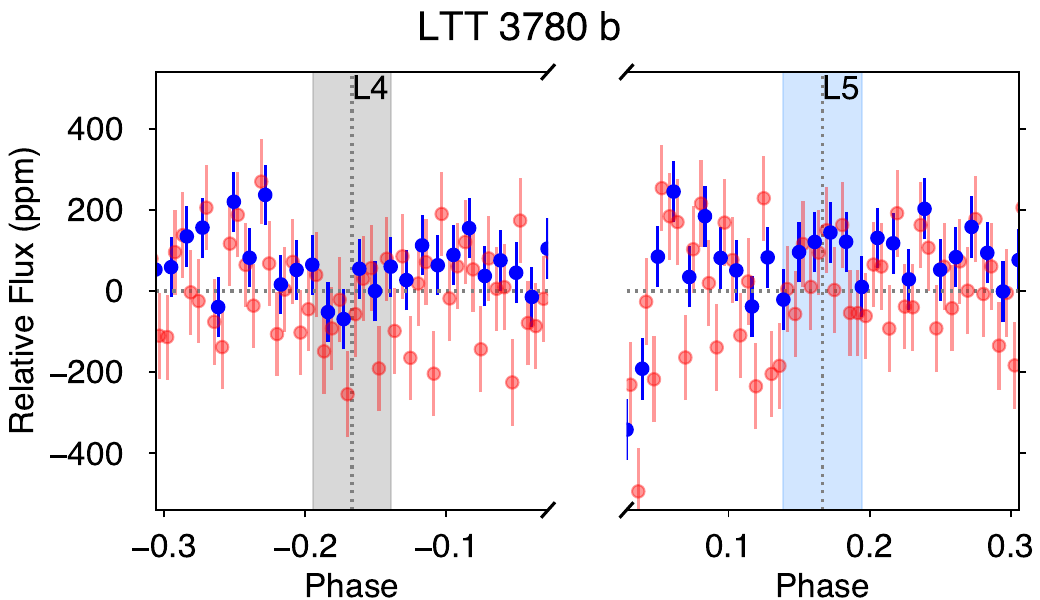}
\includegraphics[width=0.48\textwidth]{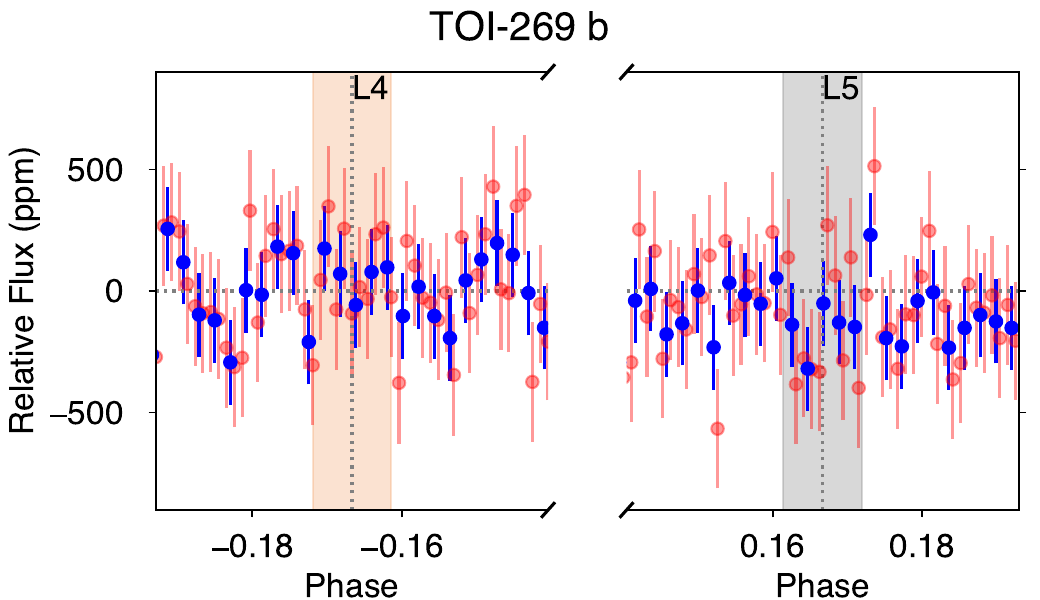}
\includegraphics[width=0.48\textwidth]{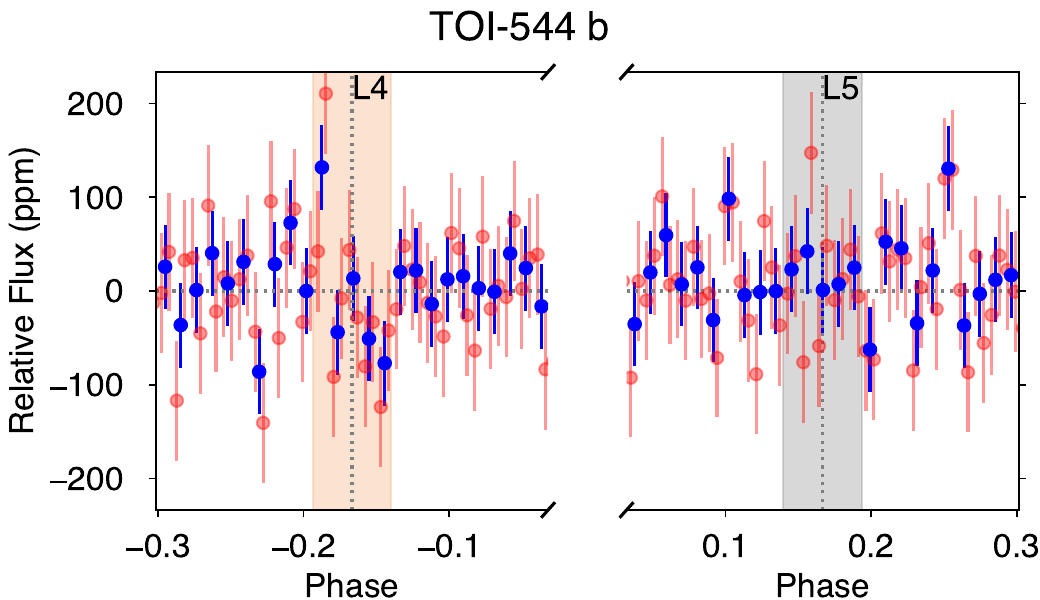}
\includegraphics[width=0.48\textwidth]{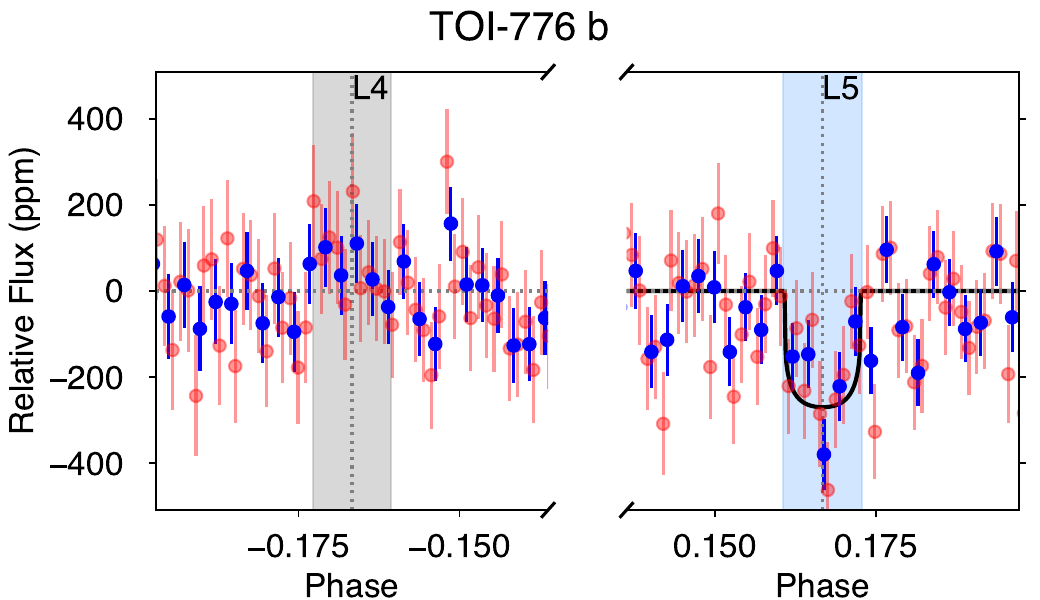}
\includegraphics[width=0.48\textwidth]{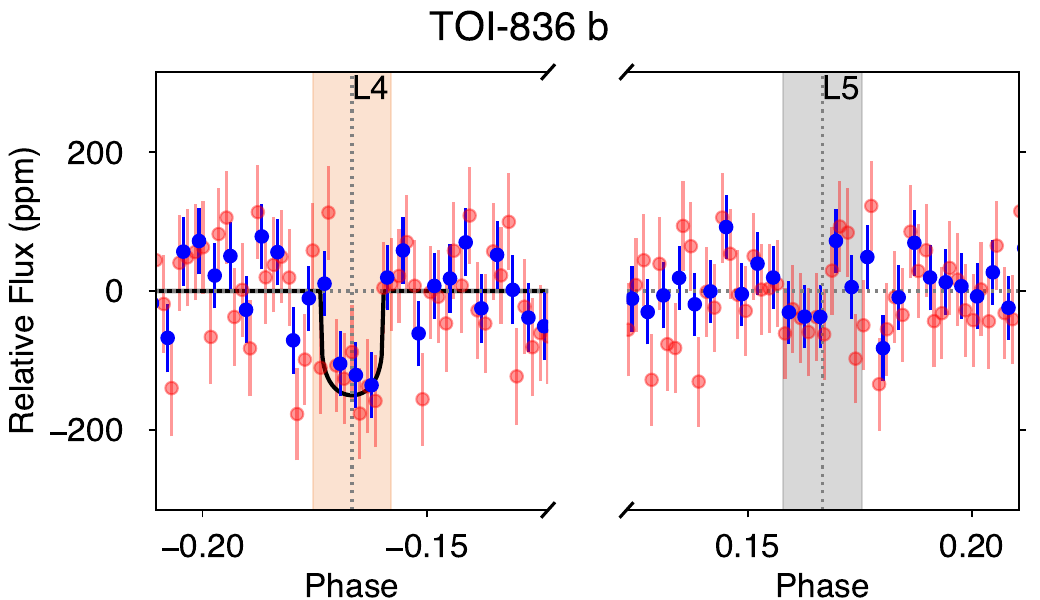}
\includegraphics[width=0.48\textwidth]{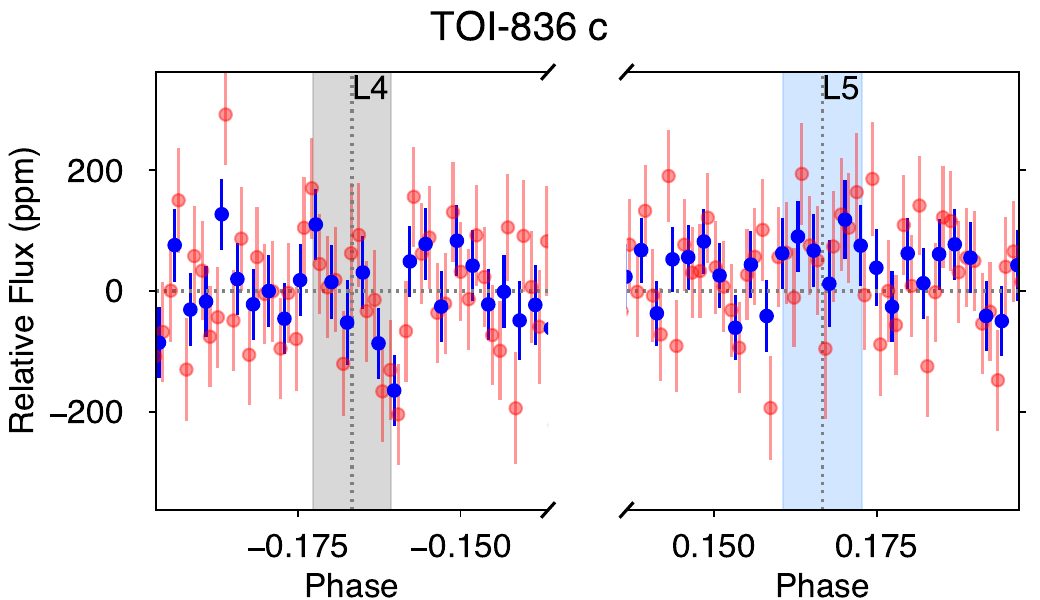}
\includegraphics[width=0.48\textwidth]{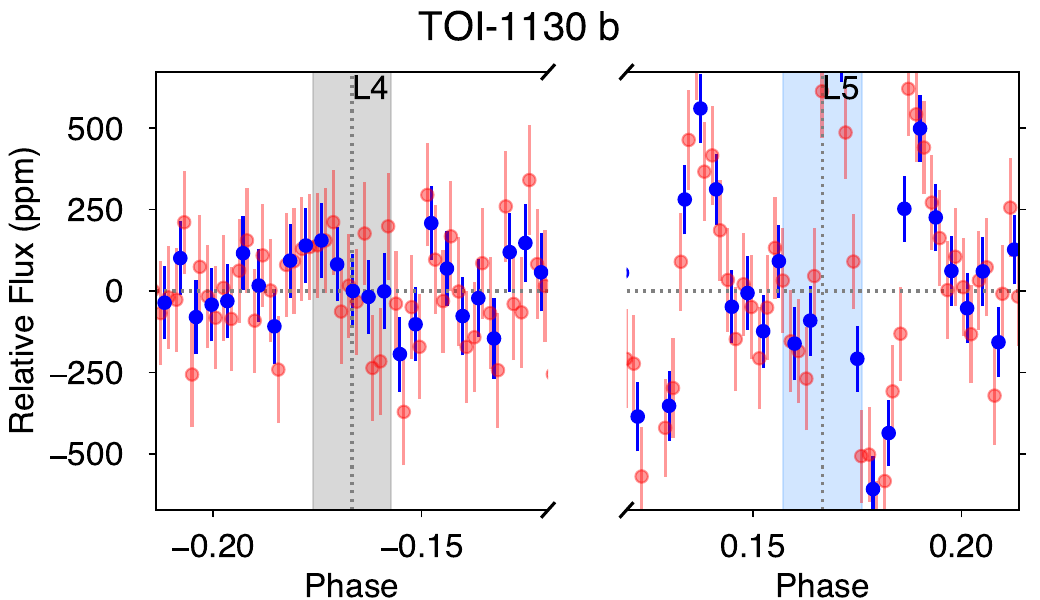}
\includegraphics[width=0.48\textwidth]{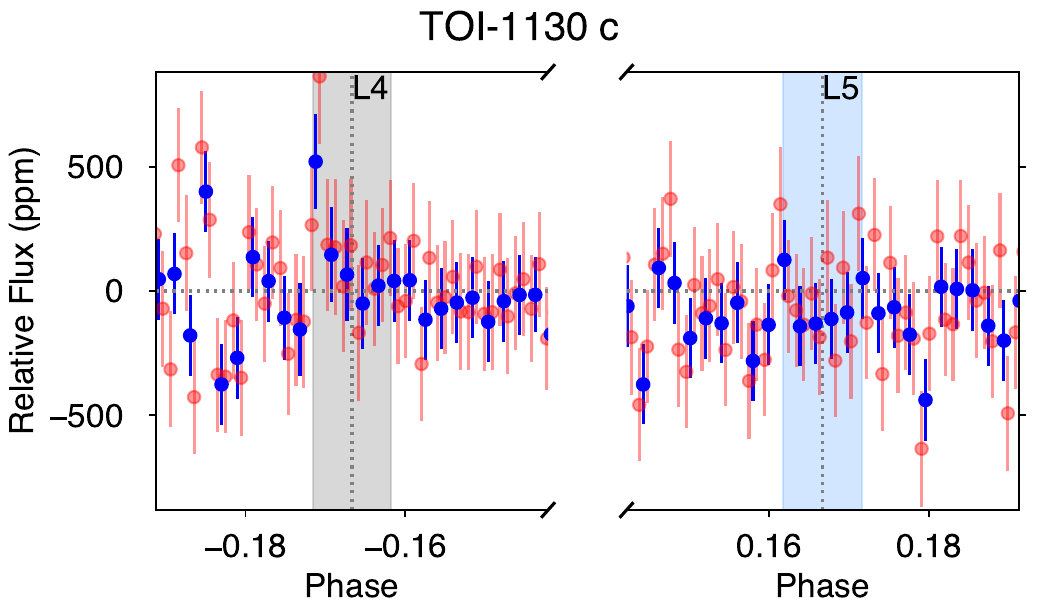}
\end{center}
 \caption{Continuation.}
 \label{fig:LC+}
\end{figure*}

\begin{figure*}[h!]
 \begin{center}
 \addtocounter{figure}{-1}
\includegraphics[width=0.48\textwidth]{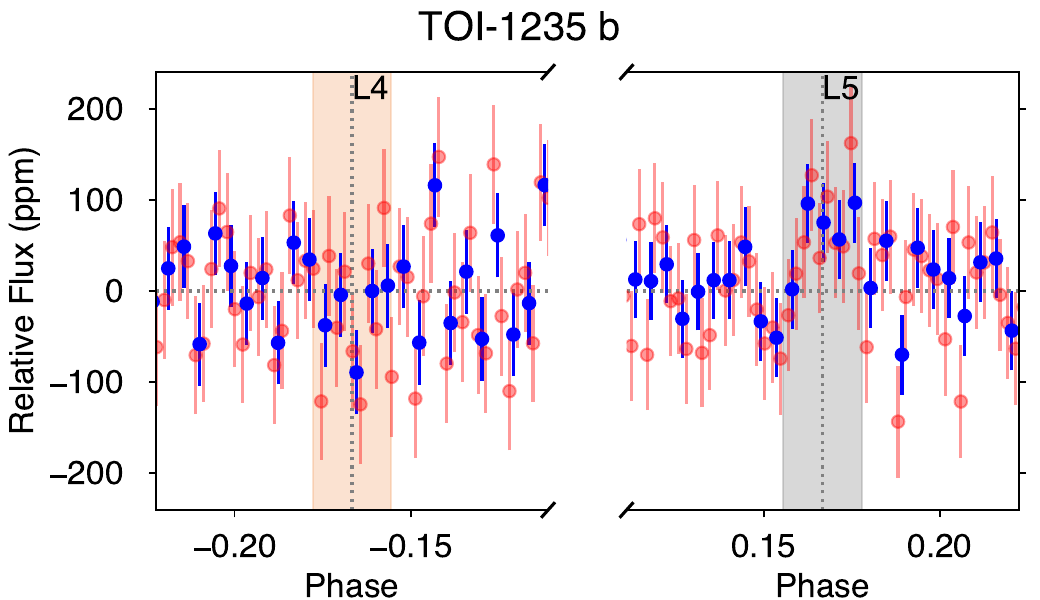}
\includegraphics[width=0.48\textwidth]{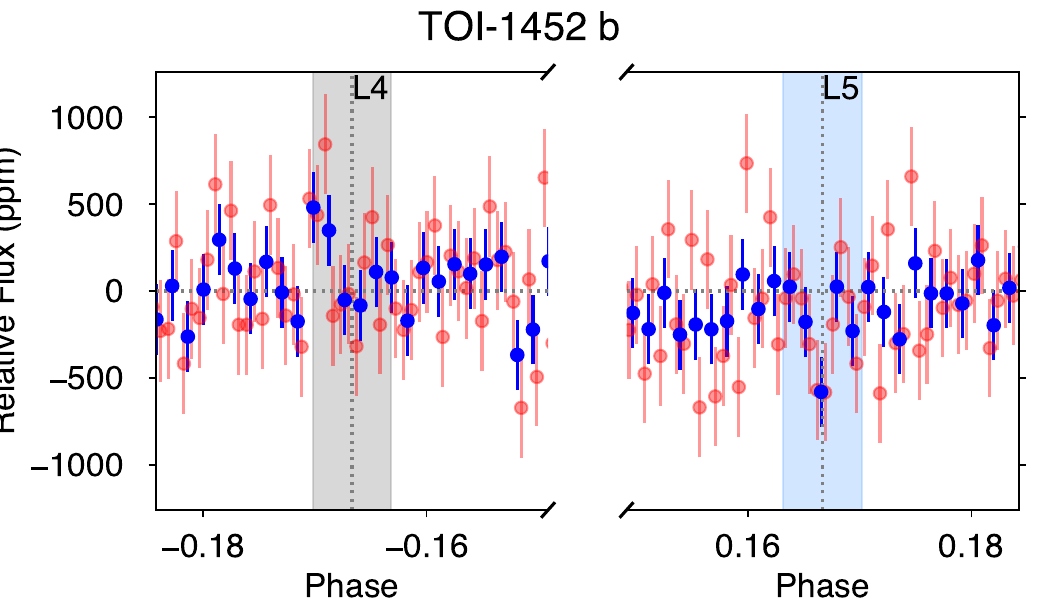}
\includegraphics[width=0.48\textwidth]{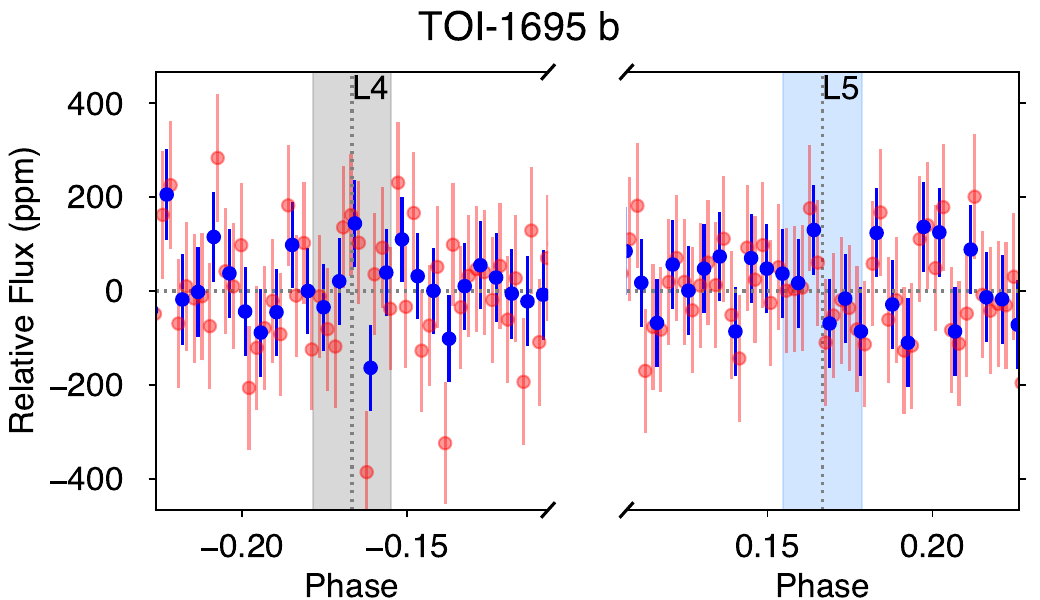}
\includegraphics[width=0.48\textwidth]{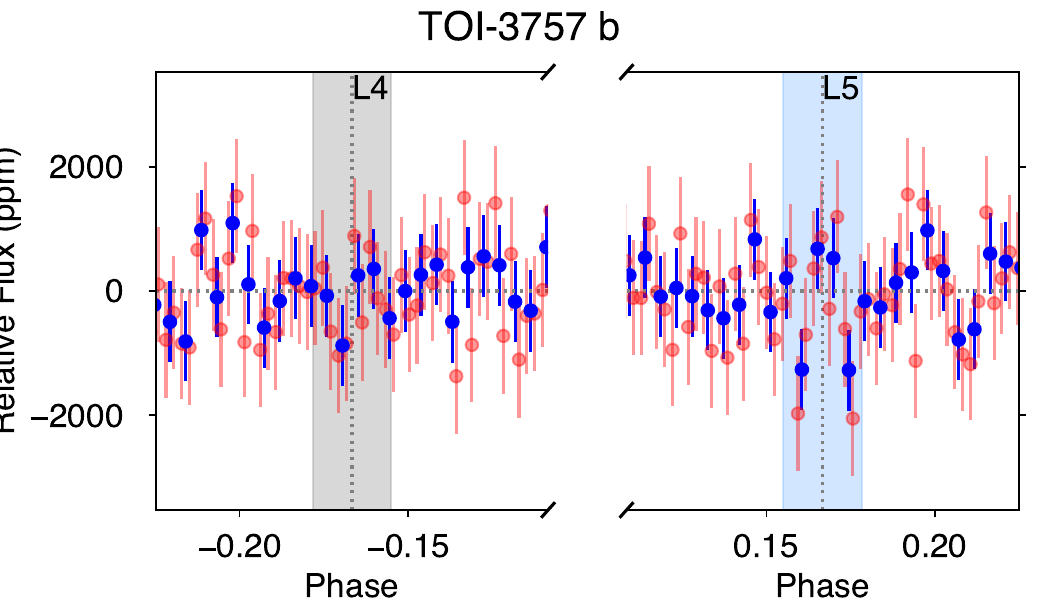}
\includegraphics[width=0.48\textwidth]{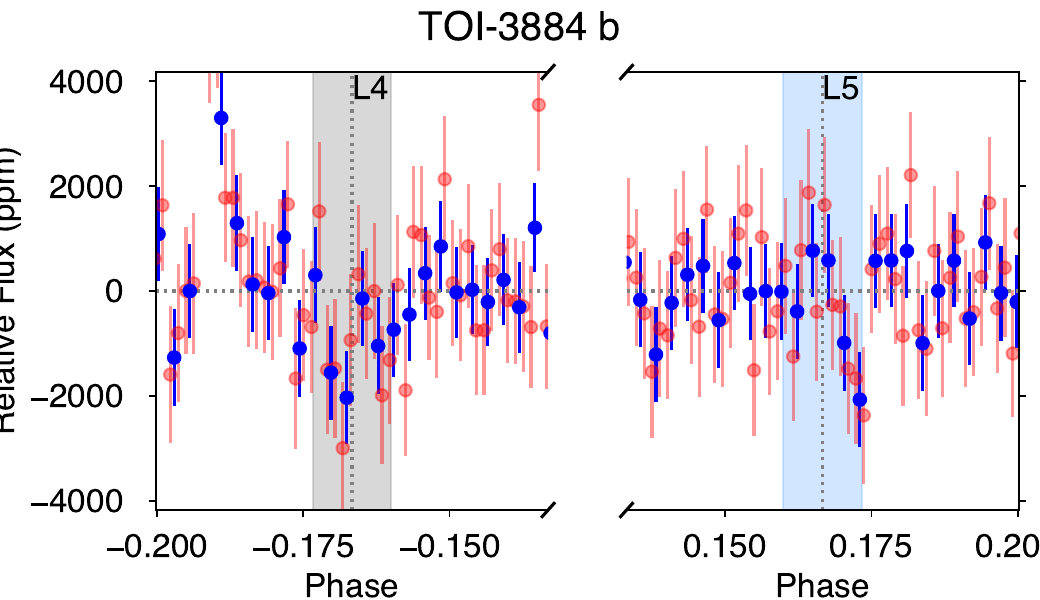}
\includegraphics[width=0.48\textwidth]{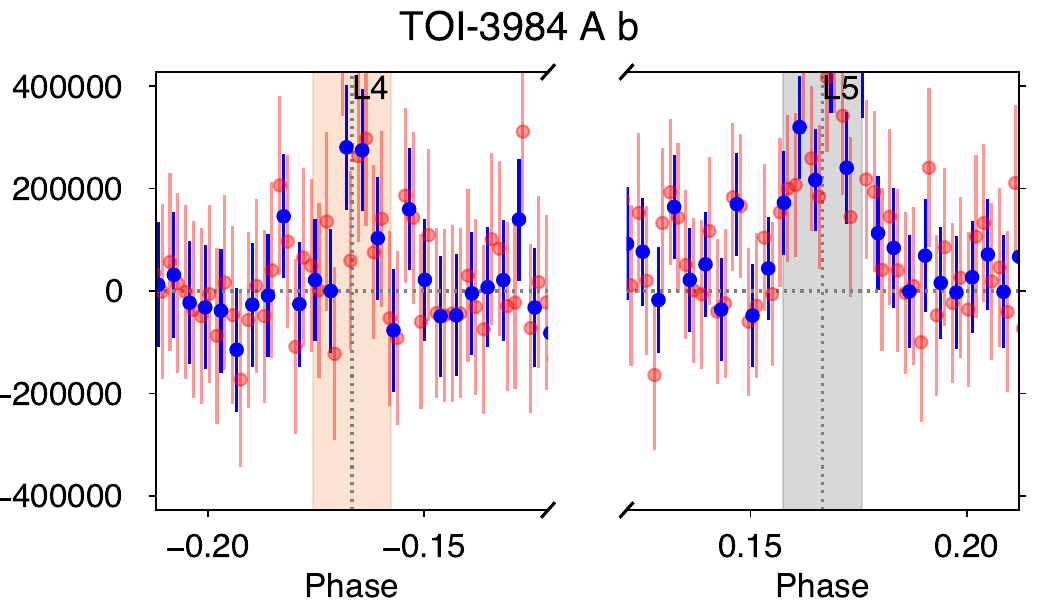}
\includegraphics[width=0.48\textwidth]{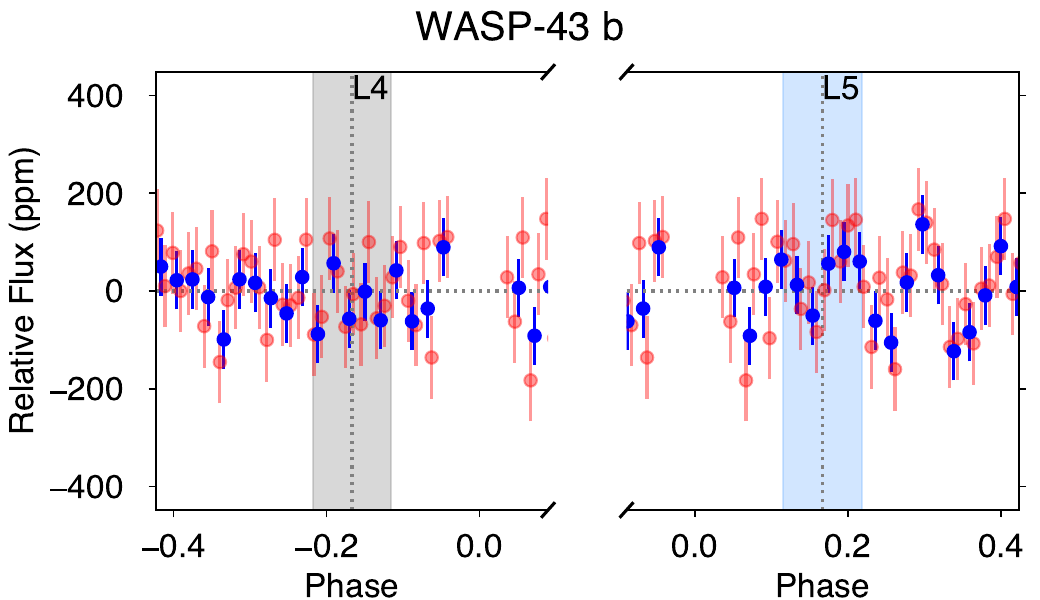}
\includegraphics[width=0.48\textwidth]{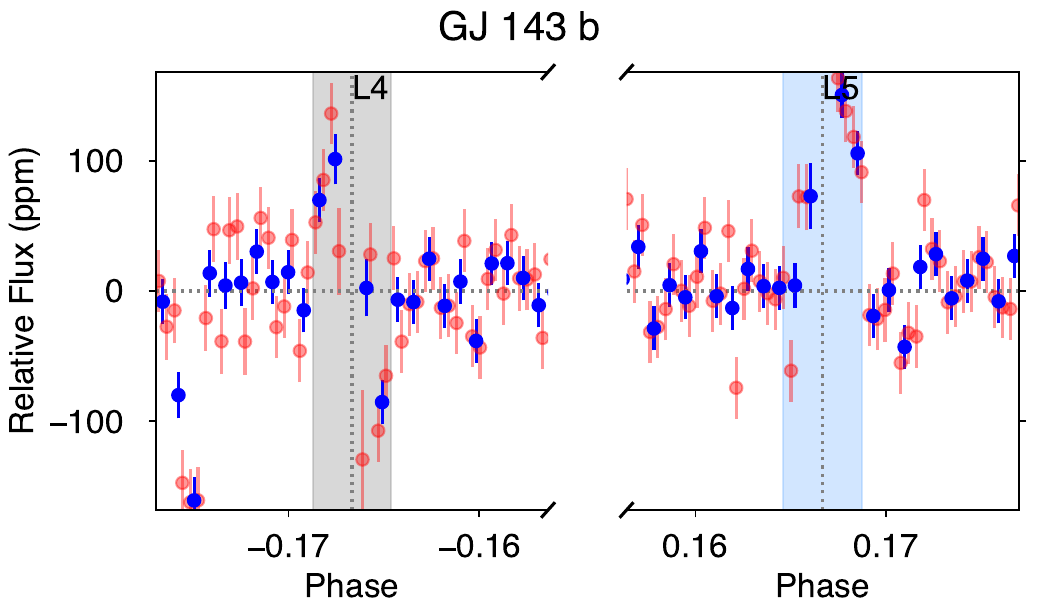}
\end{center}
 \caption{Continuation.}
 \label{fig:LC+}
\end{figure*}

\end{appendix}
\end{document}